\documentclass[manuscript]{aa}

\usepackage{graphicx}
\usepackage[intlimits,sumlimits]{amsmath}
\usepackage{deluxetable}
\usepackage{times,epsfig} 
\usepackage{amssymb}
\usepackage{mathrsfs}  
\usepackage{natbib}

\bibpunct{(}{)}{;}{a}{}{,} 
\usepackage{caption}

\usepackage{amsmath}
\usepackage[english]{babel}
\usepackage{longtable}
\usepackage{url}
\usepackage{hyperref}

\newcommand{\beqa}{\begin{eqnarray}} 
\newcommand{\eeqa}{\end{eqnarray}}

\newcommand{\bsub}{\begin{subequations}}
\newcommand{\esub}{\end{subequations}}
\newcommand{\beal}{\begin{align}}
\newcommand{\ealn}{\end{align}}

\authorrunning{Taddia et al.}
\titlerunning{Early-time light curves of SDSS-II SNe~Ib/c}

\begin{document}

\title{Early-time light curves of Type Ib/c supernovae from the SDSS-II Supernova Survey}
\author{
F. Taddia\inst{1} 
\and J. Sollerman\inst{1}
\and G. Leloudas\inst{2,3}
\and M.~D. Stritzinger\inst{4}
\and S. Valenti\inst{5,6}
\and L. Galbany\inst{7,8}
\and R. Kessler\inst{9,10}
\and D.~P. Schneider\inst{11,12}
\and J.~C. Wheeler\inst{13}}

\institute{
Department of Astronomy, The Oskar Klein Centre, Stockholm University, AlbaNova, 10691 Stockholm, Sweden.\\ (\email{francesco.taddia@astro.su.se)}
\and The Oskar Klein Centre, Department of Physics, Stockholm University, AlbaNova, 10691 Stockholm, Sweden.
\and Dark Cosmology Centre, Niels Bohr Institute, University of Copenhagen, Juliane Maries Vej 30, 2100 Copenhagen, Denmark.
\and  Department of Physics and Astronomy, Aarhus University, Ny Munkegade 120, DK-8000 Aarhus C, Denmark.
\and Las Cumbres Observatory Global Telescope Network, 6740 Cortona Dr., Suite 102, Goleta, CA 93117, USA.
\and Department of Physics, University of California, Santa Barbara, Broida Hall, Mail Code 9530, Santa Barbara, CA 93106-9530, USA.
\and Millennium Institute of Astrophysics, Universidad de Chile, Santiago, Chile.
\and Departamento de Astronom\'ia, Universidad de Chile, Casilla 36-D, Santiago, Chile.
\and Department of Astronomy and Astrophysics, The University of Chicago, 5640 South Ellis Avenue, Chicago, IL 60637, USA.
\and Kavli Institute for Cosmological Physics, The University of Chicago, 5640 South Ellis Avenue, Chicago, IL 60637, USA.
\and Department of Astronomy and Astrophysics, The Pennsylvania State University, University Park, PA 16802, USA.
\and Institute for Gravitation and the Cosmos, The Pennsylvania State University, University Park, PA 16802, USA.
\and Department of Astronomy, University of Texas at Austin, Austin, TX 78712, USA.
}


\date{Received; Accepted}

\abstract
{Type Ib/c supernovae (SNe~Ib/c) have been investigated in several single-object studies; however, there is still a paucity of works concerning larger, homogeneous samples of these hydrogen-poor transients, in particular regarding the premaximum phase of their light curves.}
{In this paper we present and analyze the early-time optical light curves (LCs, $ugriz$) of 20 SNe~Ib/c 
from the Sloan Digital Sky Survey (SDSS) SN survey II, aiming to study their observational and physical properties, as well as to derive their progenitor parameters.}
{High-cadence, multiband LCs are fitted with a functional model and the best-fit parameters are compared among the SN types.
Bolometric LCs (BLCs) are constructed for the entire sample. We also computed the black-body (BB) temperature (T$_{\rm BB}$) and photospheric radius (R$_{\rm ph}$) evolution for each SN via BB fits on the spectral energy distributions. In addition, the bolometric properties are compared to both hydrodynamical and analytical model expectations.}
{Complementing our sample with literature data, we find that SNe Ic and Ic-BL (broad-line) have shorter rise times than those of SNe~Ib and IIb. The decline rate parameter, $\Delta m_{15}$, is similar among the different subtypes. SNe~Ic appear brighter and bluer than SNe~Ib, but this difference vanishes
if we consider host galaxy extinction corrections based on colors. Templates for SN~Ib/c LCs are presented. Our SNe have typical T$_{\rm BB}$ of $\sim$10000~K at the peak and R$_{\rm ph}$ of $\sim$10$^{15}$~cm.
Analysis of the BLCs of SNe~Ib and Ic gives typical ejecta masses M$_{ej}$~$\approx$~3.6$-$5.7~$M_{\odot}$, energies E$_{K}$~$\approx$~1.5$-$1.7$\times$10$^{51}$~erg, and M($^{56}$Ni)~$\approx$~0.3~$M_{\odot}$. Higher values for E$_{K}$ and M($^{56}$Ni) are estimated for SNe~Ic-BL (M$_{ej}$~$\approx$~5.4~$M_{\odot}$, E$_{K}$~$\approx$~10.7$\times$10$^{51}$~erg, M($^{56}$Ni)~$\approx$~1.1~$M_{\odot}$). For the majority of SNe~Ic and Ic-BL, we can put strong limits ($<$2$-$4 days) on the duration of the expected early-time plateau. Less stringent limits can be placed on the duration of the plateau for the sample of SNe~Ib. In the single case of SN~Ib~2006lc, a $>$5.9~days plateau seems to be detected. The rising part of the BLCs is reproduced by power laws with index $<$2. For two events (SN~2005hm and SN~2007qx), we find signatures of a possible shock break-out cooling tail.} 
{Based on the limits for the plateau length and on the slow rise of the BLCs, we find that in most of our SNe~Ic and Ic-BL the $^{56}$Ni is mixed out to the outer layers, suggesting that SN~Ic progenitors are de facto helium poor. The derived progenitor parameters ($^{56}$Ni, E$_{K}$, M$_{ej}$) are consistent with previous works.}


\maketitle

\section{Introduction}

\vspace{0.5cm}

Recent supernova (SN) surveys are dramatically changing the landscape in observational SN research. Not only are we finding many new kinds of explosions (e.g., super-luminous SNe, \citealp{galyam12}), but we are also able to study them over a wider 
parameter space in time and wavelength. One particular aspect in modern time-domain astronomy is the 
possibility to systematically discover stellar explosions at an extremely early stage. These detections have hitherto been done rather 
serendipitously, often for nearby objects. 

In this paper we focus on the early phases (premaximum light) of Type Ib/c SNe (SNe~Ib/c). These classes of SNe 
lack hydrogen lines in their spectra. Strong helium lines characterize the spectra of SNe~Ib, whereas SNe~Ic 
are also helium-poor \citep[e.g.,][]{filippenko97}. A subclass of SNe~Ic (Ic-broad line, Ic-BL) shows high 
expansion velocities ($\sim$3$\times$10$^{4}$~km~s$^{-1}$ a few days after explosion). These SNe are often associated with long-duration gamma-ray bursts \citep[GRBs, e.g.,][]{galama98}.
Transitional events called SNe~IIb present hydrogen lines in their early spectra, but at the maximum light, these objects exhibit the characteristic helium lines of SNe~Ib (e.g., SN~1993J, \citealp{filippenko93}; SN~2011dh, \citealp{ergon14}).

The sample of  SNe~Ib/c in the literature is still relatively small{\footnote{During the refereeing process, the light curves of 64 SE-SNe and optical spectra of 73 SE-SNe were released by \citet{bianco14} and \citet{modjaz14}, respectively.}}. 
\citet{drout11} present a compilation of $V$- and $R$-band light curves 
for 25 stripped-envelope (SE) SNe. There
are also several studies that present optical (and sometimes near 
infrared) light curves and spectra for single events. Recently, \citet{cano13} has compiled and modeled
the photometric data for 61 SNe~Ib/c and Ic-BL in the literature. \citet{richardson14} present absolute $B$-band peak magnitudes for 69 SNe~Ib/c and IIb.

 The early emission from the resulting SNe is powered by both the shock energy and energy from the radioactive decay of $^{56}$Ni (see, e.g., \citealp{piro13}, hereafter PN13, for a pedagogical introduction).
The first electromagnetic signal from a SN~Ib/c is the actual shock breakout (e.g., SN~2008D, \citealp{soderberg08}, although see \citealp{mazzali08} for a different interpretation). In the optical regime 
this feature is observable only in the cooling phase, and is typically short-lived for compact stars. It has been predicted that a plateau 
phase, which could be rather faint, follows this stage (\citealp{dessart12_OriginIbc,bersten13}; PN13), until the 
radioactive decay energy starts the light curve rise in earnest.
SNe~Ib/c exhibit light curves that rise to peak in just $\sim$15$-$20 days. 
The current information on the rise times of these explosions was summarized by \citet{valenti11}. 
Their compilation of 16 SNe suggested that there is a diversity in rise times 
and that helium-rich (IIb/Ib) SNe may have longer rise times, because of either  $^{56}$Ni mixed to outer layers (PN13) and/or higher ejecta masses.

The two leading scenarios for the nature of SN~Ib/c progenitor stars are those that either consider a single, 
massive (M$_{ZAMS}\gtrsim$~25~M$_{\odot}$, \citealt{massey03}) Wolf-Rayet (WR) star, 
which sheds its hydrogen envelope through a prevalent line-driven wind or, alternatively, a 
binary system where a less massive progenitor loses its hydrogen-rich envelope also via tidal stripping from a companion. Studies of the environments of SNe~Ib/c \citep{modjaz11,kuncarayakti13} could suggest
differences in local metallicity between SNe~Ib and Ic, with the former showing lower metallicities. This result would indicate that metal-line winds play an important role in distinguishing SN~Ib and Ic progenitors, favoring the single-star scenario.
However, this result is not confirmed by other works (\citealp{anderson10}; \citealp{leloudas11}; 
\citealp{sanders12}). More and better data is needed to differentiate a single-star from a binary origin on the basis of different metallicities.
Considerations of SN rates \citep{smith11_rates} and age constraints on SN locations \citep{leloudas11} do indicate that at least some SNe~Ib/c 
originate in binary systems. For SNe~IIb there is direct evidence that their
progenitors belong to binary systems (e.g., SNe~1993J, \citealp{maund04}). Recently, light-curve modeling of 
SN~Ib iPTF13bvn \citep{fremling14} and SN~IIb 2011dh \citep{ergon14,ergon14b} (both SNe have pre-explosion detection of the progenitor star) revealed low-mass progenitor stars for these transients, implying that they belong to binary systems.

In this paper we present the light curves of a sample of 20 SNe~Ib/c in their photospheric phase obtained by the Sloan Digital Sky Survey (SDSS-II) SN survey \citep{york00,sako08,frieman08,sako14}. The properties of the sample are quantified and compared among the spectral subclasses.
Progenitor properties are 
estimated from the bolometric light curves that we build from the multiband ($ugriz$) data.
The cadence of the SDSS-II SN survey allows the discovery of SNe soon after explosion, thereby acting as a forerunner to ongoing and future surveys like iPTF\footnote{\href{http://www.ptf.caltech.edu/iptf}{http://www.ptf.caltech.edu/iptf}} and ZTF\footnote{\href{http://www.ptf.caltech.edu/ztf}{http://www.ptf.caltech.edu/ztf}}
(intermediate Palomar Transient Factory and Zwicky Transient Facility; \citealp{kulkarni12a}), 
where the rationale is to increase the cadence, thereby systematically 
discovering and monitoring transients at very early stages. 
Surveys like PTF have in fact already discovered a few events soon after explosion, such 
as SN~2011dh \citep{arcavi11}. 
The detection of SNe~Ib/c in the early stages is important
for probing the explosion physics by investigating of the $^{56}$Ni mixing, as well as the
properties of the progenitors from constraints on the
stellar radius.
These discoveries have focused recent theoretical interest
toward the early light curves of SE~CC~SNe 
(PN13, \citealp{dessart12_OriginIbc}), and our aim here is to complement these works
by exploring what can be done with a larger observational sample
of such discoveries, in particular when multiband data are available.  

This paper is organized as follows. 
In Sect.~\ref{sec:data} we introduce the dataset from the SDSS-II SN survey 
used in this paper and describe the observations and reductions. The analysis of 
these light curves is given in Sect.~\ref{sec:analysis}, and then a more general discussion  is provided in Sect.~\ref{sec:discussion}. Our conclusions and future outlook are provided in Sect.~\ref{sec:conclusions}.
 
\section{Data from SDSS-II}
\label{sec:data}

\onlfig{1}{
\clearpage
\begin{figure}
\centering
\includegraphics[width=8cm]{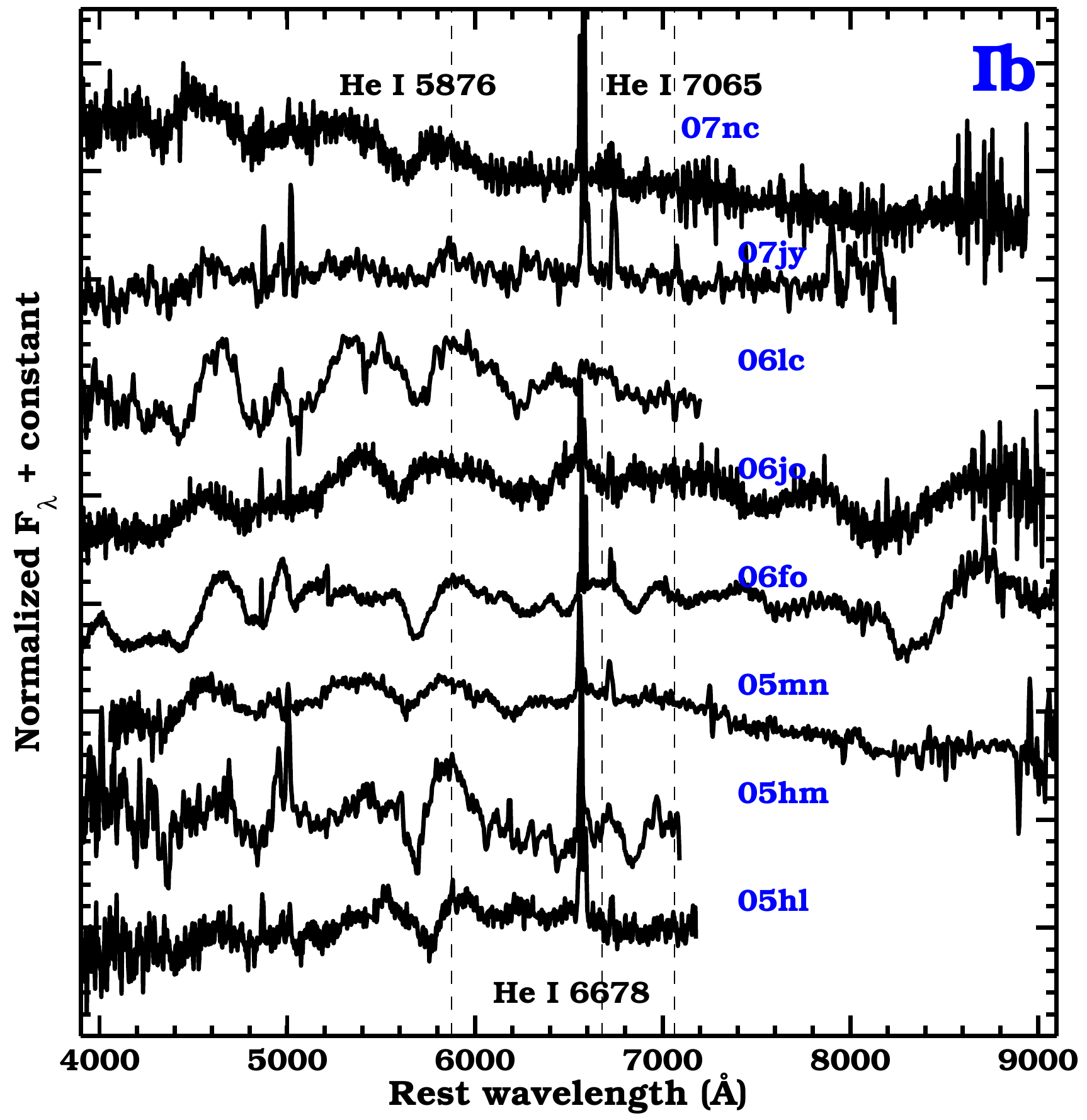} \\
\includegraphics[width=8cm]{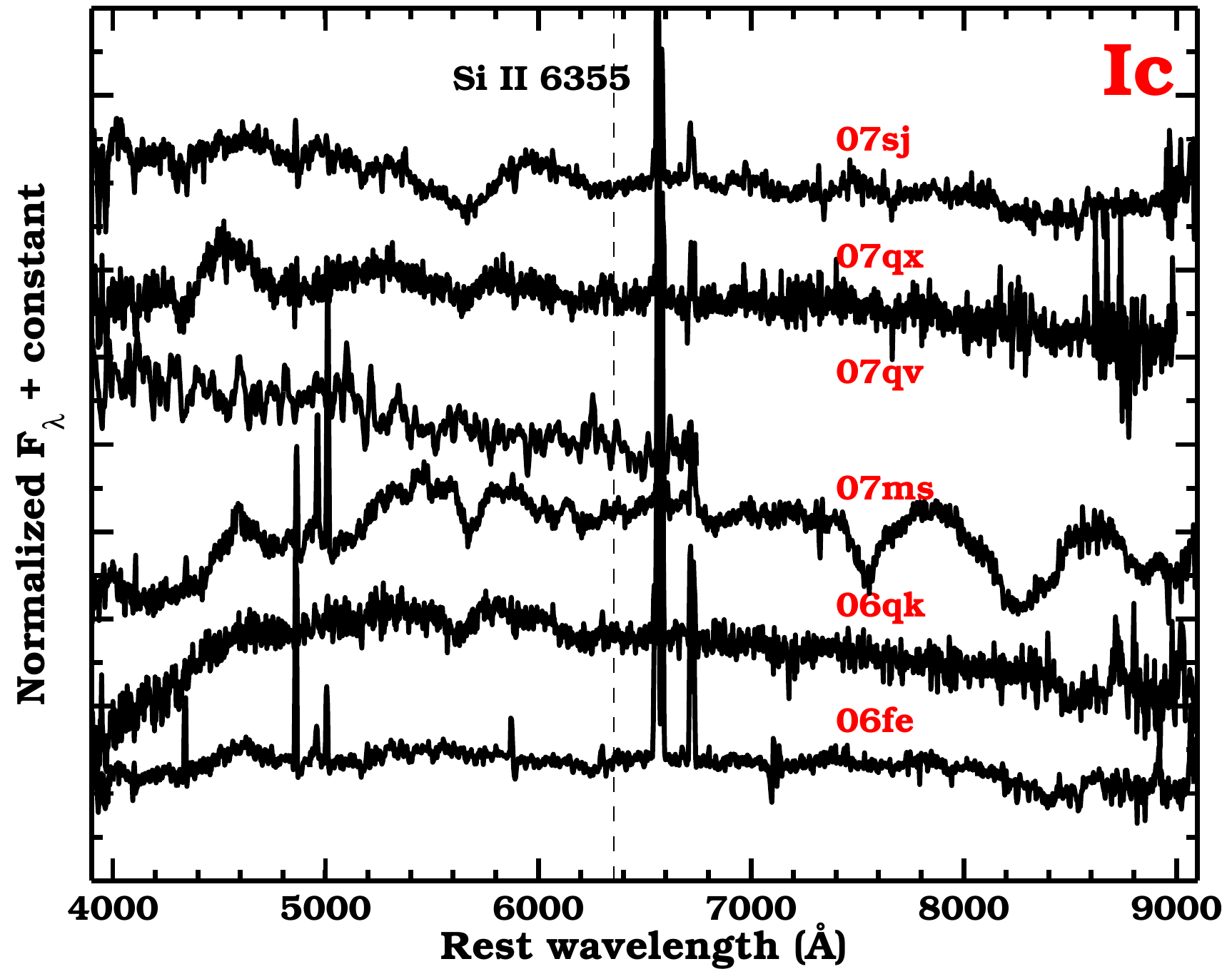} \\ 
\includegraphics[width=8cm]{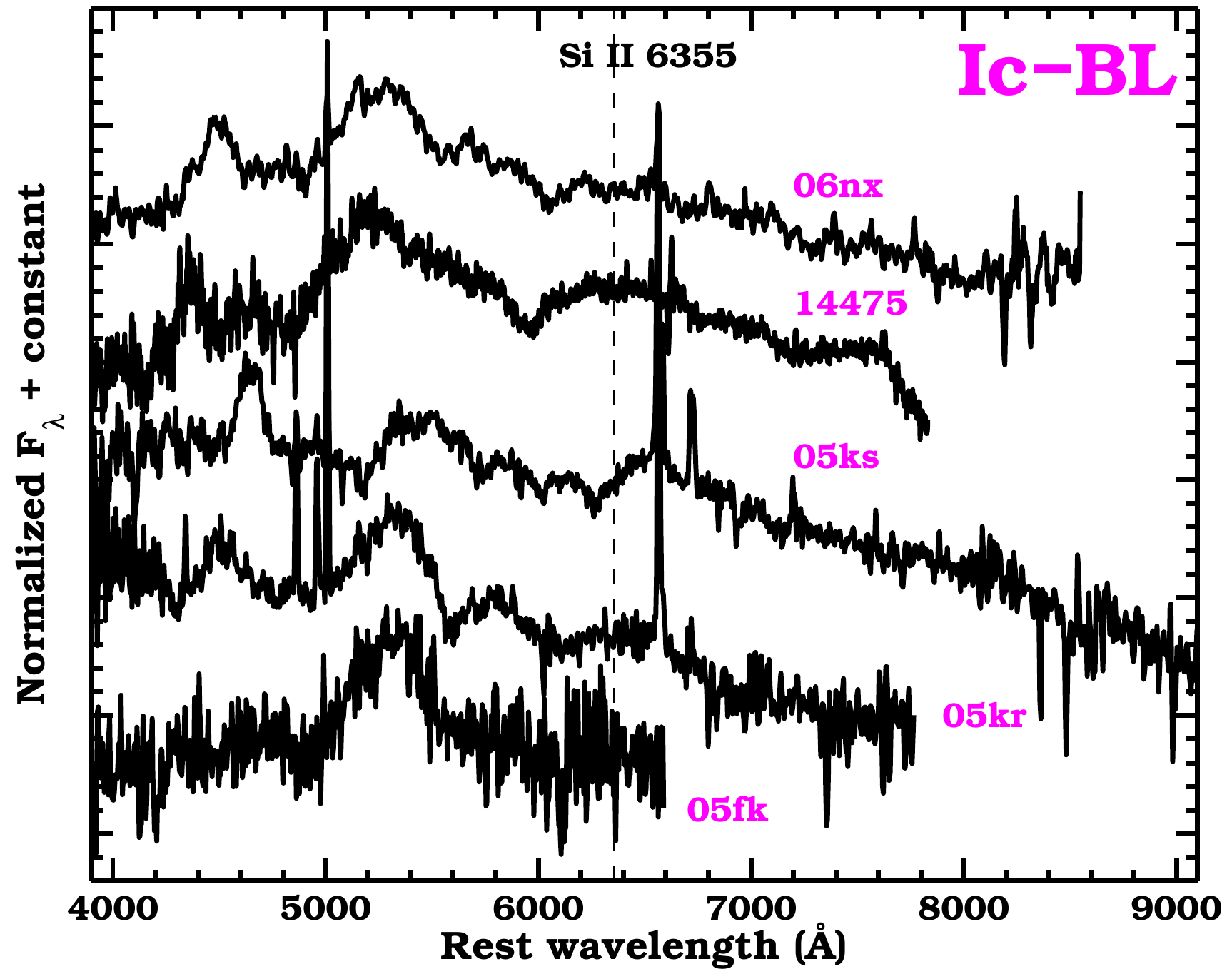}  
 \caption{{\label{spec_class}}Selected spectra (data from \citealt{sako14}) for 19 SDSS SNe~Ib/c 
(SN~2007gl was classified by \citealt{baltay07}).
The vertical dashed lines indicate the rest wavelengths of \ion{He}{i}~$\lambda\lambda$5876, 6678, 7065, and \ion{Si}{ii}~$\lambda$6355 at zero velocity. The epoch of each spectrum is marked in Fig.~\ref{lc_r} with a vertical red segment. Most of the spectra show narrow emission lines from the host galaxy.}
\end{figure}}

\begin{figure*}
\centering
\includegraphics[width=13cm,angle=0]{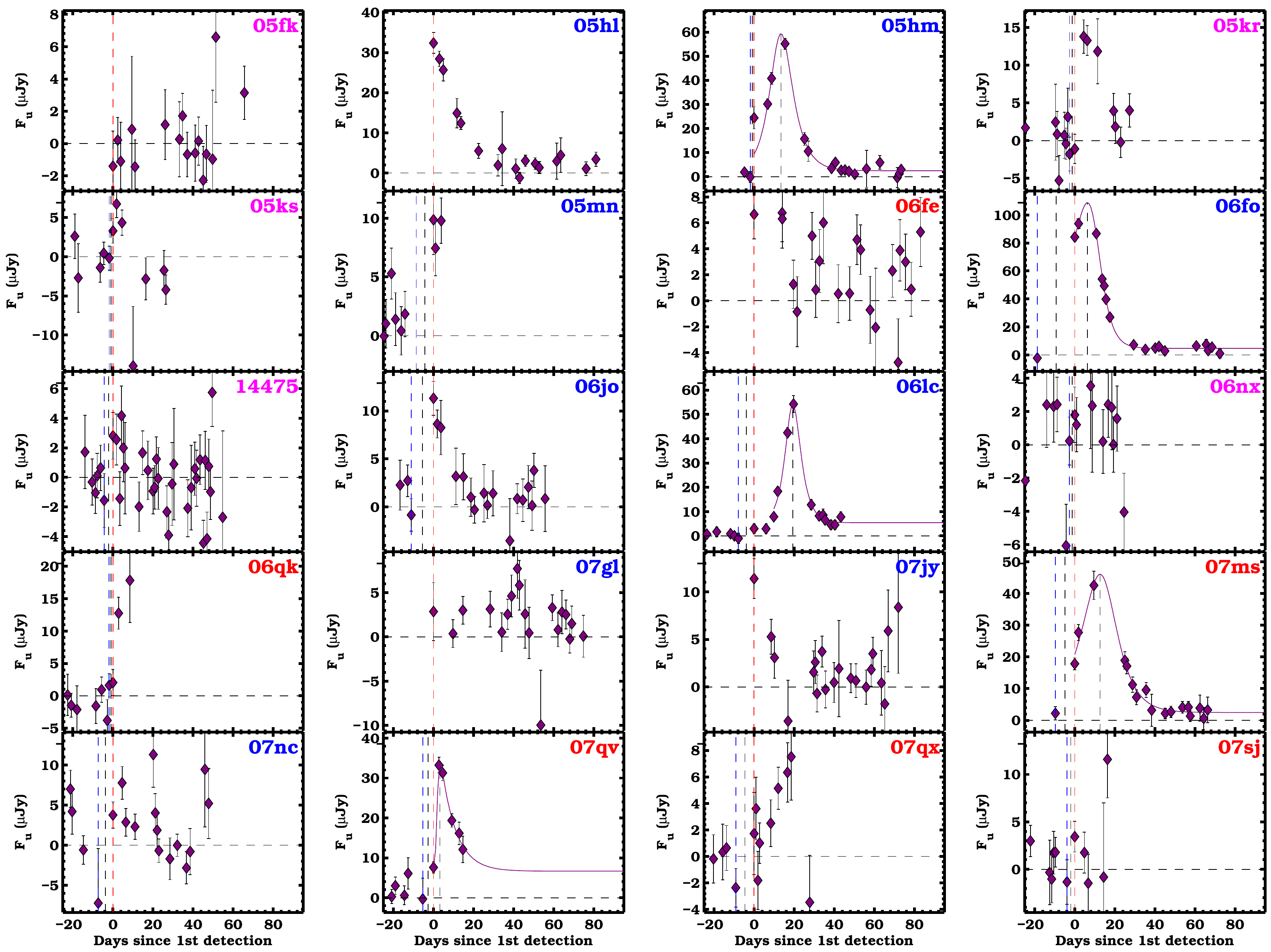}  
 \caption{ $u$-band light curves of 20 SDSS SNe~Ib, Ic, and Ic-BL. Best fits on the SN light curves observed before peak are shown by a violet solid line. Vertical blue and red dashed lines mark the last non-detection and the first detection epoch. Vertical black dashed lines indicate the derived explosion date and the peak epoch. Light curves are shown after correction for time dilation. Blue, red, and magenta labels correspond to SNe~Ib, Ic, and Ic-BL, respectively. \label{lc_u} }
\end{figure*}

\begin{figure*}
\centering
\includegraphics[width=13cm,angle=0]{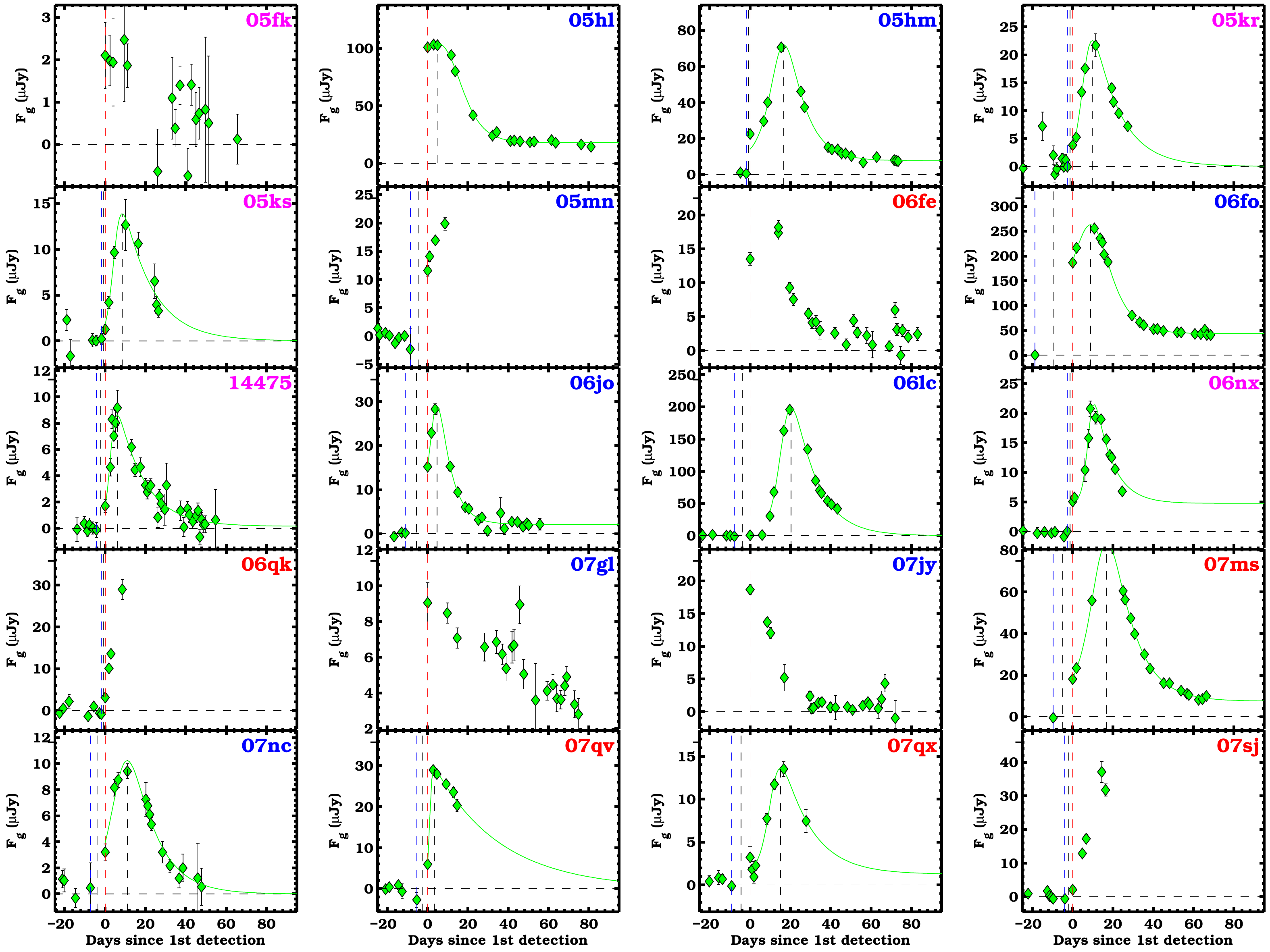}  
 \caption{ $g$-band light curves of 20 SDSS SNe~Ib, Ic, and Ic-BL. Best fits on the SN light curves observed before the peak are shown by a green solid line. Vertical blue and red dashed lines mark the last non-detection and the first detection epochs. Vertical black dashed lines indicate the derived explosion date and the peak epoch. Light curves are shown after correction for time dilation. Blue, red, and magenta labels correspond to SNe~Ib, Ic, and Ic-BL, respectively. \label{lc_g} }
\end{figure*}

\begin{figure*}
\centering
\includegraphics[width=13cm,angle=0]{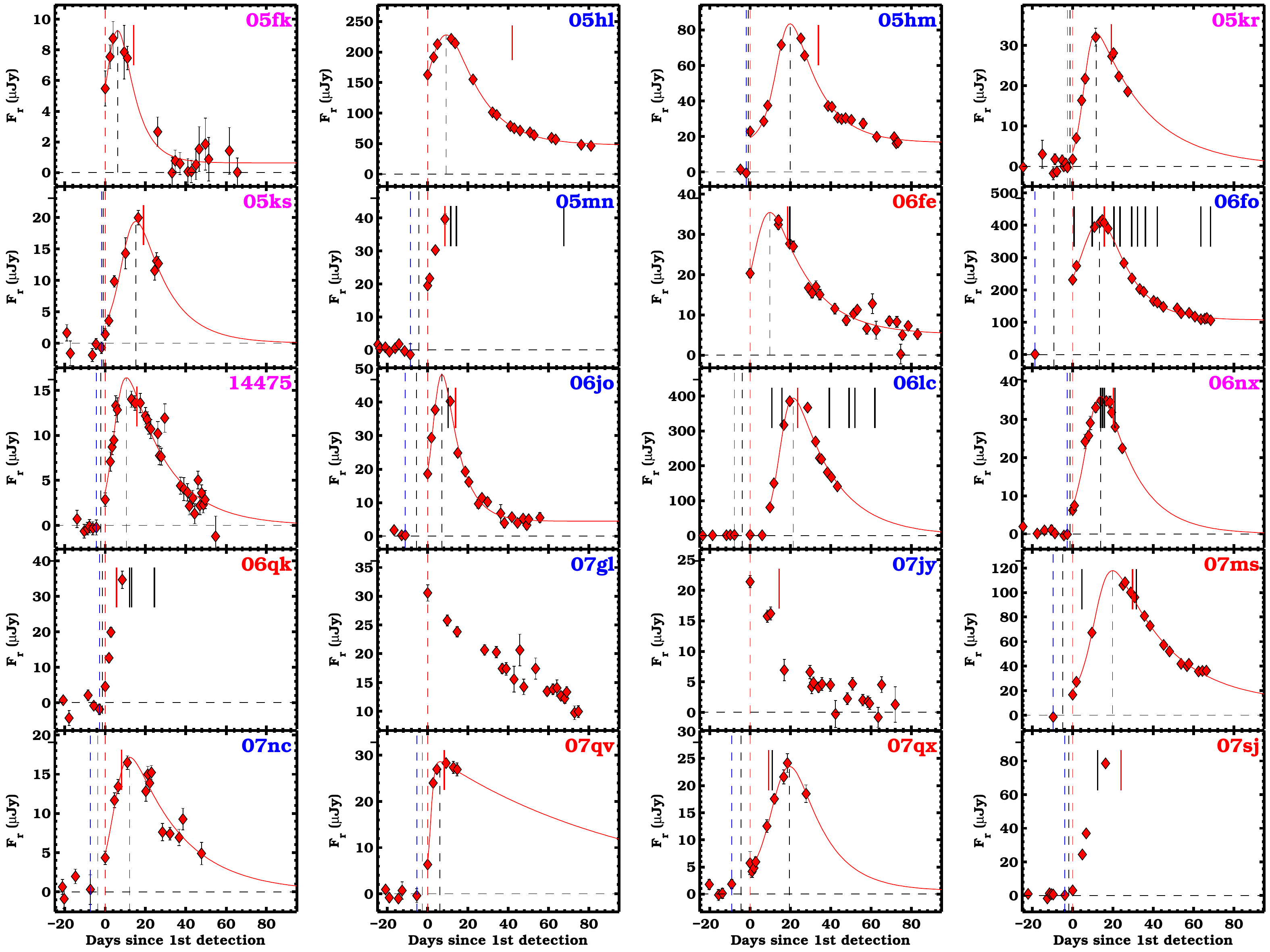}  
 \caption{ $r$-band light curves of 20 SDSS SNe~Ib, Ic, and Ic-BL. Best fits on the SN light curves observed before the peak are shown by a red solid line. Vertical blue and red dashed lines mark the last non-detection and the first detection epochs. Vertical black dashed lines indicate the derived explosion date and the peak epoch. Black and red vertical segments mark the spectral epochs; in particular, the red ones indicate the epoch of the spectra shown in Fig.~\ref{spec_class}. Light curves are shown after correction for time dilation. Blue, red, and magenta labels correspond to SNe~Ib, Ic, and Ic-BL, respectively. \label{lc_r} }
\end{figure*}

\begin{figure*}
\centering
\includegraphics[width=13cm,angle=0]{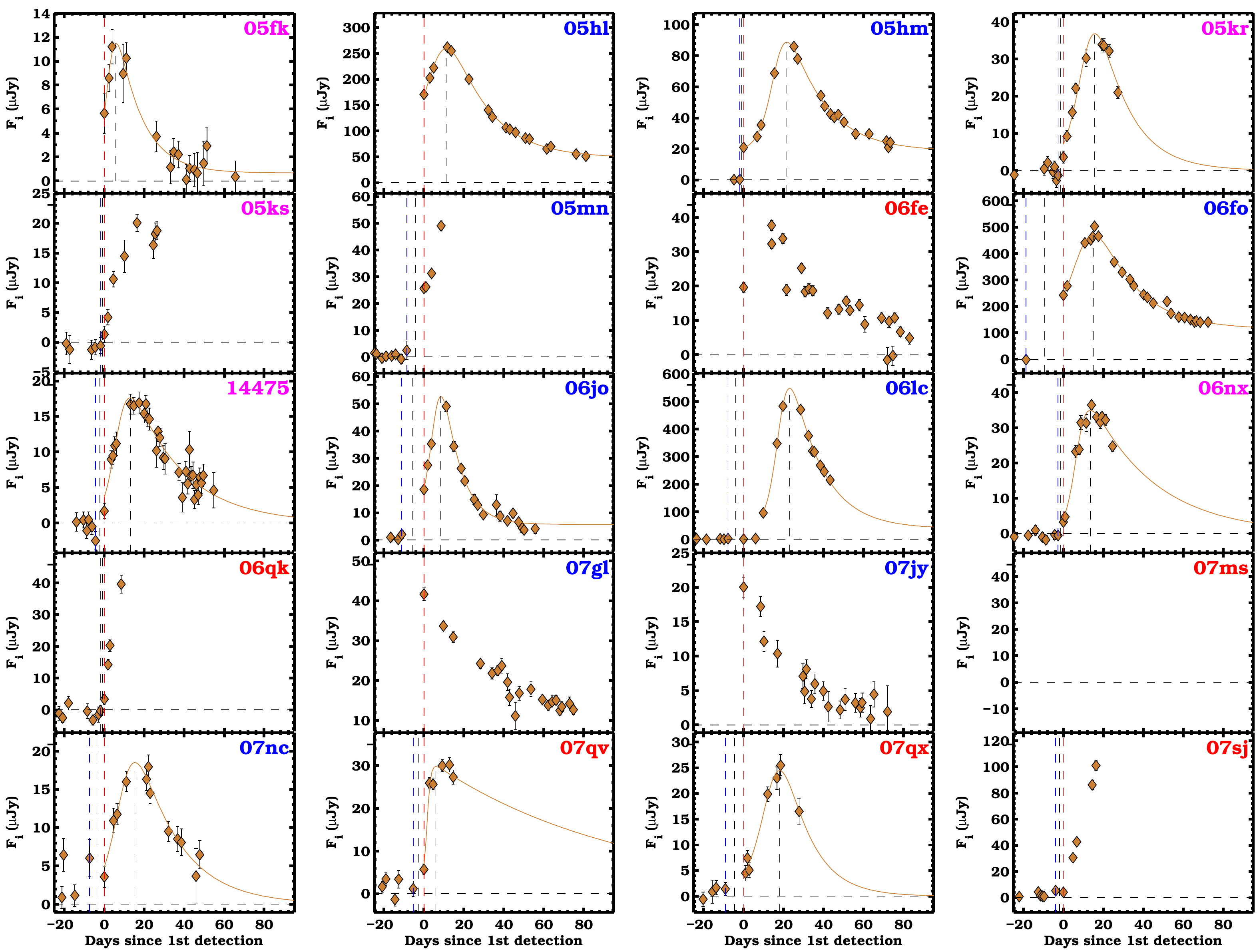}  
 \caption{ $i$-band light curves of 20 SDSS SNe~Ib, Ic, and Ic-BL. Best fits on the SN light curves observed before peak are shown by a dark yellow solid line. Vertical blue and red dashed lines mark the last non-detection and the first detection epoch. Vertical black dashed lines indicate the derived explosion date and the peak epoch. Light curves are shown after correction for time dilation. Blue, red, and magenta labels correspond to SNe~Ib, Ic, and Ic-BL, respectively. \label{lc_i} }
\end{figure*}

\begin{figure*}
\centering
\includegraphics[width=13cm,angle=0]{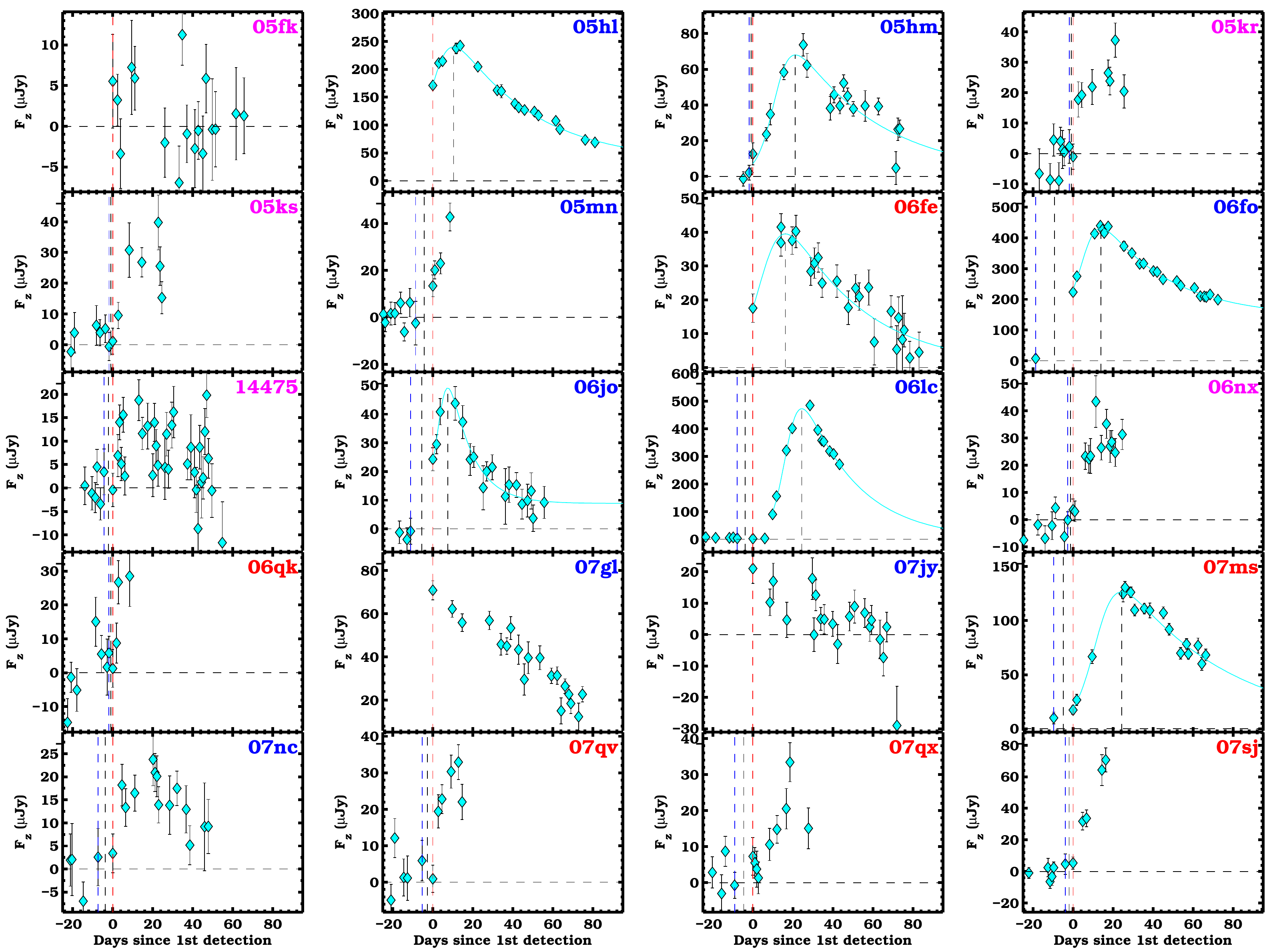}  
 \caption{$z$-band light curves of 20 SDSS SNe~Ib, Ic, and Ic-BL. Best fits on the SN light curves observed before peak are shown by a cyan solid line. Vertical blue and red dashed lines mark the last non-detection and the first detection epoch. Vertical black dashed lines indicate the derived explosion date and the peak epoch. Light curves are shown after correction for time dilation. Blue, red, and magenta labels correspond to SNe~Ib, Ic, and Ic-BL, respectively. \label{lc_z} }
\end{figure*}

The data used in this paper were obtained by SDSS-II and released by \citet{sako14}\footnote{Data are available at:\\ \ \href{http://sdssdp62.fnal.gov/sdsssn/DataRelease/index.html}
{http://sdssdp62.fnal.gov/sdsssn/DataRelease/index.html}.}.
The SDSS-II SN survey operated three fall campaigns in 2005, 2006, and 2007 using the 
2.5~m telescope \citep{gunn06} at Apache Point Observatory to scan 300~deg$^2$ of sky. The cadence was 
in principle as often as every second night, although practical conditions (e.g., weather) often decreased that 
somewhat (see below).

Each photometric observation consisted of nearly simultaneous 55~s exposures in each of the five $ugriz$ 
filters \citep{fukugita96} using the wide-field SDSS CCD camera \citep{gunn98}. High-quality light curves 
were obtained on a photometric system \citep{ivezic04} calibrated to an uncertainty of 1\% \citep{mosher12}. Details on the photometric measurements are provided by \citet{holtzman08} and \citet{sako14}.

The primary aim of the SN survey was to discover and characterize several hundred SNe~Ia at intermediate redshifts, 
in order to fill in the SN redshift desert, located between the low-
and high-$z$ samples for cosmological studies.
 Results based on the first season of SN~Ia data 
were presented in \citet{kessler09}, \citet{sollerman09}, and \citet{lampeitl10}, while \citet{betoule14} make use of the intermediate-redshift (0.05$<z<$0.4) 
SN~Ia data collected during all three SDSS-II seasons.
The SDSS-II SN survey provided a large and uniform sample of SNe, which allowed a detailed study 
of their observational properties.
 The rise times of SNe~Ia were discussed in detail by 
\citet{hayden10a,hayden10b}. The rolling search strategy and the 
high cadence of the SDSS-II SN survey assured that many of the SDSS SNe have 
well-defined, densely sampled light curves, even on the rise portion. 
The data also include pre-explosion flux limits.

Although CC~SNe were not the primary aim of the SDSS-II SN survey, 
many such events were also discovered, monitored, and spectroscopically
confirmed \citep{sako08,sako14}. A sample of nearby SNe~IIP was published by \citet{dandrea10}, 
and some individual peculiar CC~SNe have also been reported previously (e.g., \citealp{leloudas12}). 
\citet{leloudas11} discuss the spectroscopic classification of several of these objects in relation to 
a study of their host galaxies. Here we focus on a set of 20 SDSS-II SNe discovered between 2005 and 2007 in untargeted galaxies, which were spectroscopically confirmed as SNe~Ib/c. The basic information on these SNe are provided in Table\ref{sample}.

The classification of each SN (Table~\ref{sample}, 4th column) is based on spectra taken by the
SDSS-II Supernova Survey team that were 
cross-correlated to SN spectral templates \citep{zheng08}.  
After the SDSS-II SN survey classification, some of the SNe were typed again by \citet{leloudas11}, 
and we adopt the latter classification\footnote{SN~2006lc is considered an intermediate case between Ib and Ic by \citet{leloudas11}. Here we include it in the SN~Ib sample (as also done in \citealp{modjaz14}), as it shows some helium features, although weak.}.
Our sample consists of nine SNe~Ib, six SNe~Ic, and five SNe~Ic-BL. 
SN~2005em, classified as a SN~Ic by \citet{zheng08}, was typed as a SN~IIb by \citet{morrell05} and is not included in our sample. 
We reclassified SN~14475 and SN~2006nx as SNe~Ic-BL based on the best 
SNID \citep{blondin07} fits for the known host galaxy redshifts.
(The 3 other SNe~Ic-BL were classified by \citealp{zheng08}.)
Selected spectra of our targets are displayed in Fig.~\ref{spec_class}.
Many of our SNe have multiple spectra \citep{sako14}, and a spectral log is provided in Table~\ref{tab:speclog}. The spectra shown in Fig.~\ref{spec_class} were selected
based on their quality and phase (their epochs are marked by vertical red lines in 
Fig.~\ref{lc_r} and by a star in Table~\ref{tab:speclog}), 
aiming to illustrate the spectral differences among the three types (Ib, Ic Ic-BL).
We chose to only display spectra obtained around or after $r$-band maximum (which was possible  in all but a few cases) in order to 
clearly show which SNe are helium-rich and which are helium-poor. As already noted by \citet{leloudas11}, the detailed subclassification of SNe~Ib and Ic that was made from the SDSS-II spectra is sometimes problematic.

The SN redshifts, usually obtained from the host galaxy spectra \citep{zheng08}, range from 0.016 (SN~2006lc)  
to 0.264 (SN~2005fk). 
On average, the SNe~Ic and SNe~Ib discovered by SDSS-II were located at similar redshifts 
($\rm \overline{z}_{Ic}$~$=$~0.064$\pm$0.023, $\rm \overline{z}_{Ib}$~$=$~0.057$\pm$0.052), 
whereas SNe~Ic-BL were typically 
found at somewhat higher redshifts ($\rm \overline{z}_{Ic-BL}$~$=$~0.156$\pm$0.063). 

Luminosity distances were computed from the redshifts assuming WMAP (5-years) 
cosmological parameters \citep[H$_0$~$=$~70.5~km~s$^{-1}$~Mpc$^{-1}$, $\Omega_{m}=0.27$, $\Omega_{\Lambda}=0.73$,][]{komatsu08}, using the cosmology calculator by \citet{wright06} via 
NED\footnote{\href{http://ned.ipac.caltech.edu}{http://ned.ipac.caltech.edu}}.
The visual extinction due to our Galaxy is taken from \citet{schlafly11} via 
NED and is also listed 
in Table~\ref{sample}. 

The reported $g$-band absolute magnitude ($M_g^{gal}$) for each 
host galaxy was obtained through the apparent magnitude listed in \citet{hakobyan12} or 
in the SDSS catalog\footnote{\href{http://skyserver.sdss3.org}{http://skyserver.sdss3.org}}. We K-corrected the galaxy magnitudes 
using the $g-r$ color of each galaxy\footnote{\href{http://kcor.sai.msu.ru/about/}{http://kcor.sai.msu.ru/about/}} \citep{kc1,kc2}.
The absolute galaxy magnitudes range from $M_g^{gal}=-$15.5~mag to $M_g^{gal}=-$21.2~mag. The host galaxies of SNe~Ic-BL 
appear slightly fainter ($<M_g^{gal}(Ic-BL)>$~$=$~$-$18.82$\pm$0.43~mag) than those of SNe~Ib 
($<M_g^{gal}(Ib)>$~$=$~$-$19.61$\pm$0.56~mag) and Ic ($<M_g^{gal}(Ic)>$~$=$~$-$19.55$\pm$0.58~mag).

Light curves for each SN and for each filter ($ugriz$) are shown in 
Figs.~\ref{lc_u}-\ref{lc_z}. The fluxes 
are presented in $\mu$Jy, and the plots also include the pre-explosion epochs, 
where the fluxes are consistent with zero emission. 
For each light curve, the temporal axis is stretched by the factor 1$/$(1$+z$) in order to 
correct for time dilation. The times in this paper are all in the rest frame of the supernova unless
stated otherwise.
In Figs.~\ref{lc_u}-\ref{lc_z}, we also indicate the last epoch of non-detection and the first detection. The first detection corresponds to the first epoch on the light curve rise when the $g$ and/or $r$-band flux with its 1$\sigma$ error is not compatible with zero emission. The epochs of the spectra are marked by vertical segments in the $r$-band 
plot. The first spectral observation occurred before $r$-band maximum for eight out of the 20 SNe.

\onlfig{7}{
\begin{figure}
\centering
$\begin{array}{cc}
\includegraphics[width=8cm]{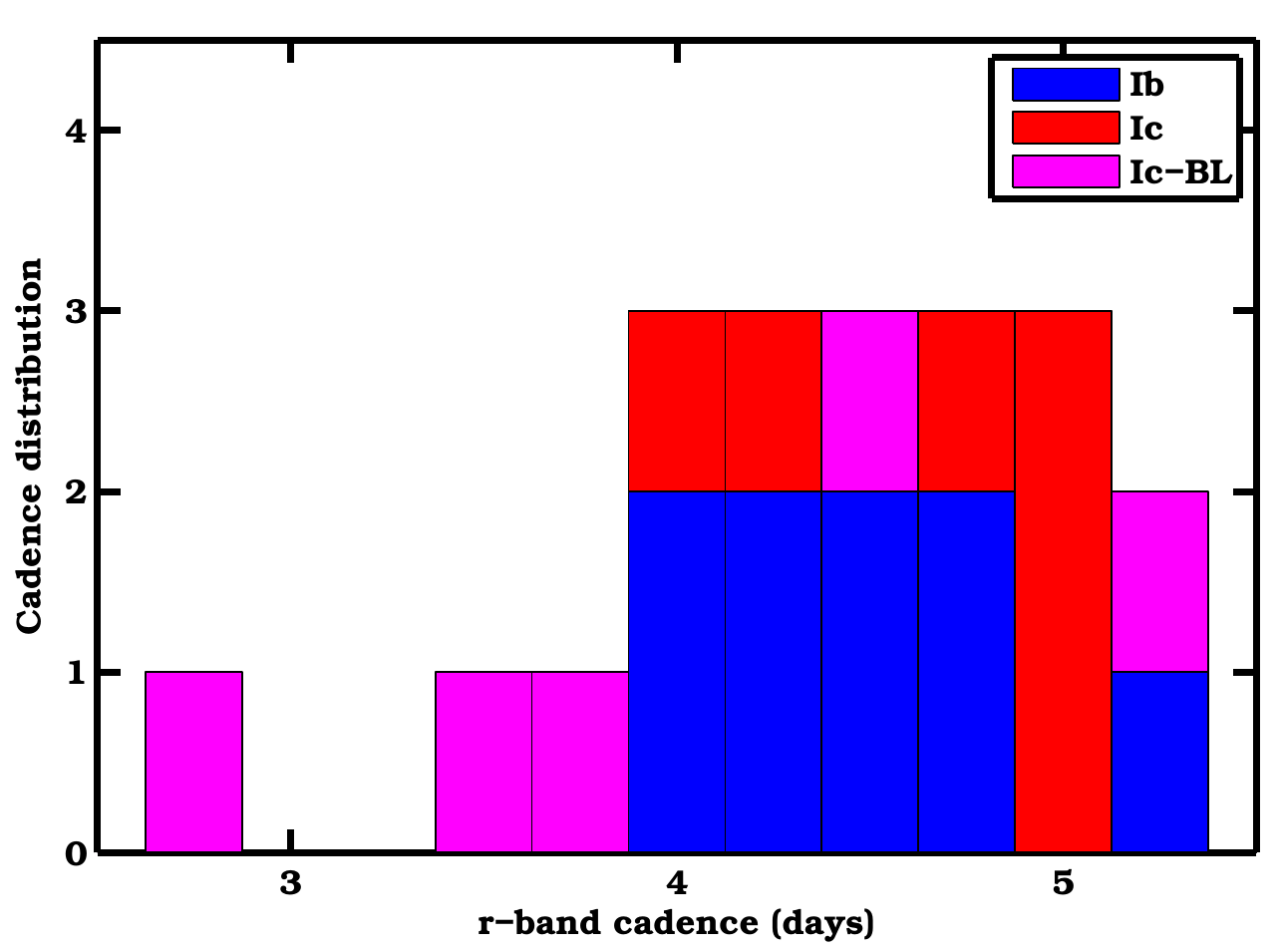}  
\includegraphics[width=8cm]{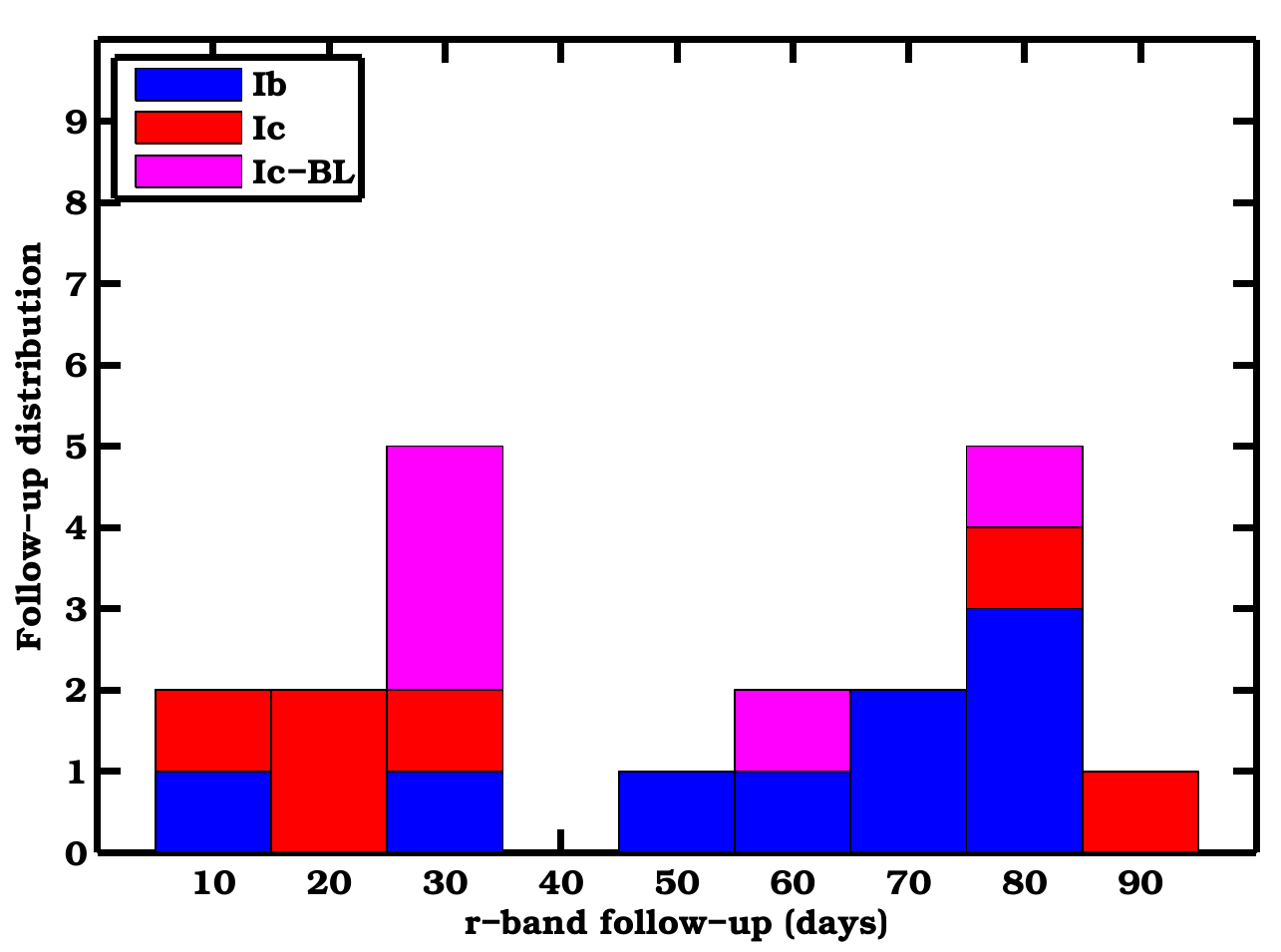}
\end{array}$
 \caption{\label{cadence}({\it Lefthand panel:}) $r$-band cadence distribution for the SDSS SN~Ib/c sample. Typical values are between 4 and 5 days. ({\it Righthand panel:}) $r$-band follow-up distribution for the SDSS SN~Ib/c sample. SNe were followed for a minimum of $\sim$10 days to a maximum of $\sim$90 days after discovery.}
\end{figure}}

The average observed cadence of the SN photometric observations is 4.4$\pm$0.6 days. In 
Fig.~\ref{cadence} (left panel) the distribution for the $r$-band cadence shows that 
SNe~Ib and Ic were observed with similar cadence.  
The righthand panel of Fig.~\ref{cadence} presents the distribution of the follow-up 
duration (in $r$ band), showing that these SNe were observed on average for 51$\pm$27 
days, with a minimum duration of nine days (SN~2005mn) and a maximum of 89 days 
(SN~2006fe). Therefore most of these SNe~Ib/c are observed in their photospheric phase.

The $r$-band light curves (Fig.~\ref{lc_r}) show that 15 objects were observed both before and after maximum and that 12 of them also have pre-explosion images (15 objects in total have pre-explosion images).
In summary, this SDSS data set of SNe~Ib/c is well-suited to studying these fast transients given its unique high cadence, its multiband coverage and the availability of pre-explosion images. 
 
\section{Analysis}
\label{sec:analysis}

\subsection{Light curve fit}
\label{sec:fit}

\begin{figure}[h]
\centering
\includegraphics[width=9cm,angle=0]{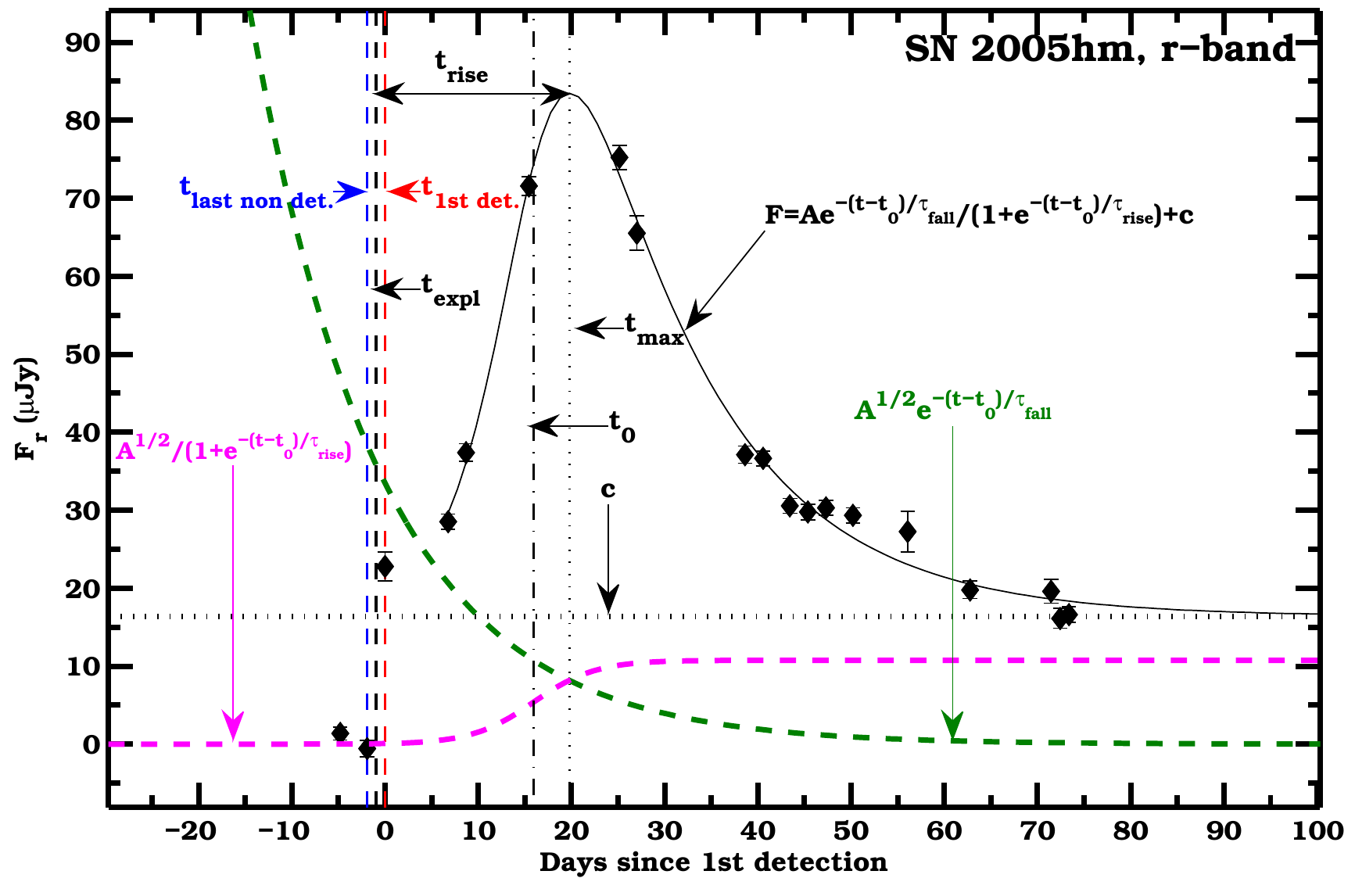}  
 \caption{ $r$-band light curve of SN 2005hm fit by the model given in Eq.~\ref{fitformula}. We label the different parameters that we discuss in the paper and show the main components of the model that produce a fast rising and exponential declining light curve. The light curve is shown after correction for time dilation.\label{fitexample}}
\end{figure}

To quantitatively compare the SN light curves, we performed fits with the phenomenological model 
used by \citet{bazin11}. In this model the flux $\rm F(t)$ in each filter is given by

\begin{equation}
\rm F(t)\;=\; A \frac{e^{-(t-t_0)/\tau_{\rm fall}}}
{1+e^{-(t-t_0)/\tau_{\rm rise}}} \;+ c.
\label{fitformula}
\end{equation}

\noindent To illustrate the different parameters that Eq.~\ref{fitformula} includes and the light curve properties that we derive and discuss in the next sections, 
Fig.~\ref{fitexample} presents an example of this model fit to
the $r$-band light curve of SN~2005hm.
  
The model is shown by a black solid line,
which smoothly fits the flux at each epoch after first detection. 
In this specific case, we excluded the epoch of first detection from the fit 
(indicated by a vertical red dashed line), because it may belong to the early shock break-out cooling tail\footnote{The same was done for the first epoch of SN~2007qx, and the first two epochs of SN~2006lc.}. 
The last non-detection is denoted by a vertical dashed blue line. The explosion
epoch is assumed here to be the mean between the last non-detection and the first detection epochs 
($\rm t_{\rm expl}=\frac{1}{2}(t_{\rm~last~non~det.}+t_{\rm 1st~det.})$), and it is shown by a vertical black dashed line
in the figure. 
The uncertainty in this quantity corresponds to half the time interval between those two epochs 
($\rm \Delta t_{\rm expl}=\frac{1}{2}(t_{\rm 1st~det.}-t_{\rm last~non~det.})$).
The light-curve rise time t$_{\rm rise}$ (see the double-sided arrow in the top left corner of Fig.~\ref{fitexample}) 
is then defined as the time interval between this assumed explosion epoch and the time of the peak (t$_{\rm max}$, 
see the black dotted vertical line). The epoch of maximum is measured from the fit and is given by 
$\rm t_{\rm max}^k=t_0+\tau_{\rm rise} \ln (\tau_{\rm fall}/\tau_{\rm rise}-1)$. 

The light curve model can be understood as the combination of two main factors: a function (magenta in Fig.~\ref{fitexample}) with asymptotic values zero (at $\rm t=-\infty$) and $\rm \sqrt{A}$ (at $\rm t=\infty$) that modulates 
an exponential declining function (green in Fig.~\ref{fitexample}) with normalization $\sqrt{A}$. 
The parameter $\rm t_0$ represents both the inflection point for the modulating function and the epoch when 
the exponential function is equal to its normalization.
The product of these two factors plus an additional positive constant $\rm c$ 
(horizontal black dotted line in Fig.~\ref{fitexample}) reproduce the general light curve shape very well, 
with a fast rise followed by an exponential decline. 
The time scales of rise and fall are set by the parameters $\rm \tau_{\rm rise}$ and $\tau_{\rm fall}$. 
The lower the value of $\rm \tau_{\rm rise}$, the steeper the rise of the light curve 
(given that all the other parameters are constant). 
Since at late times (when $\rm t-t_0>>\tau_{\rm rise}$) the model approaches an exponential decline, 
the late time slope is determined by $\rm \tau_{\rm fall}$; i.e., 
the higher the value of $\rm \tau_{\rm fall}$, 
the slower the decline of the light curve (given the same normalization parameter).
We emphasize that this formula does not have any physical motivation. It is, however, 
useful to obtain good fits even for poorly sampled data and therefore used to quantify many of the properties of the SDSS-II SE-SN light curves.

The fit was applied to all objects whose maximum was covered by observations. Each fit was made on the epochs 
following and including discovery. 
We present the best fit to each light curve with solid, colored lines in Figs.~\ref{lc_u}-\ref{lc_z} and 
the fit parameters are presented in Table~\ref{fitparam}. 
The error on each parameter was estimated by fitting N~$=$~1000 Monte Carlo simulated light 
curves\footnote{For each point of the light curve we produced N normally-distributed flux values, with the central value of the normal distribution equal to the flux at that epoch and $\sigma$ equal to the flux uncertainty.} and taking the standard deviation of the resulting N parameter values. 
When the parameter errors were considerably larger than the parameters themselves, we considered the best fit to be 
unreliable and we excluded the results from Table~\ref{fitparam}. This occurred in a 
few cases (e.g., in the $z$ band for SNe~2007nc and 2007qv) where the light curves are poorly 
sampled and/or have large photometric uncertainties.
Beside the fit parameters, Table~\ref{fitparam} also presents the epoch of maximum light 
(t$_{\rm max}$), the observed flux at maximum (F$_{\rm max}$),
the difference in magnitude between maximum epoch and $+15$ and $-10$ days since maximum 
($\Delta$m$_{15}$, $\Delta$m$_{-10}$), the 
rise time t$_{\rm rise}$ (see Sect.~\ref{sec:risetime}), t$_{\rm expl}$, the peak apparent magnitude (m$_{\rm max}$), and the peak absolute magnitude (M$_{\rm max}$). In the following, we discuss these parameters for the different filters and SN types.

\subsection{Light curve properties}
\label{sec:lightcurveprop}

\subsubsection{Peak epoch as a function of the effective wavelength, t$_{\rm max}(\lambda_{eff})$}

\begin{figure}
\centering
\includegraphics[width=9cm]{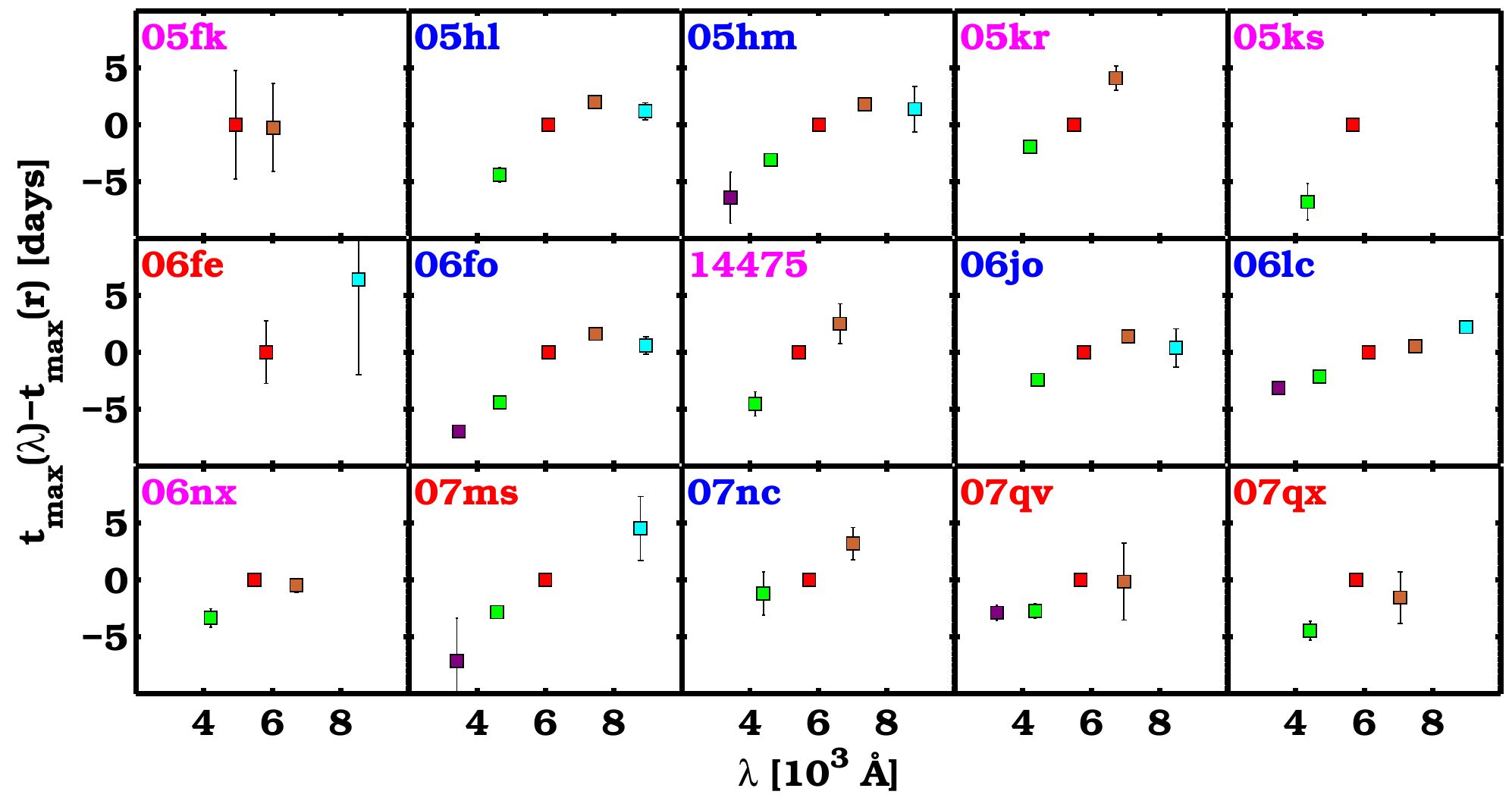}  
\includegraphics[width=9cm]{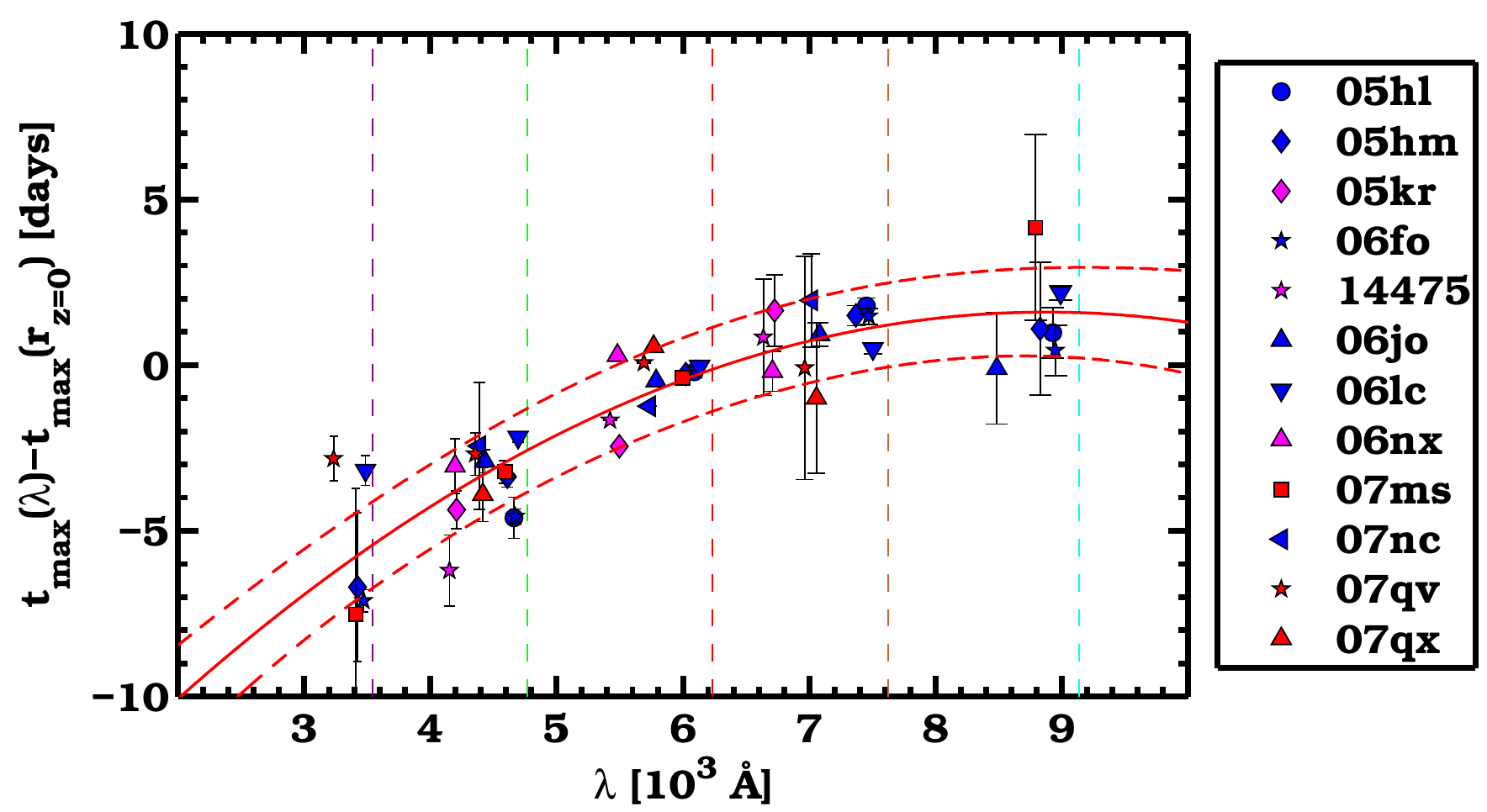}  
 \caption{\label{lcvslambda1}  ({\it Top panel:}) Epoch of the peak (t$_{\rm max}$) for each light curve with respect to  t$_{\rm max}$ in the $r$ band as a function of the effective wavelength of the filters in the rest frame for 15 SNe~Ib/c in the SDSS sample (violet, green, red, dark yellow, and cyan represent $ugriz$ filters). Blue, red, and magenta labels correspond to SNe~Ib, Ic, and Ic-BL, respectively. ({\it Bottom panel:}) Same as in the top panel, but t$_{\rm max}$ for each light curve and for each object is plotted with respect to the interpolated t$_{\rm max}$ at the effective wavelength of $r$ band at redshift $z~=$~0 (6231~\AA). Three out of 15 objects (SNe~2005fk, 2005ks, 2006fe) are excluded from the fit since the interpolation to 6231~\AA\ is not possible or reliable (e.g., for SN~2006fe, we do not have the $i-$band maximum epoch). The 2nd-order polynomial fit to the data (and plotted in red) corresponds to t$_{\rm max}$($\lambda$)$-$t$_{\rm max} (r_{z=0})=a\lambda^{2}  +  b\lambda+c$, where $a=-$2.47$\times10^{-7}$, $b=+$4.38$\times10^{-3}$, $c=-$17.86.  t$_{\rm max}$ is expressed in days and $\lambda$ in \AA. This fit allows predicting peak epochs in the whole optical range given a measured peak in a single band, with little uncertainty ($\sigma$~$=$~1.3~days). Vertical dashed lines indicate the effective wavelengths for the $ugriz$ filters.}
\end{figure}

The phenomenological fit to the light curves allows a comparison of the light curve properties of the different events.
A first result from the fit is that each SN peaks first in the $u$ band, then in the $g$ band, and successively 
later in the redder bands. (The $z$ band sometimes peaks at the same time or slightly earlier than the $i$ band.) This systematic property is shown in the the top panel of Fig.~\ref{lcvslambda1}, where we plot 
the time of the peak epoch 
for each light curve with respect to the epoch of $r_{\rm max}$ as a function of the 
effective wavelength ($\lambda_{eff}$) for each filter in the rest frame\footnote{Effective wavelengths for the SDSS filters are listed at \href{http://skyserver.sdss.org/dr1/en/proj/advanced/color/sdssfilters.asp}{http://skyserver.sdss.org/dr1/en/proj/advanced/color/sdssfilters.asp}}. 
The time delay between $u_{\rm max}$ and $z_{\rm max}$ is close to five to ten  
rest-frame days, depending on the object.

The bottom panel of Fig.~\ref{lcvslambda1} 
displays the peak epochs of each light curve with respect to the peak 
epoch interpolated to $\lambda_{eff}(r_{z=0}) = 6231$~\AA\ 
as a function of the effective wavelength of the filter 
for each object. This comparison reveals
a common behavior in the time of the peak in the different bands, which we
 can describe with a polynomial obtained through a fit to the data. The polynomial 
coefficients are reported in the caption of Fig.~\ref{lcvslambda1}. 
Using this fit, once the epoch of $r_{\rm max}$ is known for any SN~Ib/c,  
we can estimate the peak epochs in the other optical bands with a fairly small uncertainty ($1\sigma=\pm$1.3~days).
The sequential order of the peak epochs from the bluer to the redder bands reflects the progressive 
cooling of the photospheric temperature of these objects, as shown in Sect.~\ref{sec:temp}. 
This behavior holds for both SNe~Ib and Ic and
has been noticed in several single-object papers (e.g., SN~2004aw, \citealp{taubenberger06}; SN~2007Y, \citealp{stritzinger09}; SN~1999ex \citealp{stritzinger02}; SN~2003jd \citealp{valenti08}; SN~2007gr \citealp{hunter09}; SN~2009bb \citealp{pignata11}; SN~2003dh \citealp{lipkin04}; SN~2010bh \citealp{cano11}), but not previously generalized for a larger SE~SN sample \citep[with the exception of the recent paper by][]{bianco14}. 
SNe~Ia do not show such a simple relation; for example, \citet{contardo00} show that SNe~Ia $I$-band
 light curves peak before those in the $B$ band.

\subsubsection{Rise times, t$_{\rm rise}$}
\label{sec:risetime}

\begin{figure}
\centering  
\includegraphics[width=9cm]{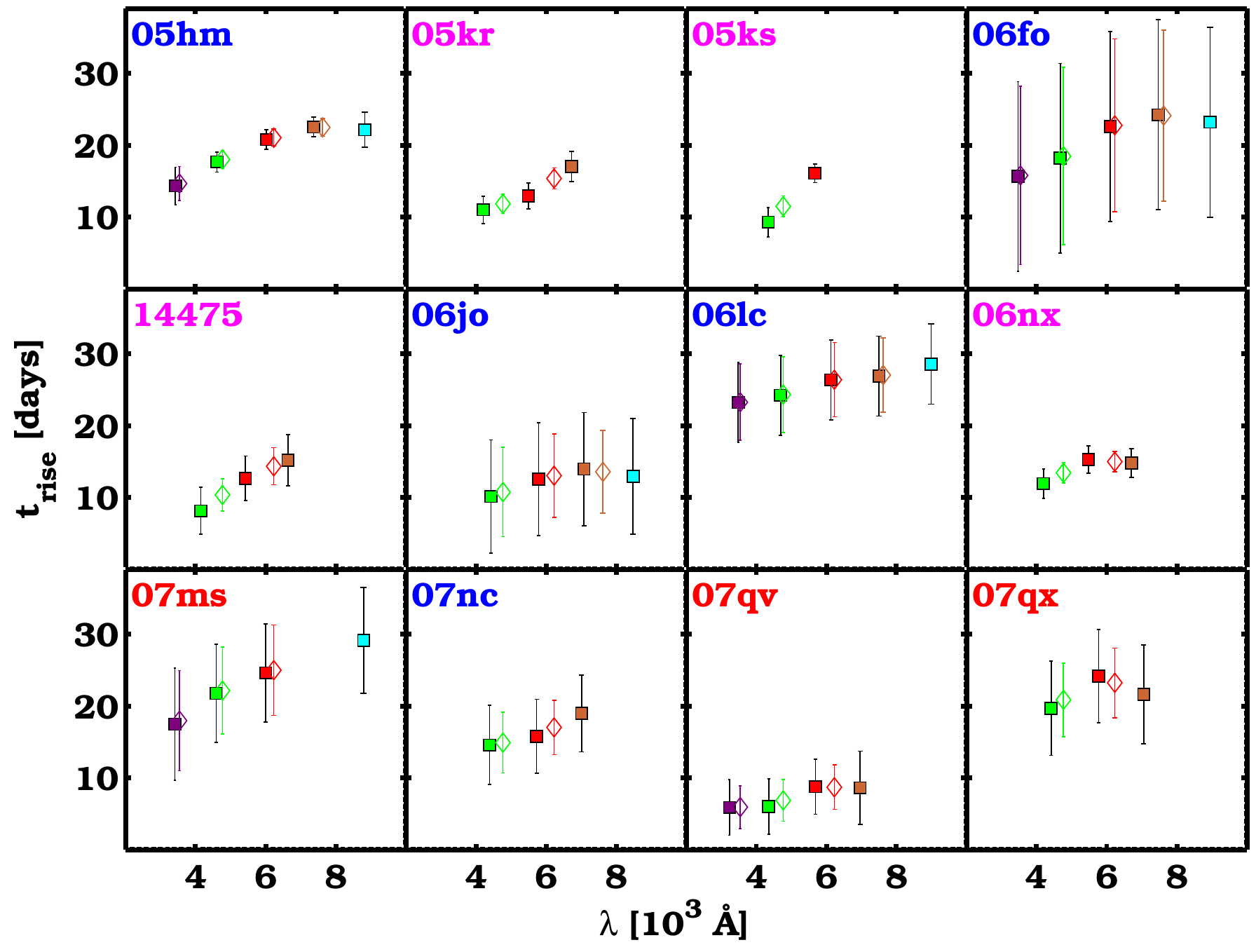}\\
\includegraphics[width=9cm]{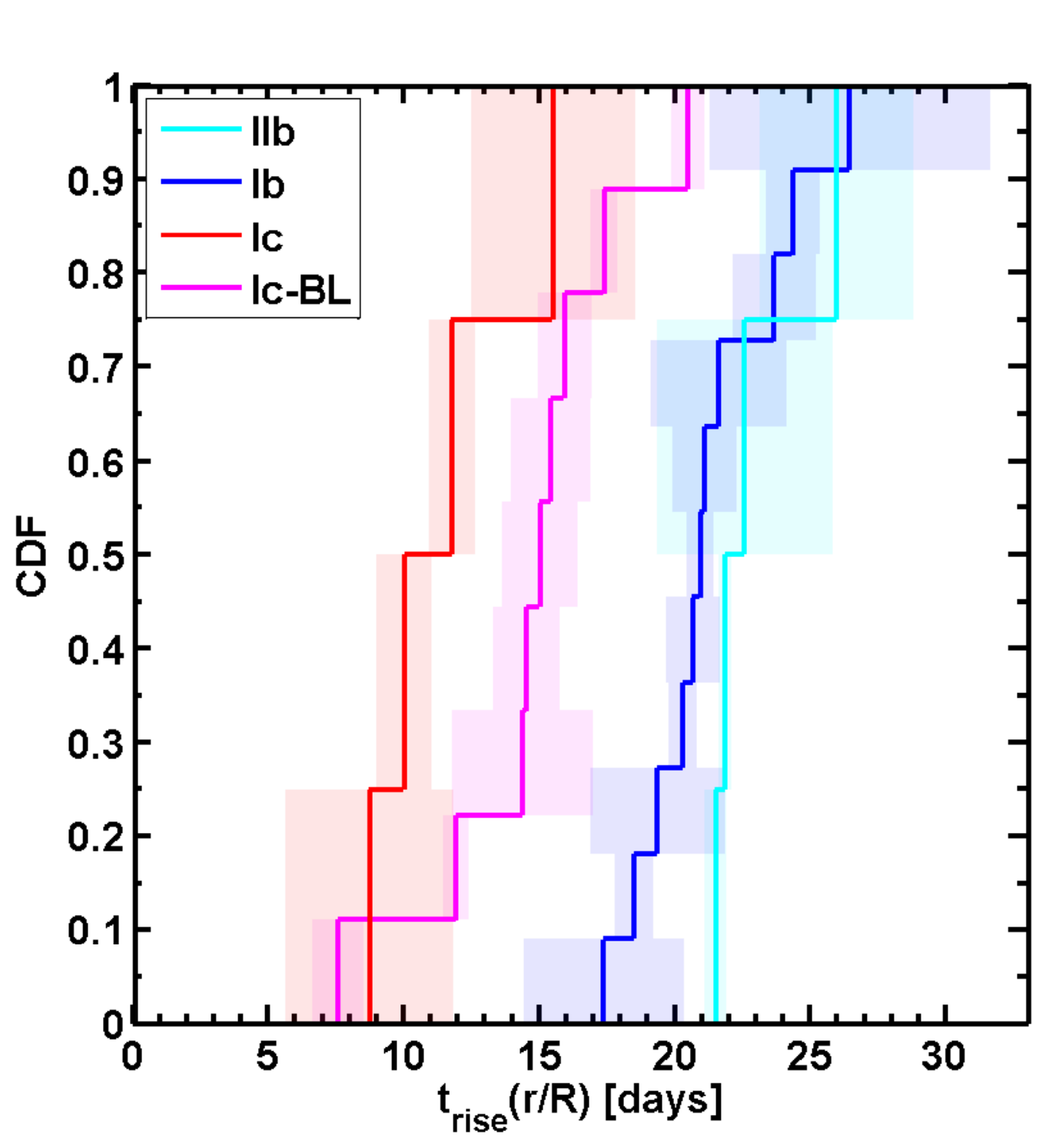}
 \caption{\label{rise_vs_lambda} ({\it Top panel:}) Rise times for 12 SNe~Ib/c in the SDSS sample as a function of the effective wavelength of each filter in the rest frame (violet, green, red, dark yellow, and cyan represent $ugriz$ filters). Bluer bands rise faster than redder bands. Blue, red, and magenta labels correspond to SNe~Ib, Ic, and Ic-BL, respectively. ({\it Bottom panel:}) Cumulative distribution functions for the $r$/$R$-band rise times of SNe~Ib (11), Ic (4), Ic-BL (9), and IIb (4), from the SDSS sample and from the literature (see Table~\ref{tab:risetimeR}). Helium-poor SNe (Ic, Ic-BL) rise to maximum in a shorter time than helium-rich transients (Ib, IIb), and this result is statistically significant (p-value~$<$~0.05). The shaded areas indicate the uncertainties for the rise times.}
\end{figure}

 With accurate estimates of the epoch of maximum light for the majority of the 
SNe, we can establish the rise time in the different bands for those
SNe that have pre-explosion images and therefore a constraint on the explosion epoch t$_{\rm expl}$. 
As explained in Sect.~\ref{sec:fit}, t$_{\rm expl}$ is defined as the average 
between the epochs of last non-detection and first detection, with a conservative
uncertainty as half the time interval between them.
The difference between t$_{\rm expl}$ and the peak epoch (t$_{\rm max}$) 
corresponds to the rise time (t$_{\rm rise}$); this is reported in the last 
column of Table~\ref{fitparam} for each SN in each band. 
The uncertainty on t$_{\rm rise}$ is the square root of the sum of the squared errors for 
t$_{\rm expl}$ and t$_{\rm max}$, where the uncertainty on the explosion epoch largely dominates. This definition of t$_{\rm expl}$ ignores a possible 
faint-plateau phase
between core collapse and light curve rise, 
which has been 
predicted by the theoretical models of \citet{dessart12_OriginIbc}.
We discuss the accuracy and implications of this assumption in 
Sects.~\ref{sec:explo_error}, \ref{sec:earlyplateau}, and \ref{sec:nimix}.

In Fig.~\ref{rise_vs_lambda} (top panel), the rise time at the different $\lambda_{eff}$ 
for each filter in the SN 
rest frame is shown (marked by colored squares) for 12 SNe. 
To allow for a direct comparison of the rise times of different objects, 
we linearly interpolated the observed rise times to the 
effective wavelengths of $ugri$ filters at redshift $z$~$=$~0 (marked by empty 
diamonds).
Given the cadence of the SDSS-II SN survey observations, the uncertainty on the explosion dates, hence on the rise times, 
is relatively large for most of the objects. However, for six objects we know the $r_{z=0}$-band rise time with an 
uncertainty $\lesssim$ 3~days. We use these objects, together with 
other events in the literature with well-constrained rise times in the $r$/$R$-band light curves,  
to compare the rise times of SNe~Ib, Ic, Ic-BL, and also IIb. This comparison is shown in the bottom panel of 
Fig.~\ref{rise_vs_lambda}, where the cumulative distribution function of the rise times 
is given for the different SN types. (We also indicate the errors on the rise times with shaded area.) 
All the $r$/$R$-band rise times are also reported in Table~\ref{tab:risetimeR} with their errors.

Our result confirms what the data presented in \citet{valenti11} suggested: 
helium-poor SNe (Ic and Ic-BL) rise to maximum faster than helium-rich SNe (Ib and IIb).  
We found $<t_{\rm rise}(IIb,Ib,Ic,Ic-BL)>=22.9\pm0.8,21.3\pm0.4,11.5\pm0.5,14.7\pm0.2$~rest frame days, 
where the quoted uncertainties are the standard deviations in the rise time divided 
by the square root of the number of objects in each sample (4, 11, 4, and 9 SNe~IIb, Ib, Ic, and Ic-BL, respectively). We note that the sample of SNe~Ic is rather small.

The difference between SNe~Ib and Ic is statistically significant. A Kolmogorov-Smirnov (K-S) test comparing 
these two populations gives a p-value~$=$~0.001. Aiming to estimate the robustness of the p-value, 
we made 10$^{4}$ Monte Carlo simulations of the 
rise times\footnote{For each event we produced 10$^{4}$ normally distributed random rise times with central values of the normal distribution equal to the measured rise time of the SN and $\sigma$ equal to the rise-time uncertainty.} 
and performed 10$^{4}$ K-S tests on the resulting couples of SN~Ib and SN~Ic rise-time simulations. 
 In this exercise 68\% of the simulated p-values were included within 0.005, 
confirming the statistical significance of the difference, we measured between these two populations. The bottom panel of Fig.~\ref{rise_vs_lambda} also shows that SNe~Ib and SNe~Ic-BL display a statistically significant difference. In fact, we found p-value~$=$~0.001(0.001), 
where the number in parenthesis corresponds to the value within which 68\% of 
the simulated p values are included. 
Statistically significant differences were also found when we compared SNe~IIb to Ic, 
p-value~$=$~0.011(0.011), and SNe~IIb to Ic-BL, p-value~$=$~0.002(0.009). The peculiar SN~Ic 2011bm \citep{valenti12} was 
excluded from the rise time comparison, which has an unusual $\sim$40 days rise time.

\subsubsection{$\Delta$m$_{15}$ and $\Delta$m$_{-10}$}
\label{sec:dm15}

\begin{figure}
\centering  
\includegraphics[width=9cm]{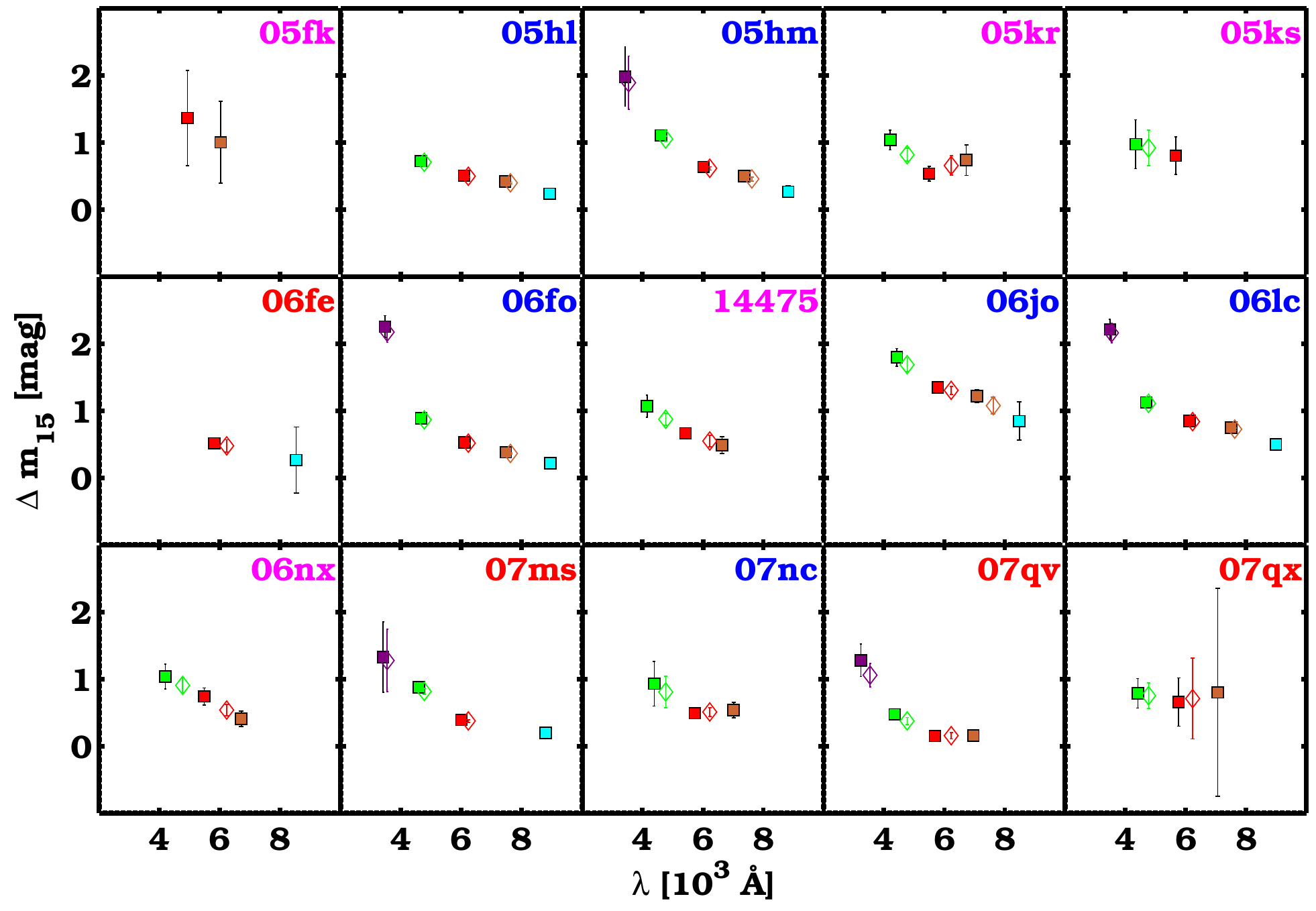}\\
\includegraphics[width=9cm]{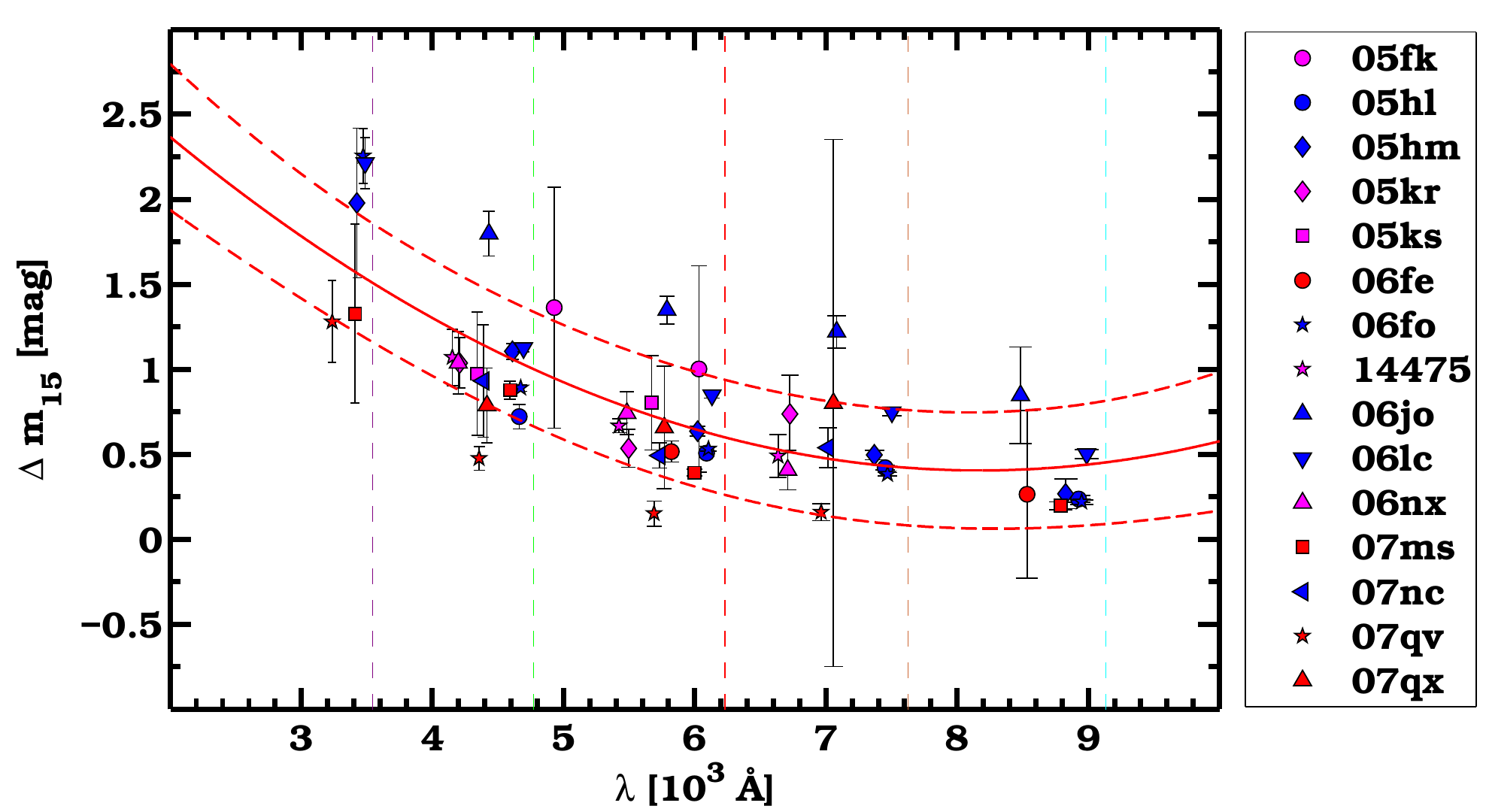}\\
\includegraphics[width=9cm]{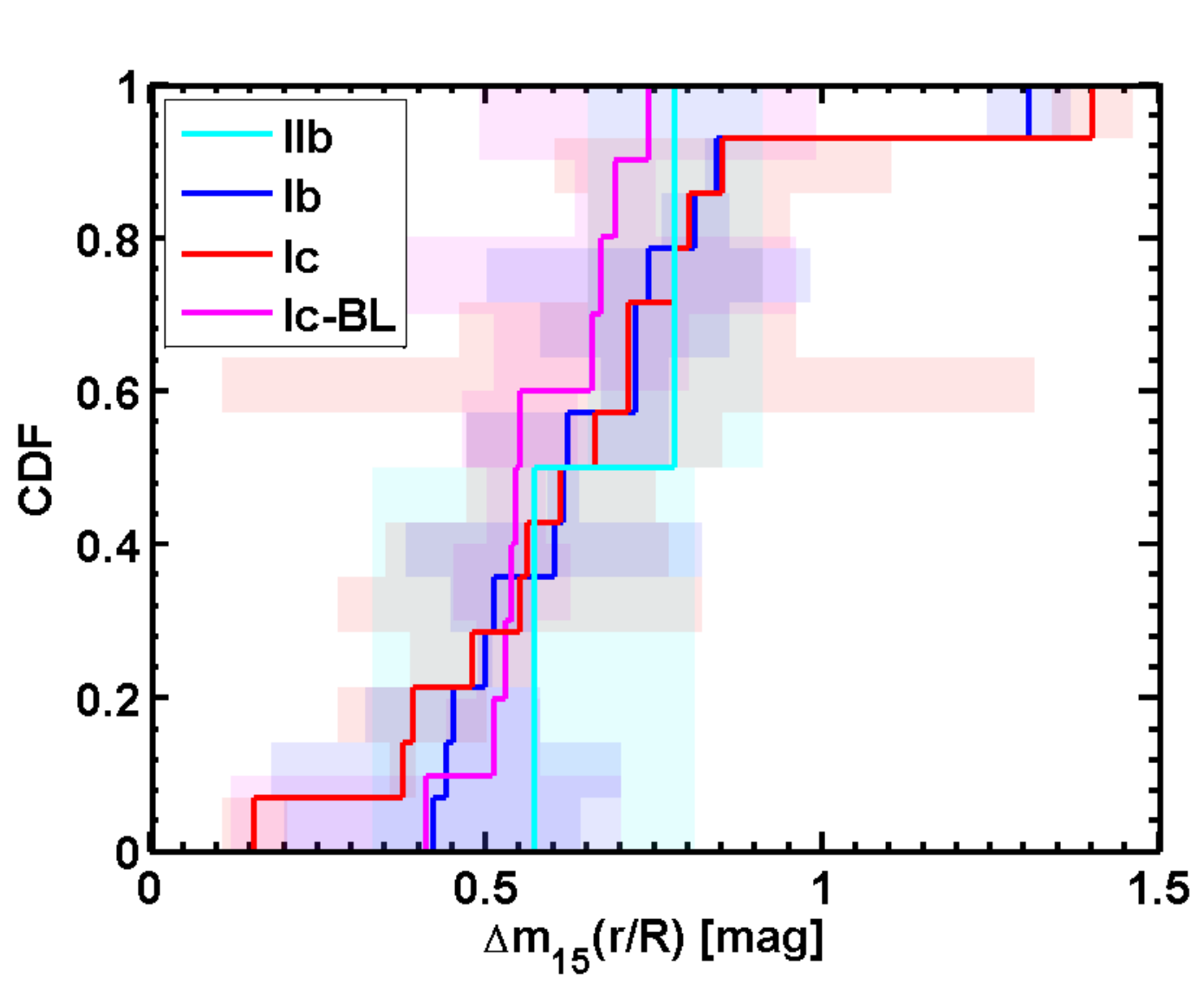}
 \caption{\label{dm15comp} ({\it Top panel:}) $\Delta$m$_{15}$ for 15 SNe~Ib/c in the SDSS sample as a function of the effective wavelength of each filter in the SN rest frame (violet, green, red, dark yellow and cyan represent $ugriz$ filters). Bluer bands decline faster than redder bands. 
({\it Central panel:}) 
 Same as in the top panel, the 2nd order polynomial fitted to the data (and plotted in red) corresponds to $\Delta$m$_{15}$($\lambda$)~$=$~ a$\lambda^2+$b$\lambda+$c, where
 a~$=$~5.14$\times10^{-8}$, b~$=~-$8.41$\times10^{-4}$ and c~$=$~3.84.
   $\Delta$m$_{15}$ is expressed in magnitudes and $\lambda$ in \AA. Typical 1$\sigma$ uncertainties on $\Delta$m$_{15}$($\lambda$) are on the order of 0.4~mag. Vertical dashed lines indicate the effective wavelengths for the $ugriz$ filters.
 ({\it Bottom panel:}) Cumulative distribution functions for the $r$/$R$-band $\Delta$m$_{15}$ of SNe~Ib (14), Ic (14), Ic-BL (10) and IIb (2), from the SDSS sample and from \citet{drout11}. SDSS $\Delta$m$_{15}$ values are those interpolated at $\lambda_{\rm eff}(r_{z=0})$. Shaded areas indicate the uncertainties for $\Delta$m$_{15}$. There is no significant difference among the distributions of the different classes, with p-values~$>$~0.05.}
\end{figure}

\begin{figure}
\centering  
\includegraphics[width=9cm]{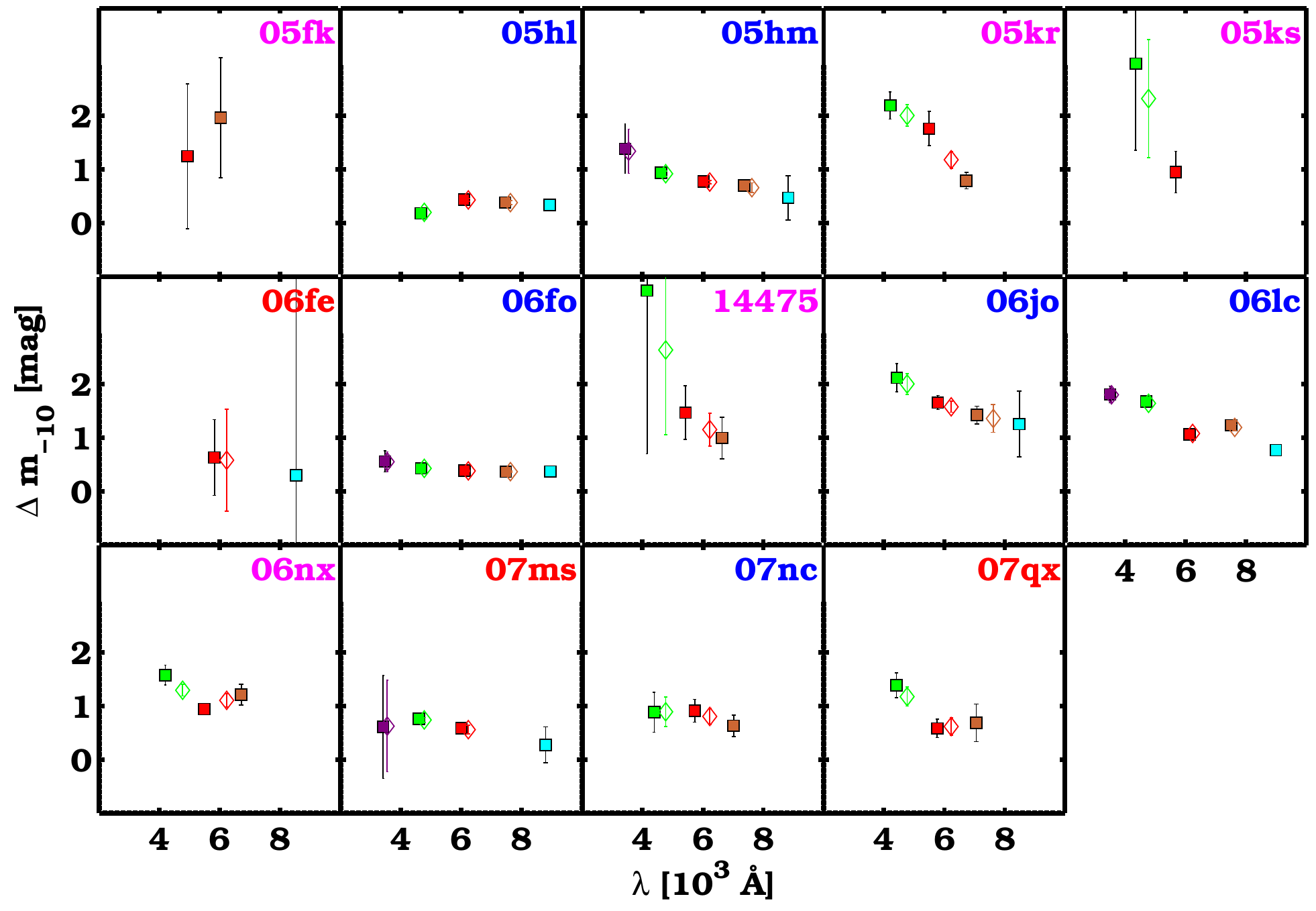}  \\
\includegraphics[width=9cm]{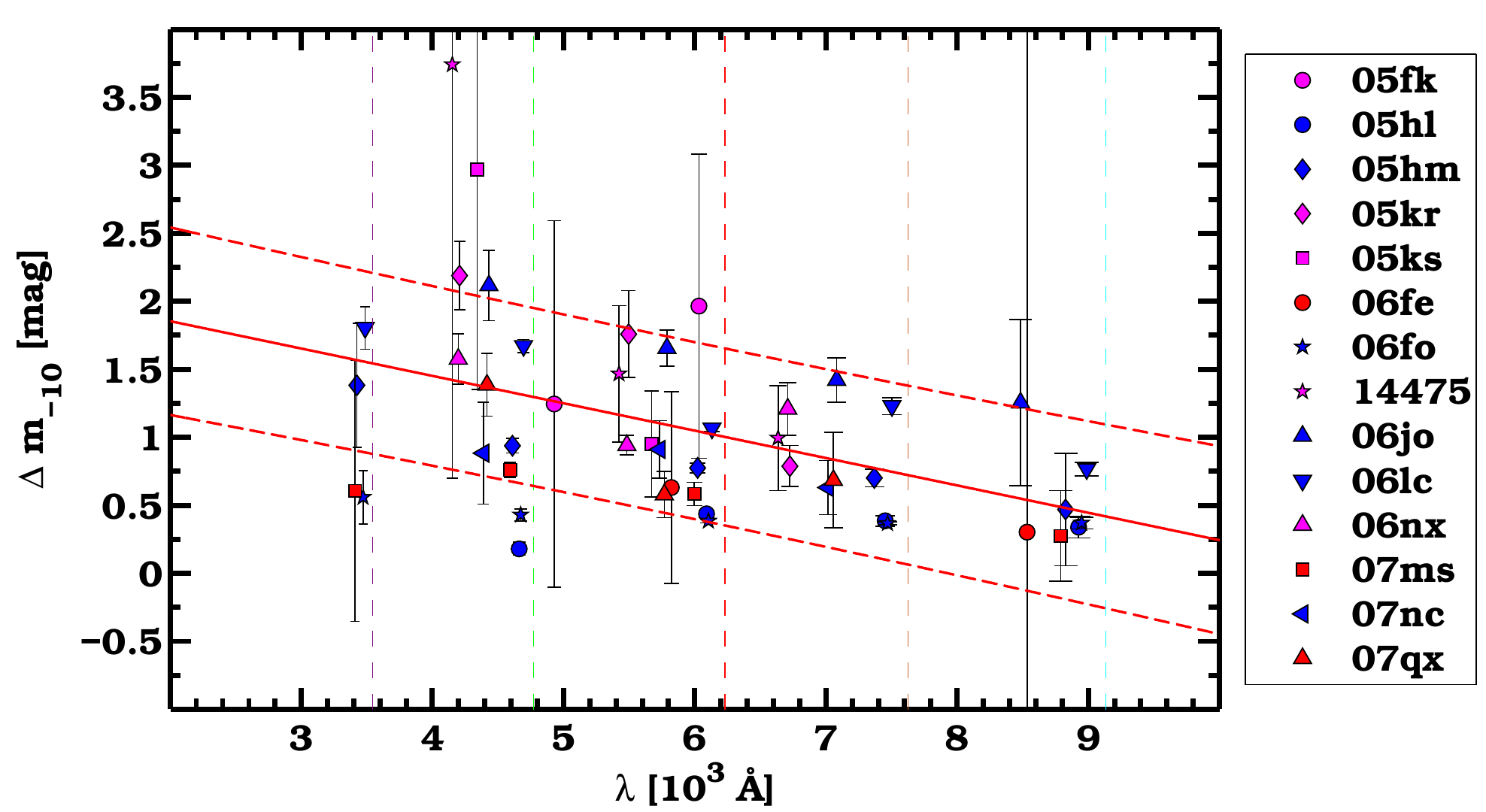}\\   
\includegraphics[width=9cm]{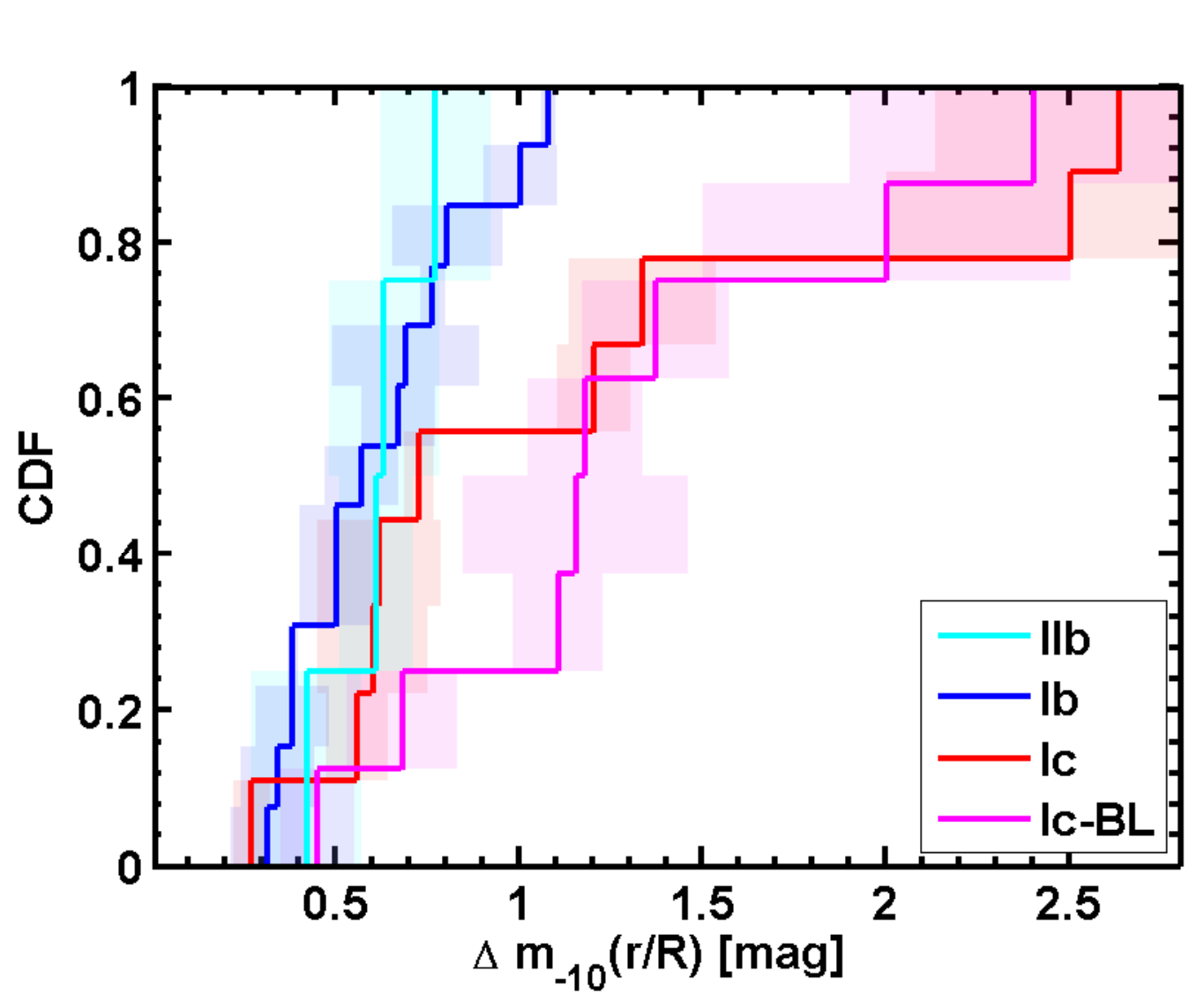}
 \caption{\label{dm10comp} ({\it Top panel:}) $\Delta$m$_{-10}$ for 14 SNe~Ib/c in the SDSS sample as a function of the effective wavelength of each filter in the SN rest frame (violet, green, red, dark yellow and cyan represent $ugriz$ filters). Bluer bands rise faster than redder bands for most of the objects. Blue, red and magenta labels correspond to SNe~Ib, Ic and Ic-BL, respectively.
({\it Central panel:}) 
 Same as in the top panel, the line fitted to the data (and plotted in red) corresponds to
$\Delta$m$_{-10}$($\lambda$)~$=$~$-$2.01$\times10^{-4}\lambda+$2.26,
  with $\Delta$m$_{-10}$ expressed in magnitudes and $\lambda$ in \AA. Uncertainties on $\Delta$m$_{-10}$($\lambda$) are $\sim$0.66~mag.
  ({\it Bottom panel:}) Cumulative distribution functions for the $r$-band $\Delta$m$_{-10}$ of SNe~IIb (4), Ib (13), Ic (9), and Ic-BL (8) from the SDSS sample and from the literature. Shaded areas indicate the uncertainties for $\Delta$m$_{-10}$. SNe~Ic and Ic-BL show higher values of $\Delta$m$_{-10}$ when compared to SNe~Ib (p-values$=$0.18 and 0.003, respectively).}
\end{figure}

Besides 
the rise time, another important parameter that characterizes the light curve shapes 
of radioactively powered SNe 
is $\Delta$m$_{15}$, i.e., the difference in magnitude between the value at peak and at 15 days 
(in the rest frame) past maximum [$\Delta$m$_{15}=-$2.5log$_{10}$(F(t$_{\rm max}+$15)/F$_{\rm max}$)].
It is particularly useful for SNe~Ia, where its correlation to peak brightness allows for their use as standardizable candles \citep{phillips93}. GRB~SNe (\citealp{li14a}; \citealp{cano14b}) have also been recently suggested as standardizable candles showing luminosity--decline relations (\citealp{li14b}; \citealp{cano14c}).
From the fits on the light curves we derived $\Delta$m$_{15}$ for 
15 SNe in the sample, and plot the measured values as a function of the effective wavelength of each filter in Fig.~\ref{dm15comp} (top panel). 
It is clear that $\Delta$m$_{15}$ is larger for the bluer bands than for the redder ones. A similar trend is observed in the optical light curves of SNe~Ia \citep{folatelli10}. 
When we overplot all the $\Delta$m$_{15}$ values as a function of wavelength in the central panel of 
Fig.~\ref{dm15comp}, the common $\Delta$m$_{15}$ behavior for all the SDSS-II SNe~Ib/c becomes evident. 
If we fit a polynomial to these data, an analytical expression for $\Delta$m$_{15}$ as a function of 
the effective wavelength of the filters can be derived 
(Fig.~\ref{dm15comp}, see the caption for the polynomial expression). 

Because we have all the $\Delta$m$_{15}$ values in different filters, we linearly interpolated the observed $\Delta$m$_{15}$ at the effective wavelength of the $gri$ filters at $z$~$=$~0, to allow for a direct comparison between the different SN types. 
In the bottom panel of Fig.~\ref{dm15comp}, we show that the $\Delta$m$_{15}$ distribution for the $r_{z=0}$ band of SNe~Ib does not differ significantly from that of SNe~Ic or Ic-BL. Here we consider the SDSS-II data and the $\Delta$m$_{15}$($R$) from Table 5 of \citet{drout11}, where a couple of SNe~IIb are also reported. We included a total of 2, 14, 14, and 9 SNe~IIb, Ib, Ic, and Ic-BL, respectively. 
The $\Delta$m$_{15}$ distributions for SNe~Ib, Ic, and Ic-BL are also similar when we 
consider the $g_{z=0}$ instead of the $r_{z=0}$ band (including only the SDSS-II data).

The width of the light curves on the rising part is evaluated by measuring $\Delta$m$_{-10}$, i.e., 
the difference in magnitude between the value at peak and what was measured ten days before maximum 
[$\Delta$m$_{-10}=-$2.5log$_{10}$(F(t$_{\rm max}-$10)/F$_{\rm max}$)]. 
This provides an estimate for the steepness  of the rise, which is 
independent of knowledge of the exact explosion date, and the potential presence of an early plateau phase.  
In Fig.~\ref{dm10comp} (top panel) we show the rest frame $\Delta$m$_{-10}$ values for each SN in the sample
as a function of the effective wavelength of the filters (colored squares). 
We also interpolate (empty symbols) these values at the $gri$ effective wavelengths for $z~=$~0 in order to 
obtain $\Delta$m$_{-10}$ values suitable for a comparison among the different SN types.

The $\Delta$m$_{-10}$ values are higher at short 
wavelengths than in the redder bands; i.e., the SNe rise is steeper in the bluer bands. 
This is confirmed by overplotting the data for all 14 SNe in the central panel of Fig.~\ref{dm10comp}. 
A linear fit to the data is shown and its expression is reported in the corresponding caption. 
The scatter around the fit is $\sim$0.66~mag.
We compare the $\Delta$m$_{-10}$ values interpolated at $r$-band effective wavelengths among the 
different SN types (Fig.~\ref{dm10comp}, bottom panel). Here we consider the $\Delta$m$_{-10}$ values of our SNe (excluding those with extrapolated values of $\Delta$m$_{-10}$) and those of other SNe~Ib/c (and IIb) whose $r/R-$band light curves are available in the literature (see Table~\ref{tab:risetimeR}).
SNe~Ic and SNe~Ic-BL exhibit a steeper rise compared to SNe~Ib and IIb,
consistently with the different rise times. The difference between SNe~Ib and Ic is marginally significant (p-value~$=$~0.18), whereas the K-S test between SNe~Ib and Ic-BL gives p-value~$=$~0.003.

Armed with the interpolated values of $\Delta$m$_{-10}$ and $\Delta$m$_{15}$ at the effective wavelength of the $gri$ filters, we searched for a correlation between these two parameters. 
Apart from the single object SN~2006jo, which displays high values for both parameters in all bands,
there is little evidence of any strong correlation between the shape of the rising part and the falling part of the light curve. The rise time of SNe~Ia is also not strongly correlated with the fall,
even though the slowest decliners tend to be among the fastest risers \citep{hayden10a}.

\subsubsection{Light curve templates}
\label{sec:template}

\begin{figure}
\centering
\includegraphics[width=6cm]{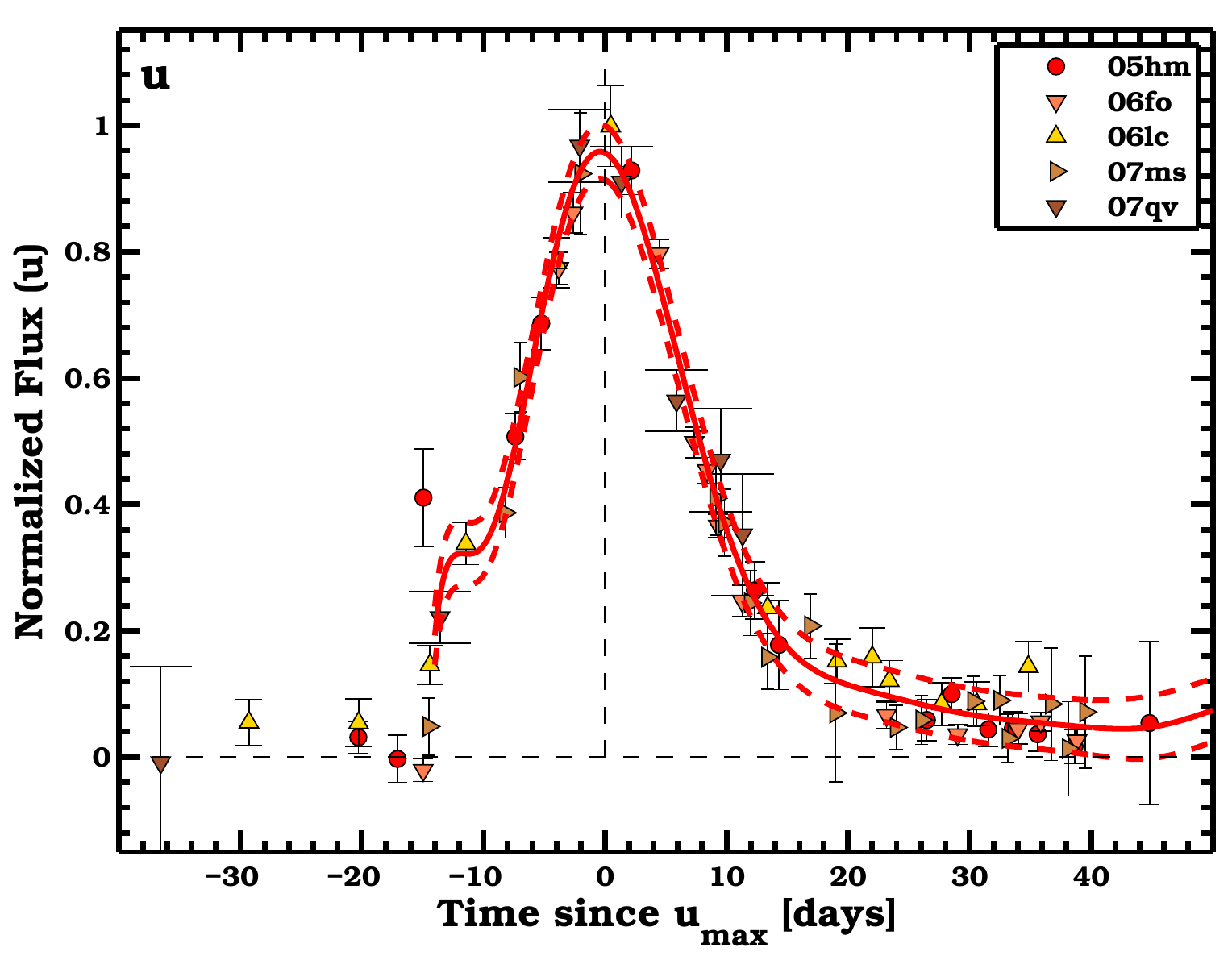}\\  
\includegraphics[width=6cm]{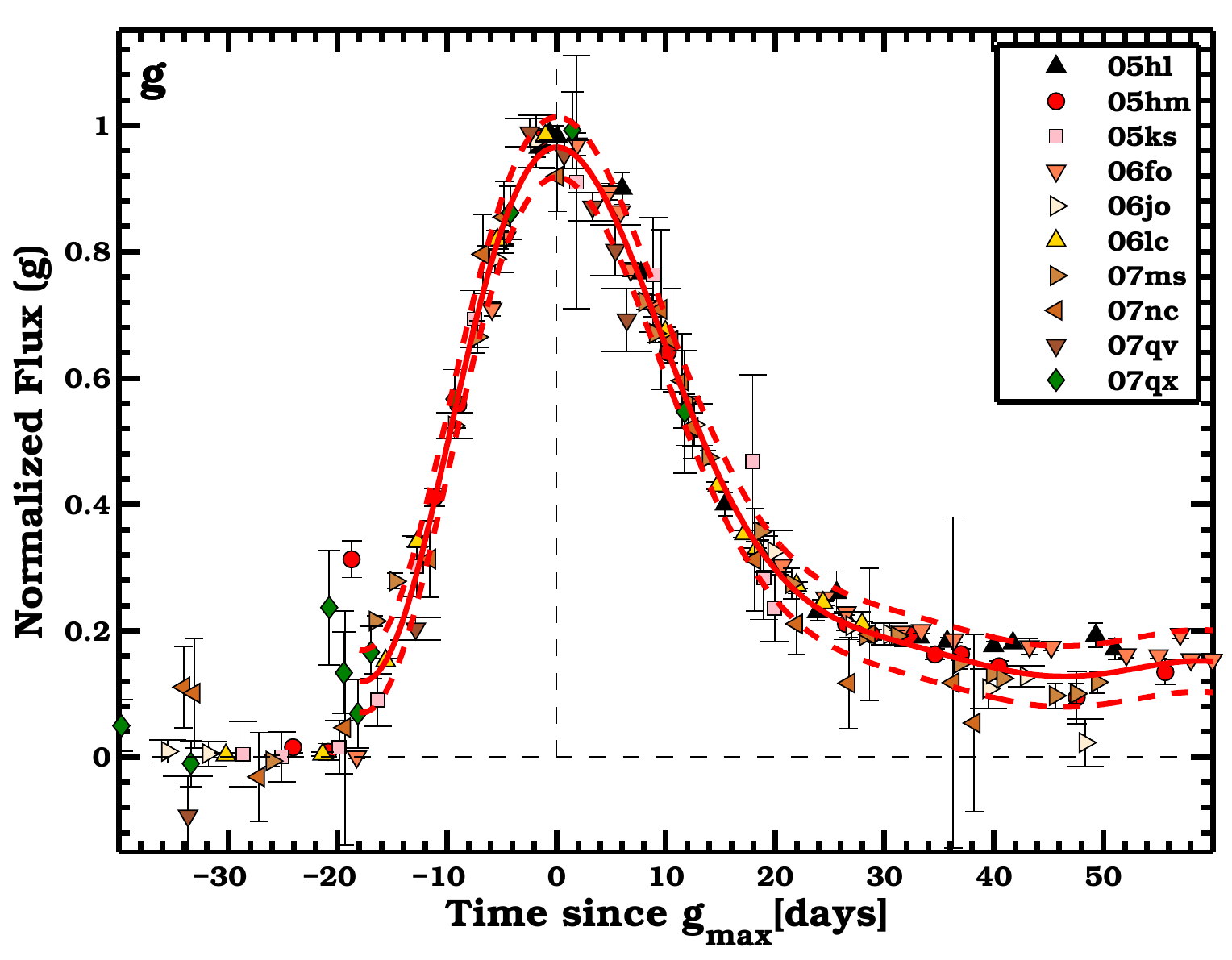}\\  
\includegraphics[width=6cm]{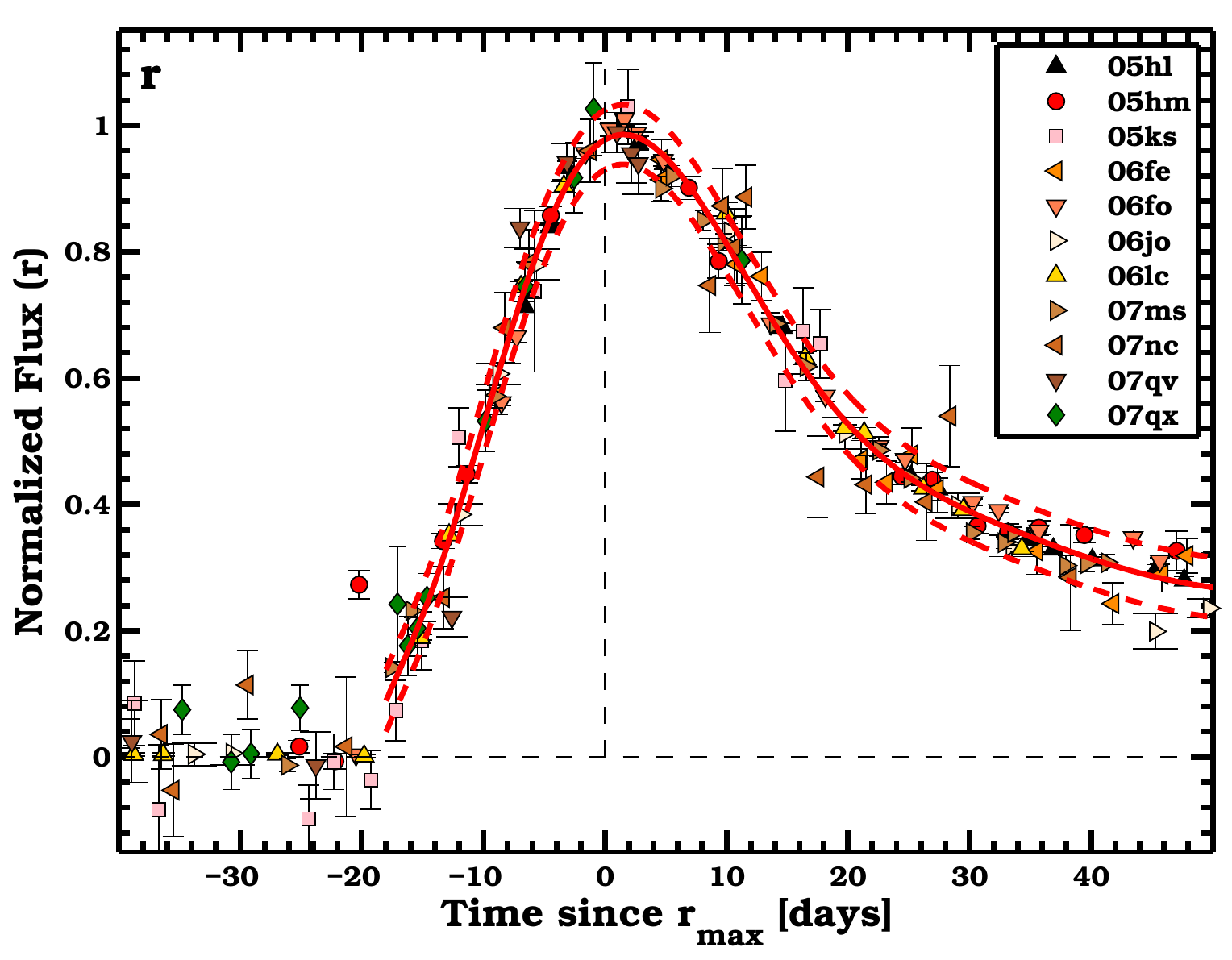}\\  
\includegraphics[width=6cm]{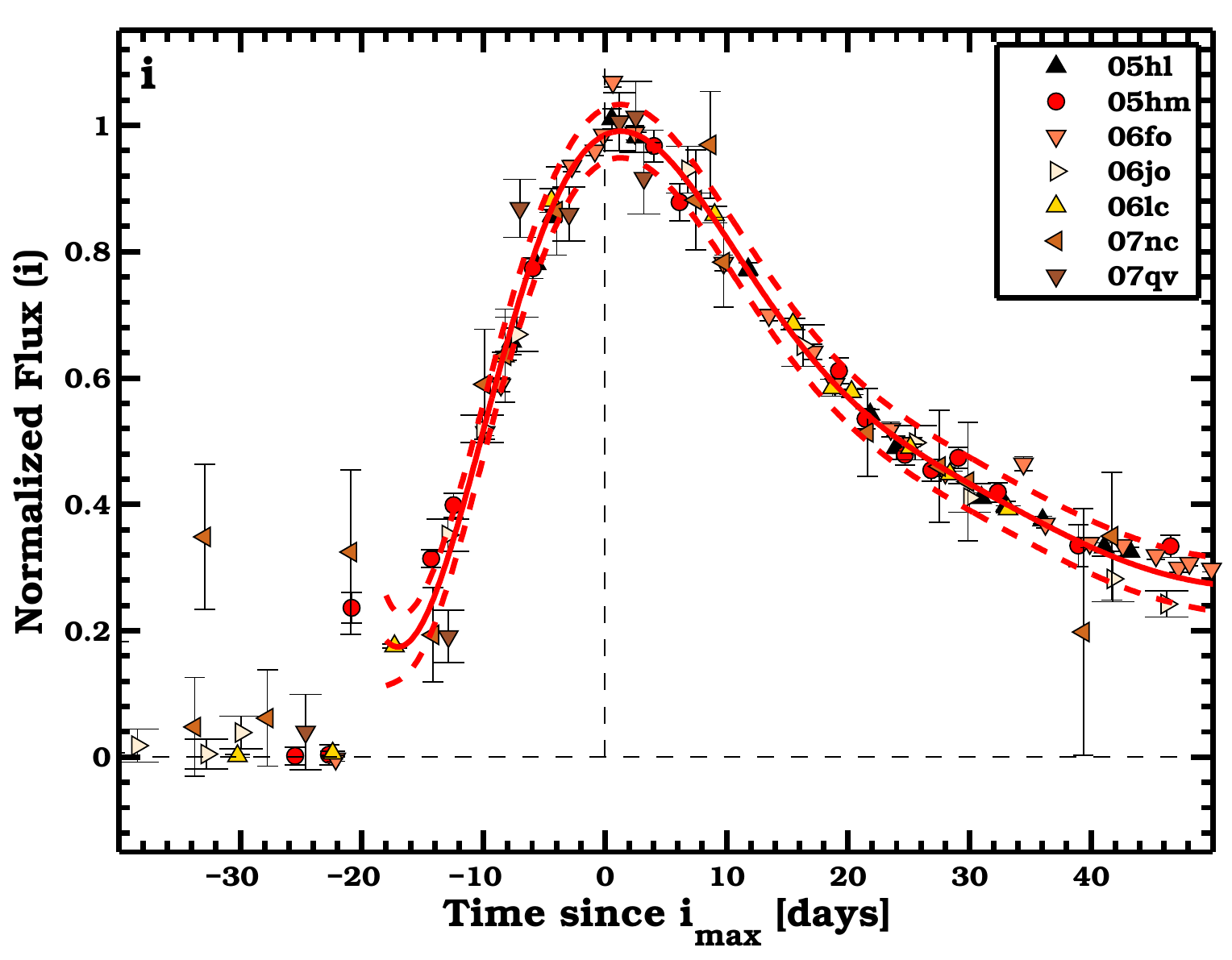}\\ 
\includegraphics[width=6cm]{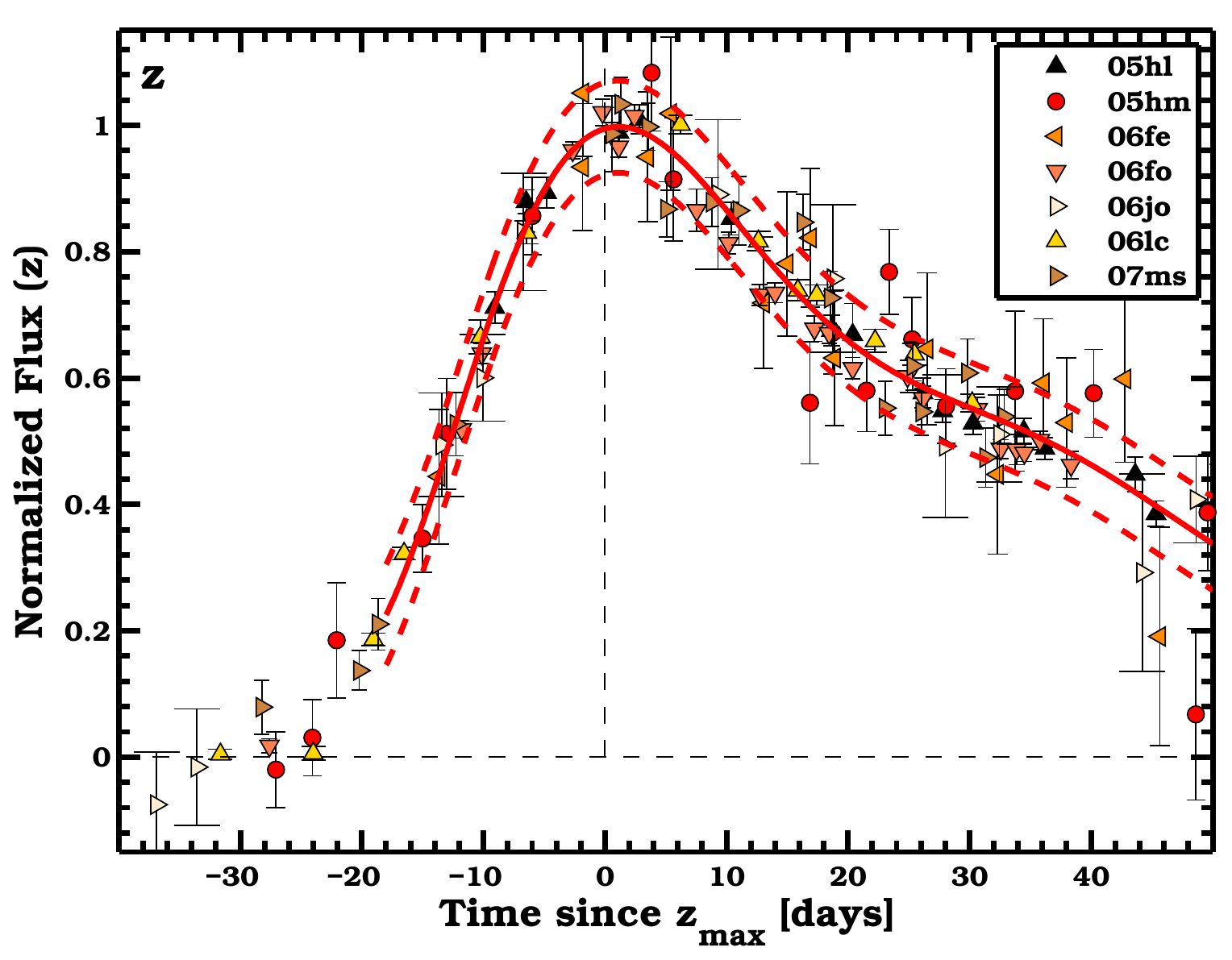}\\  
 \caption{\label{stretch}Normalized and double-stretched $ugriz$ light curves of SNe~Ib/c from the SDSS sample. Only the objects at redshift $z<$0.1 were selected. We show each object individually and fit the data with a polynomial (12th order). The best fit is shown by a red solid line, uncertainties by dashed red lines. We make these templates available in electronic form
at the CDS via anonymous ftp to cdsarc.u-strasbg.fr (130.79.128.5)
or via \href{http://ttt.astro.su.se/~ftadd/template.txt}{http://cdsweb.u-strasbg.fr/cgi-bin/qcat?J/A+A/}.}
\end{figure}

The previous analysis 
showed that the light curve shapes for SNe~Ib 
and Ic/Ic-BL are similar overall. Using the light curve fit 
and the derived parameters for most of the objects, 
we construct SN~Ib/c light curve templates in the $ugriz$ bands.  
We selected only objects at redshift $z<$0.1 in order to avoid the need 
for large k-corrections.
Each light curve was normalized by the flux at the peak, 
and the time axis was shifted so that the epoch of the peak is at zero.
Following the approach by \citet{hayden10a} 
for SNe~Ia, we applied a double 
stretch to the $ugriz$-band light curves. We measured the half width at half 
maximum (HWHM) before and after the peak for each light curve and took 
the ratio between each HWHM and its average value in the sample as 
the stretch factor, both before and after the peak. 
The result for the $ugriz$ bands is shown in Fig.~\ref{stretch}. 
A single stretch based on the full 
width at half maximum (FWHM) of the light curves
would lead to a higher dispersion on the rising part of the 
normalized light curves, which is obviously reduced by applying a 
double-stretch correction. For the $r$ band, with a double stretch, we obtain a fit with 
RMS~$=$~0.046 (in normalized flux units), whereas the RMS is 0.052 (in normalized flux units)
 when a single stretch is applied.
We fit the double-stretched and normalized light curves with 
polynomials to obtain templates, which are made available 
online\footnote{\href{http://ttt.astro.su.se/~ftadd/template.txt}{http://ttt.astro.su.se/$\sim$ftadd/template.txt}}.
It is confirmed that the bluer bands rise and decline faster than the 
redder bands. The best fit is shown in Fig.~\ref{stretch}. 
At early times, only a few points are clearly off with respect to the best fit 
and were excluded (in the $ugr$ bands). These points belong to 
SNe~2005hm and 2007qx and are likely to be signatures of the shock 
break-out cooling tail observed in these events (see Sect.\ref{sec:radius}).

\subsubsection{Summary of the light curve properties}

We list some main results that we determined through the analysis described in the previous sections:

\begin{itemize}

\item{Both SNe~Ib and SNe~Ic/Ic-BL peak first in the bluer and then successively in 
the redder bands.}

\item{SNe~Ic and Ic-BL rise to maximum in a shorter time than SNe~Ib and IIb, 
a result that is statistically significant in the $r$ band.}

\item{The steepness of the light curves after maximum ($\Delta$m$_{15}$) 
is similar for the different SE~SN types.}

\item{The steepness of the light curves on the rise ($\Delta$m$_{-10}$) 
is greater for SNe~Ic-BL and Ic than for SNe~Ib and IIb.}

\item{$\Delta$m$_{15}$ and $\Delta$m$_{-10}$ are larger in the bluer bands.}

\end{itemize}

\subsection{Light-curve absolute maxima and bolometric properties}

\subsubsection{Host extinction estimates}
\label{sec:hostext}

\begin{figure*}
\centering
$\begin{array}{cc}
\includegraphics[height=7cm]{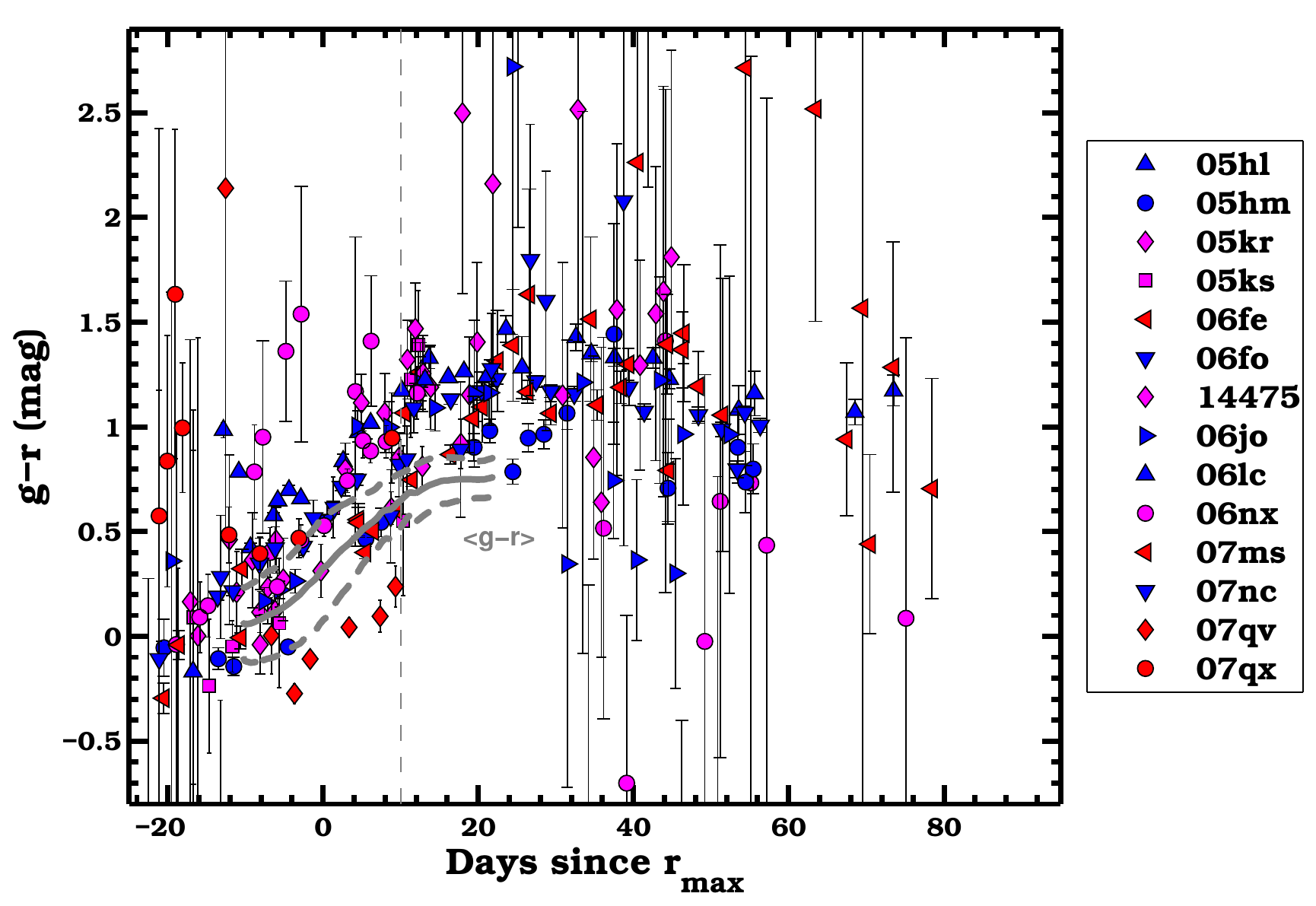}  
\includegraphics[height=7cm]{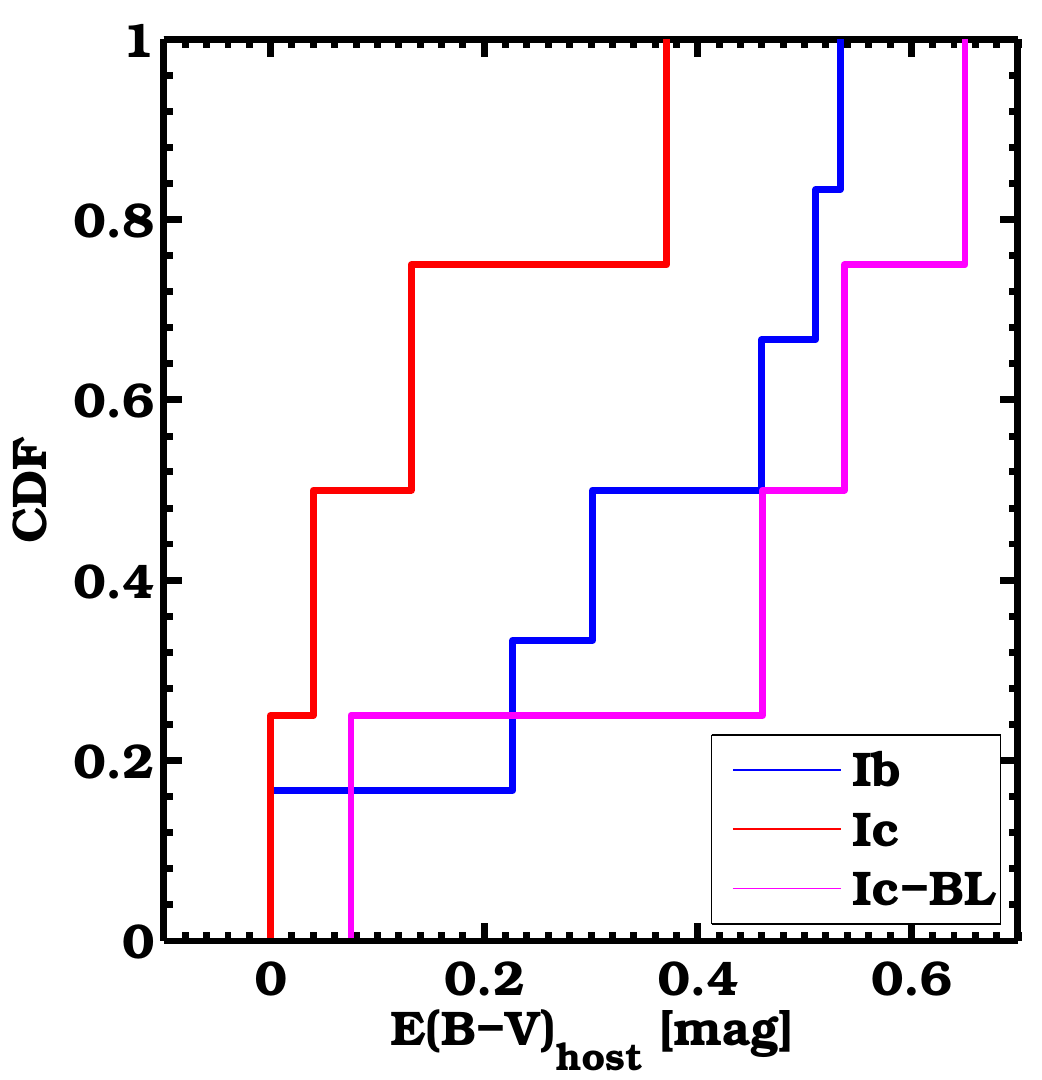}
\end{array}$
 \caption{\label{wien}(Lefthand panel:) $g-r$ color (corrected only for the Galactic extinction) as a function of time since $r_{\rm max}$ for 14 SDSS~SNe~Ib/c. 
 The average extinction-corrected $<g-r>$ color of seven SE~SNe (see Sect.~\ref{sec:hostext} for details) is shown by a solid gray line. Its value at 10 days (this epoch is labeled by a vertical dashed gray line, where $g-r$~$=$~0.64$\pm$0.13~mag) is assumed as the intrinsic colour for computing the host extinction for each SN. A \citet{cardelli89} reddening law with $R_V=3.1$ is adopted in the computation.
(Righthand panel:) Cumulative distribution functions of $E(B-V)_{\rm host}$ for SNe~Ib, Ic, and Ic-BL. The corrections for SNe~Ib and Ic-BL appear larger than those needed for SNe~Ic. The uncertainty on each $E(B-V)_{\rm host}$ is 0.2~mag.}
\end{figure*}


Lacking high resolution spectra for our objects, we cannot use the relations
between extinction and \ion{Na}{i}~D equivalent width (EW)
suggested in the literature \citep[e.g.,][]{poz12}. Our low resolution spectra are not of 
sufficient quality to measure the EW of the unresolved \ion{Na}{i}~D lines (which is 
anyway a poor proxy of the extinction, \citealp{poz11}). Therefore, following the 
approach by \citet{drout11}, we based our host extinction estimates on SN colors. In 
particular we make use of $g-r$ as this color is the closest to $V-R$, which is the one 
used by \citet{drout11}. Also, $g-$ and $r-$band light curves show higher quality 
compared to those obtained with other filters in our dataset.

First we determine the average, galaxy and host extinction-corrected $<g-r>$ 
from seven SE~SNe in the literature with well-known host extinction, available 
$g-$band and $r-$band photometry, and well-known $r-$band peak epoch. We 
stress that, as in \citet{drout11}, the
host extinction value reported in the literature for each object is in itself uncertain, given
that most of them are obtained through the EW of unresolved \ion{Na}{i}~D lines. We 
select objects belonging to the different SE-SN sub-classes: SNe~Ib~2007Y 
\citep{stritzinger09}, 2008D \citep{soderberg08}, 2009jf \citep{valenti11}, 
SNe~Ic~2009bb \citep{pignata11}, 20011bm \citep{valenti12}, and SNe~IIb~2010as 
\citep{folatelli14}, 2008ax \citep{pastorello08}. The $<g-r>$ evolution along with its standard deviation are shown by grey lines in 
Fig.~\ref{wien} (lefthand panel).
The dispersion of $<g-r>$ has its minimum (0.13~mag) about ten days past 
$r_{\rm max}$, similar to what \citet{drout11} show for their $<V-R>$. Therefore we chose the value of $<g-r>$ at this epoch ($<g-r>_{10}$~$=$~0.64$\pm$0.13~mag) as 
the intrinsic color. 

We then interpolate at ten days past $r_{\rm max}$ all the $g-r$ profiles of our SDSS 
SNe (corrected only for the Galactic extinction), which are shown in Fig.~\ref{wien} 
(lefthand panel). 
The difference in color ($E(g-r)$) at this phase between each SN and $<g-r>$ is assumed to only be produced by the effect of host extinction. We assume a \citet{cardelli89} reddening law with R$_V=$3.1 to obtain $E(B-V)_{\rm host}$ from $E(g-r)$. The derived values for $E(B-V)_{\rm host}$ are reported in Table~\ref{sample}. The uncertainty on the derived $E(B-V)_{\rm host}$ is 0.2~mag, as derived from the dispersion of $<g-r>$.
 SNe~Ic exhibit smaller host-extinction estimates compared to those of SNe~Ib and Ic-BL (p-value$=$0.25 and 0.11), as shown in Fig.~\ref{wien}.
Two events are found to be unreddened (SNe~2005hm and 2007qv; the latter is found to be intrinsically bluer than typical SE~SNe).

In the following, we do not simply add the Galactic and the host-galaxy color excess to compute the extinction in each filter. Instead, we compute the rest-frame extinction separately, because it depends on the redshift. This is important for the most distant objects of our sample.

\begin{figure*}
\centering
\includegraphics[width=13cm]{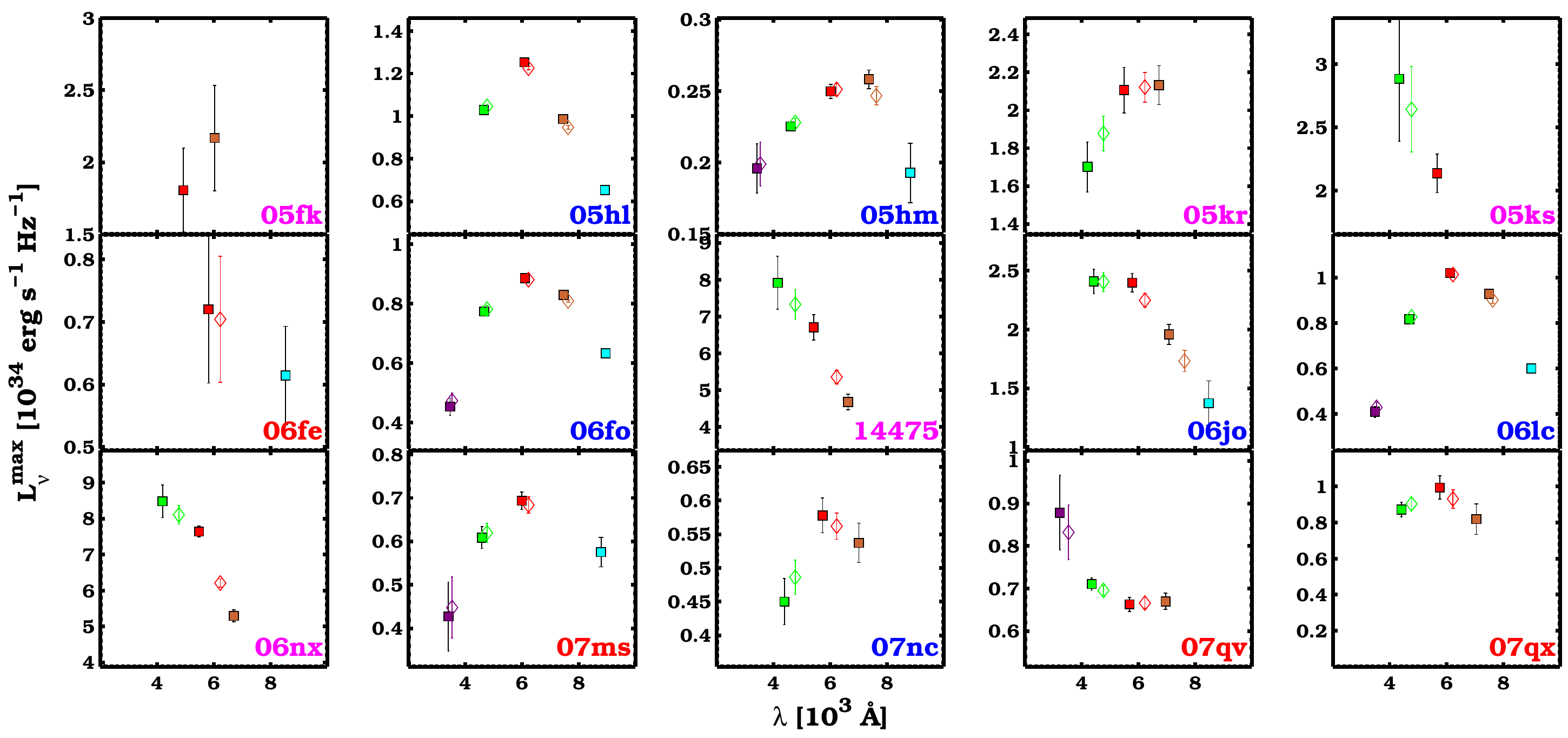} 
\includegraphics[width=13cm]{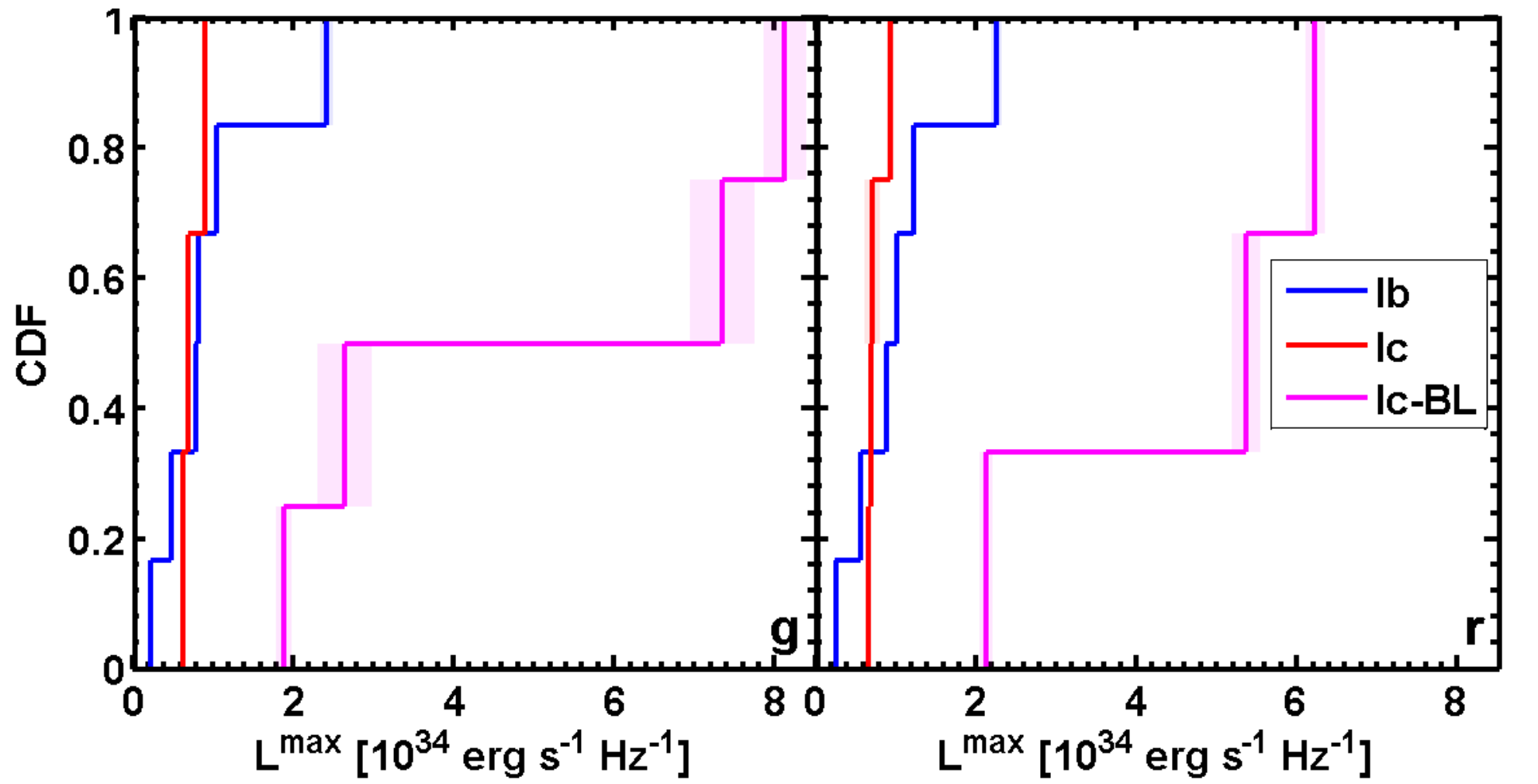}  
 \caption{\label{Fmax_nohost} 
(\textit{Top panel:}) Peak luminosity per unit frequency as a function of the effective wavelength of the filter in the rest frame for 15 SNe~Ib/c in the SDSS sample (with host-extinction corrections). Violet, green, red, dark yellow, and cyan correspond to $ugriz$ filters. Blue, red, and magenta labels correspond to SNe~Ib, Ic and Ic-BL, respectively.
 (\textit{Bottom panels:}) Cumulative distributions of the peak luminosity per unit frequency in $g_{z=0}$ (left) and $r_{z=0}$ (right) bands for SNe~Ib (6), Ic (3-4) and Ic-BL (4-3) in the SDSS sample. 
Shaded areas indicate the uncertainties for F$_{\rm max}$.
 SNe~Ic-BL are clearly brighter than SNe~Ib and Ic.}
\end{figure*}

\subsubsection{Light curve absolute maxima, L$_\nu^{\rm max}$}
\label{sec:lcpeak}

The fit of the light curves presented in Sect.~\ref{sec:fit} also provides the value of the observed peak fluxes 
(see F$_{\rm max}$ in Table~\ref{fitparam}).
With the known luminosity distance (D$_L$) and total extinction for each object (see Table~\ref{sample}), 
we compute the peak luminosities (per unit frequency, L$_\nu^{\rm max}$) for each light curve and 
SN (L$_{\nu_e}^{\rm max}=4\pi D_L^2 (1+z)^{-1} $F$_{\nu_o}^{\rm max}$).
Here $\nu_o$ corresponds to the observed effective frequency of the SDSS filter, while  $\nu_e$ corresponds to the rest frame of the SN. To compare the
results among objects with different redshifts, we interpolate the L$_\nu^{\rm max}$
at the effective wavelengths of the $gri$ filters. L$_\nu^{\rm max}$ as a function of wavelength is shown in 
the top panel of Fig.~\ref{Fmax_nohost}, where the empty diamonds represent the interpolated fluxes.
The cumulative distributions of L$_\nu^{\rm max}$ in the $g$ and $r$ bands are shown for SNe~Ib, Ic and Ic-BL 
in the bottom panel of Fig.~\ref{Fmax_nohost}. These plots show that
SNe~Ic-BL are clearly more luminous than SNe~Ib and Ic in both bands, 
with high significance (p-value$\leq$0.05 in both bands).  SNe~Ic and Ib have a similar distribution of L$_\nu^{\rm max}$. This result is similar to what was found by \citet{cano13}.

If we exclude the host-extinction corrections, SNe~Ic-BL are still more luminous than both SNe~Ic and SNe~Ib, and SNe~Ic appears slightly brighter than SNe~Ib (p-value$=$0.20, 0.13 for the $g$ and $r$ bands, respectively). A similar trend was visible 
for the $V$ and $R$ bands in the sample by \citet[][their figure~10]{drout11}.

\subsubsection{Pseudo-bolometric luminosity}
\label{sec:bolonoext}

\begin{figure*}
\centering
\includegraphics[width=17.5cm,angle=0]{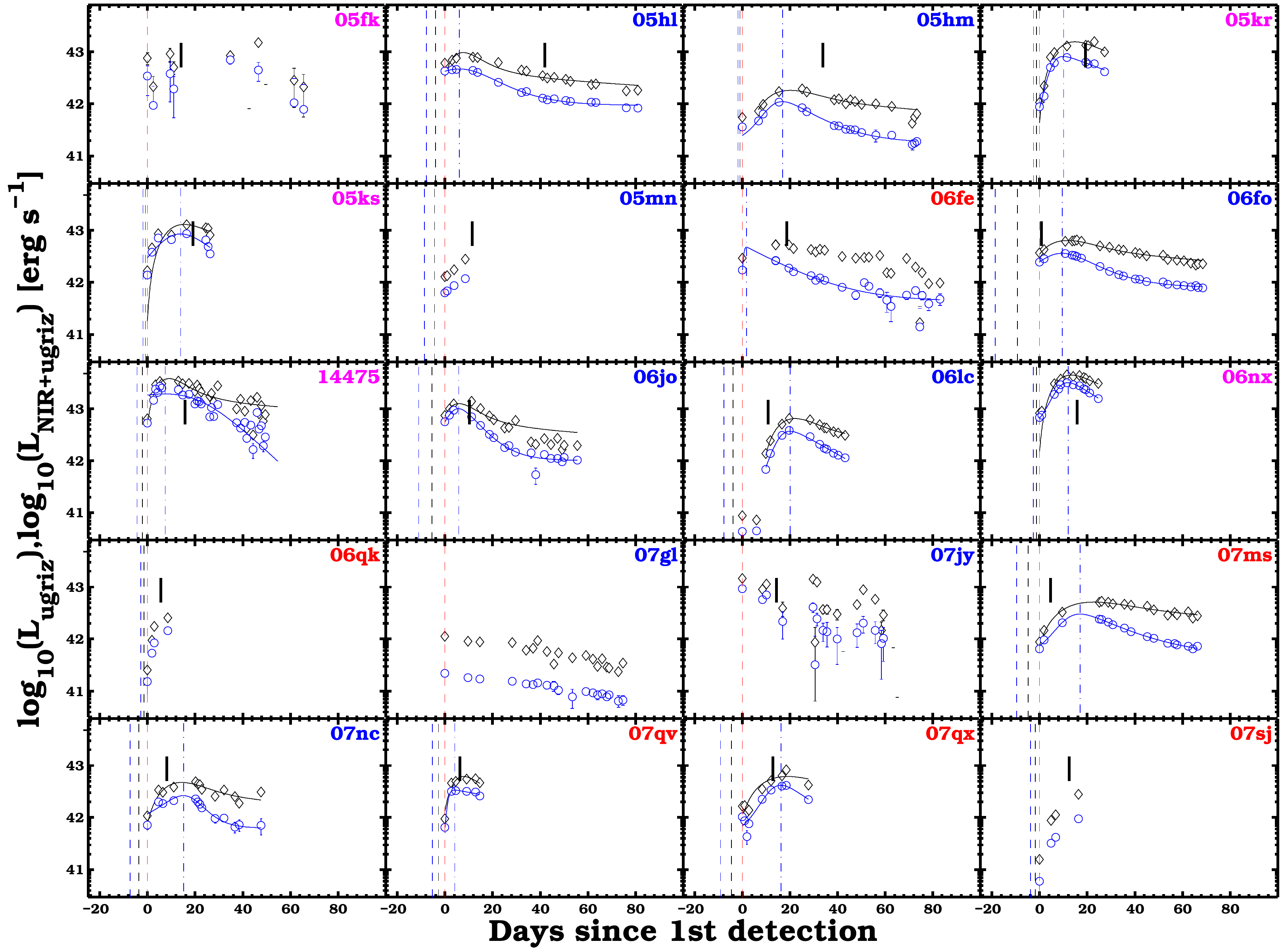}  
 \caption{\label{boloHEno_fit} Pseudo-bolometric (L$_{ugriz}$, blue circles) and bolometric (L$_{{\rm NIR}+ugriz}$, black diamonds) light curves for the sample of SNe~Ib/c from SDSS. Discovery epochs are marked by red dashed lines, the last non-detections by dashed blue lines. Each pseudo-bolometric light curve has been fit with Eq.~\ref{fitformula}, and the best fit is shown by a solid blue line. The peak epoch is marked by a blue/black dashed-dotted line. Each bolometric light curve has been fit with the Arnett model  (see Sect.~\ref{sec:NiEM}) when the explosion date and the maximum epoch are known, and the best fit is shown by a black solid line. The epoch of the first spectrum is marked by a black segment. Blue, red, and magenta labels correspond to SNe~Ib, Ic, and Ic-BL, respectively.}
\end{figure*}

\begin{figure*}
\centering
$\begin{array}{ccc}
\includegraphics[width=6cm]{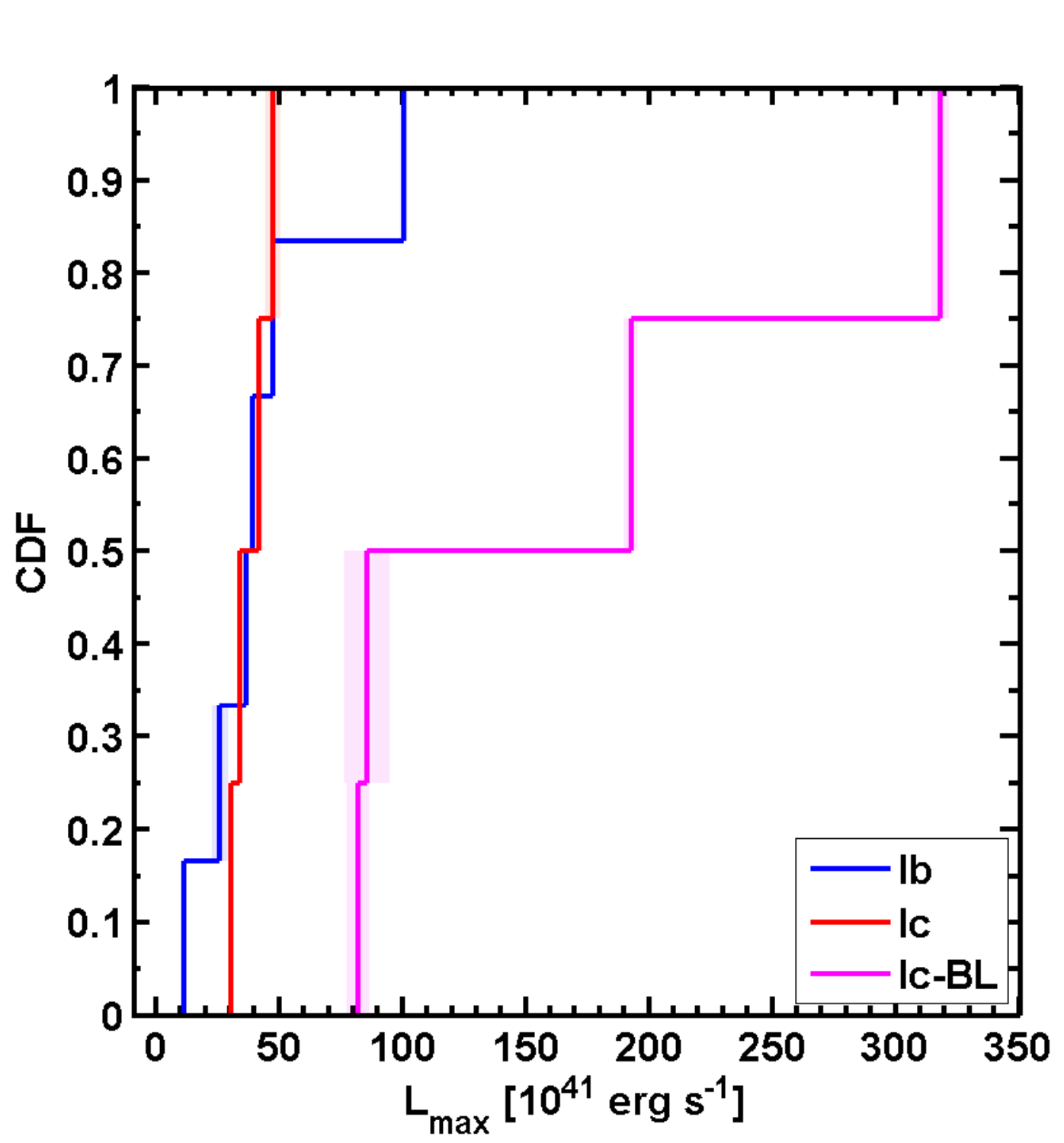}&
\includegraphics[width=6cm]{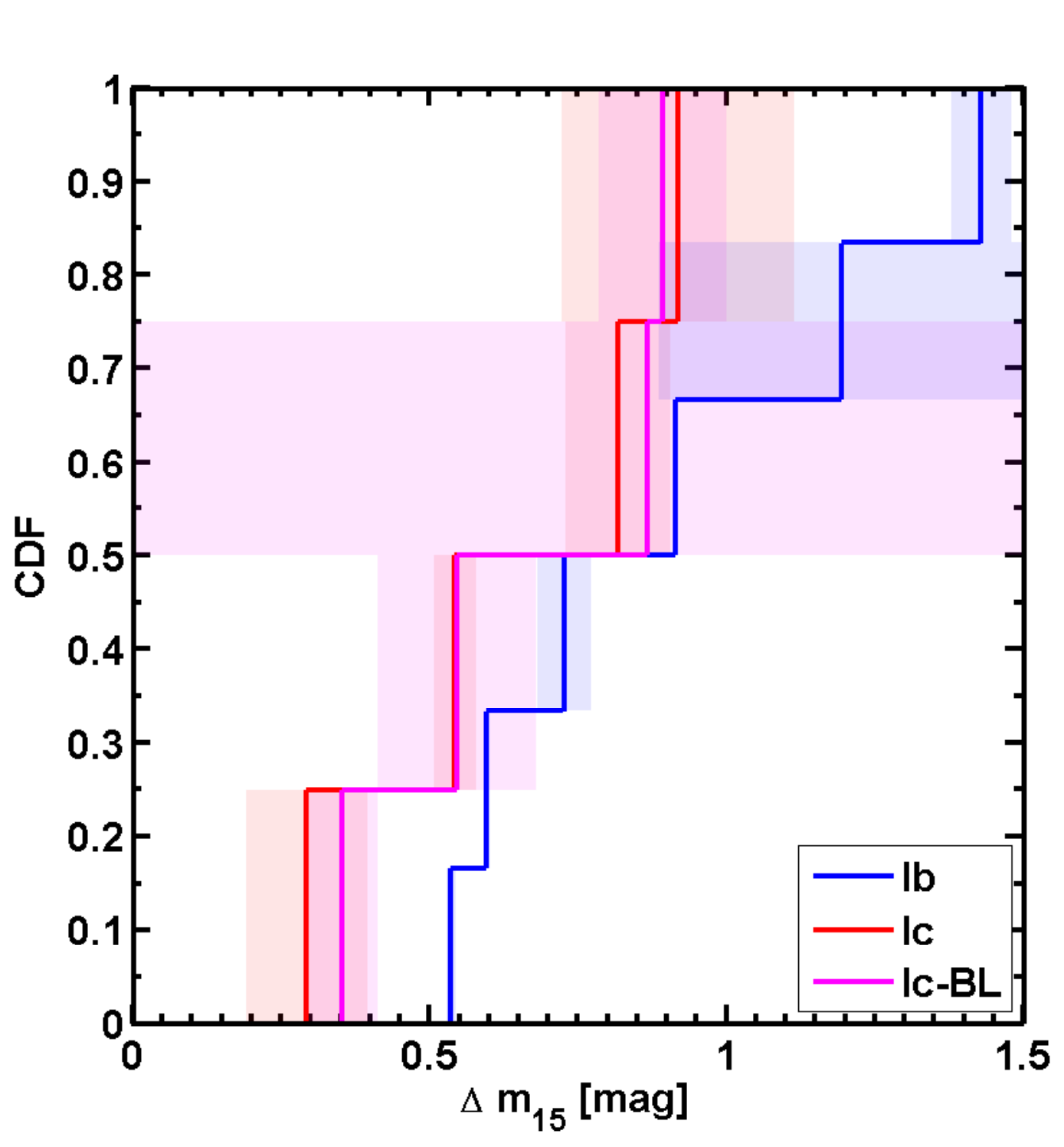} & 
\includegraphics[width=6cm]{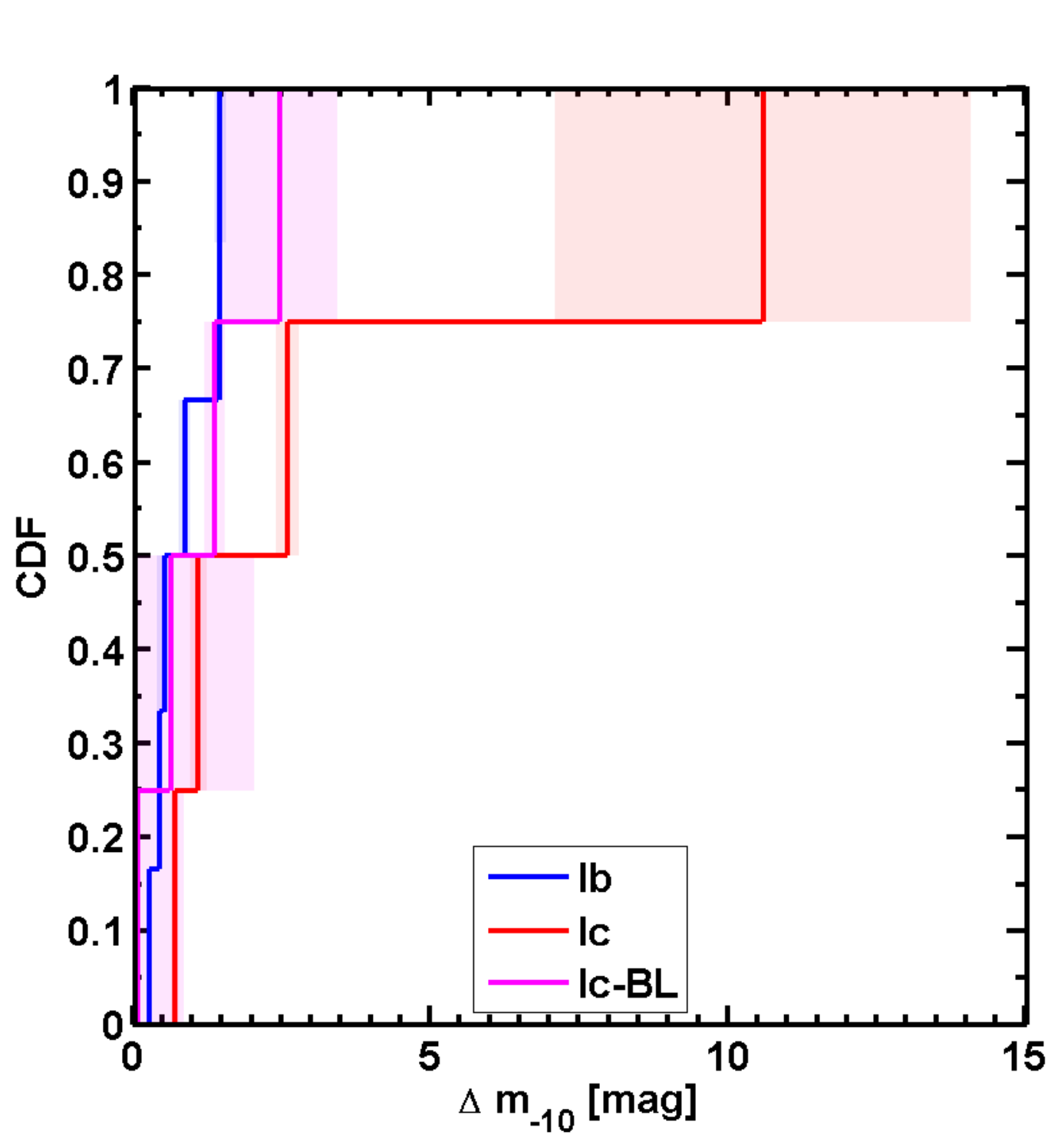}\\
\end{array}$
 \caption{\label{Lmax_cdf_noext} ({\it Lefthand panel:}) Cumulative distributions for the pseudo-bolometric peak luminosity with host-extinction correction. SNe~Ib and Ic show similar distributions, SNe~Ic BL are more luminous. ({\it Center panel:}) $\Delta$m$_{15}$ of SNe~Ib, Ic and Ic-BL.  There is no significant difference among the three SN classes. ({\it Righthand panel:}) $\Delta$m$_{-10}$ of SNe~Ib, Ic and Ic-BL. Helium-poor SNe show higher values compared to those of SNe~Ib. Shaded areas indicate the uncertainties for each SN.}
\end{figure*}

Using D$_L$, 
multi-band fluxes, $E(B-V)_{\rm MW}$ and $E(B-V)_{\rm host}$, we computed pseudo-bolometric $ugriz$ light curves for each SN. Each broadband flux was corrected for the  extinction computed at its corresponding wavelength assuming a \citet{cardelli89} reddening law (R$_V~=~$3.1). 
For each epoch, the $ugriz$ SED was de-redshifted, multiplied by 1$+z$, and integrated between rest frame 3500~\AA\ and 9000~\AA, in order to compare the same SED portion for each event.
The resulting flux was then multiplied by 4$\pi$D$_L^2$ to obtain the pseudo-bolometric luminosity (L), and is displayed in Fig.~\ref{boloHEno_fit}.
Each pseudo-bolometric light curve observed before and after maximum was fit with the expression in Eq.~\ref{fitformula}, and the resulting best fit is shown. This fit allowed us to measure the peak luminosity and its epoch, as well as the parameters derived for the single-band light curves. In Table~\ref{fitparam_bolo} we report these parameters for 14 pseudo-bolometric light curves.

Comparing the host extinction-corrected L$_{\rm max}$ between SNe~Ib and Ic, we found that the two distributions
overlap (Fig.~\ref{Lmax_cdf_noext}, lefthand panel).
SNe~Ic-BL are more luminous than the other types, with high statistical significance (p$<$0.05).
This result was also found by \citet{drout11} and \citet{cano13}.
If the host-extinction corrections are neglected, SNe~Ic appear brighter than SNe~Ib, consistent with the individual filtered peak fluxes but with higher statistical significance (p-value~$=$~0.07).


There is no visible difference between the SN types 
when we compare the pseudo-bolometric 
$\Delta$m$_{15}$ (Fig.~\ref{Lmax_cdf_noext}, bottom left panel). 
The pseudo-bolometric $\Delta$m$_{-10}$ comparison (bottom right panel of Fig.~\ref{Lmax_cdf_noext})) reveals that SNe~Ic 
and Ic-BL have slightly higher values than those of SNe~Ib (same trend as in Fig.~\ref{dm10comp}, this time with p-value~$=$~0.43, respectively).
These results do not strongly depend on the extinction corrections.

\subsubsection{Temperature}
\label{sec:temp}

\begin{figure*}
\centering
\sidecaption
\includegraphics[width=12cm]{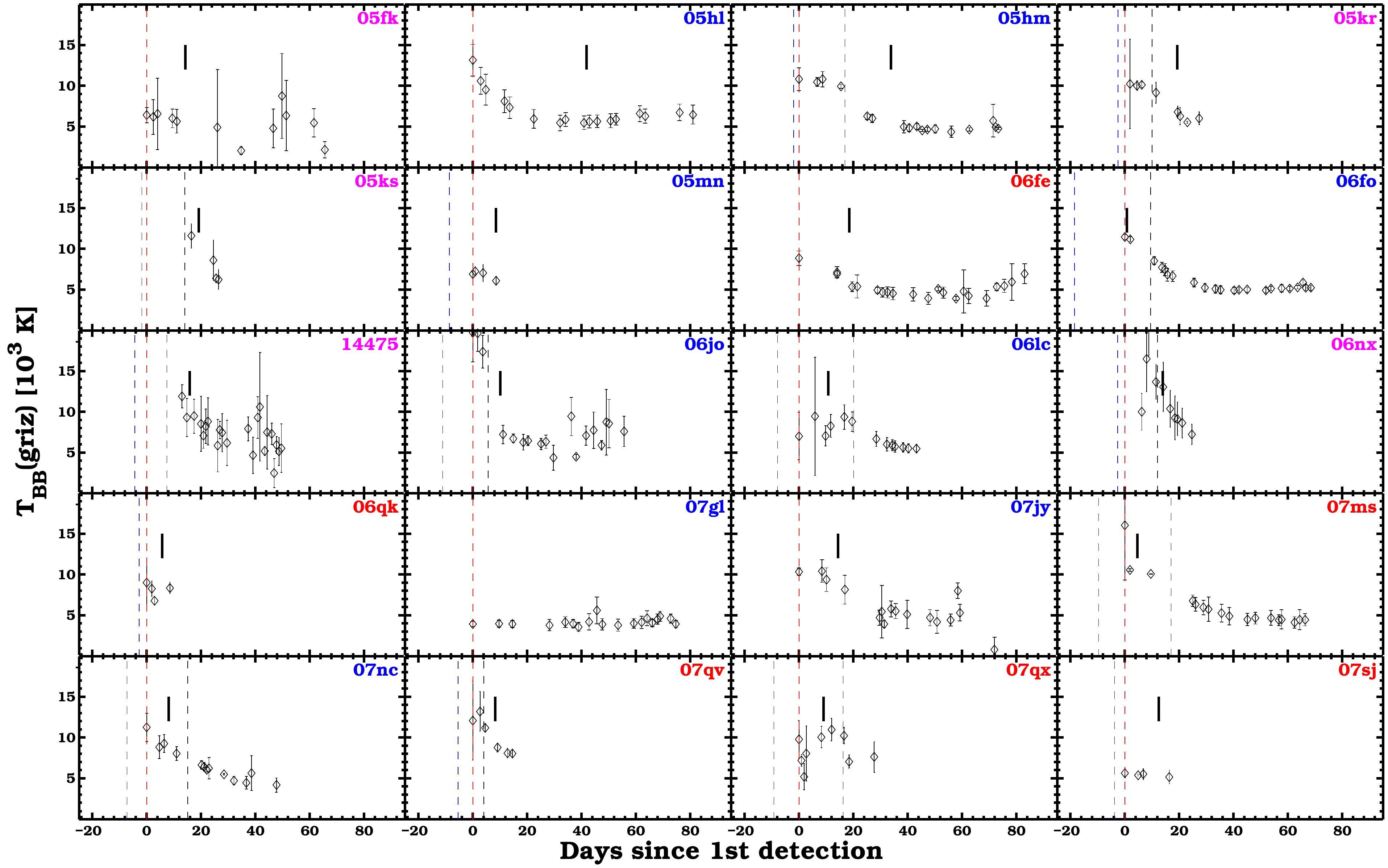}  
 \caption{\label{T_noext}Temperature evolution for the 20 SDSS SNe~Ib/c from the BB fits on the $griz$ SEDs. Discovery epochs are marked by red dashed lines, the last non-detections by dashed blue lines. The epoch of the first spectrum is marked by a black segment. Here we include the host-extinction corrections. Maximum luminosity epochs are marked by black, vertical dashed lines. Blue, red, and magenta labels correspond to SNe~Ib, Ic, and Ic-BL, respectively.}
\end{figure*}

\begin{figure*}
\centering
$\begin{array}{cc}
\includegraphics[height=7cm]{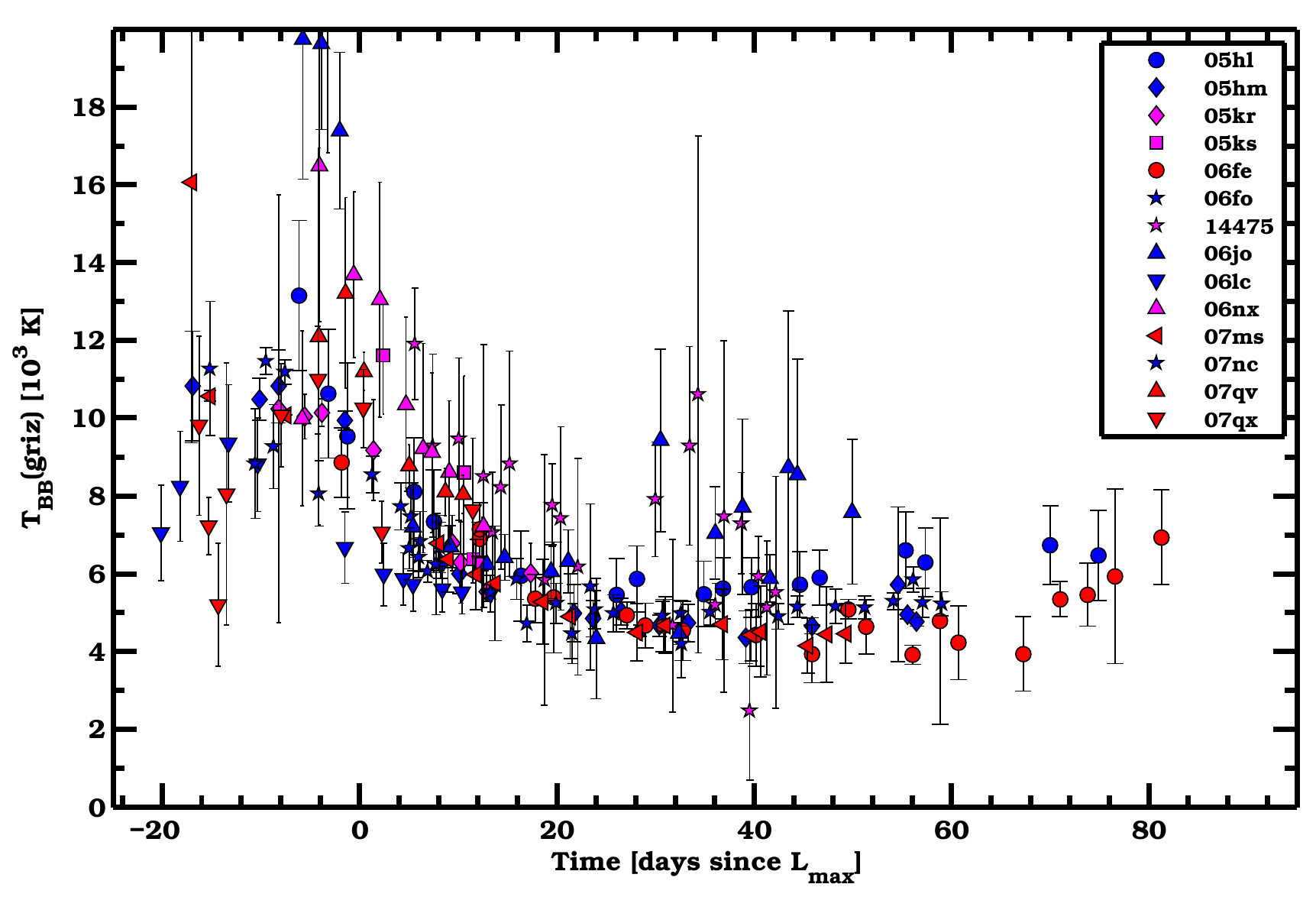}  
\includegraphics[height=7cm]{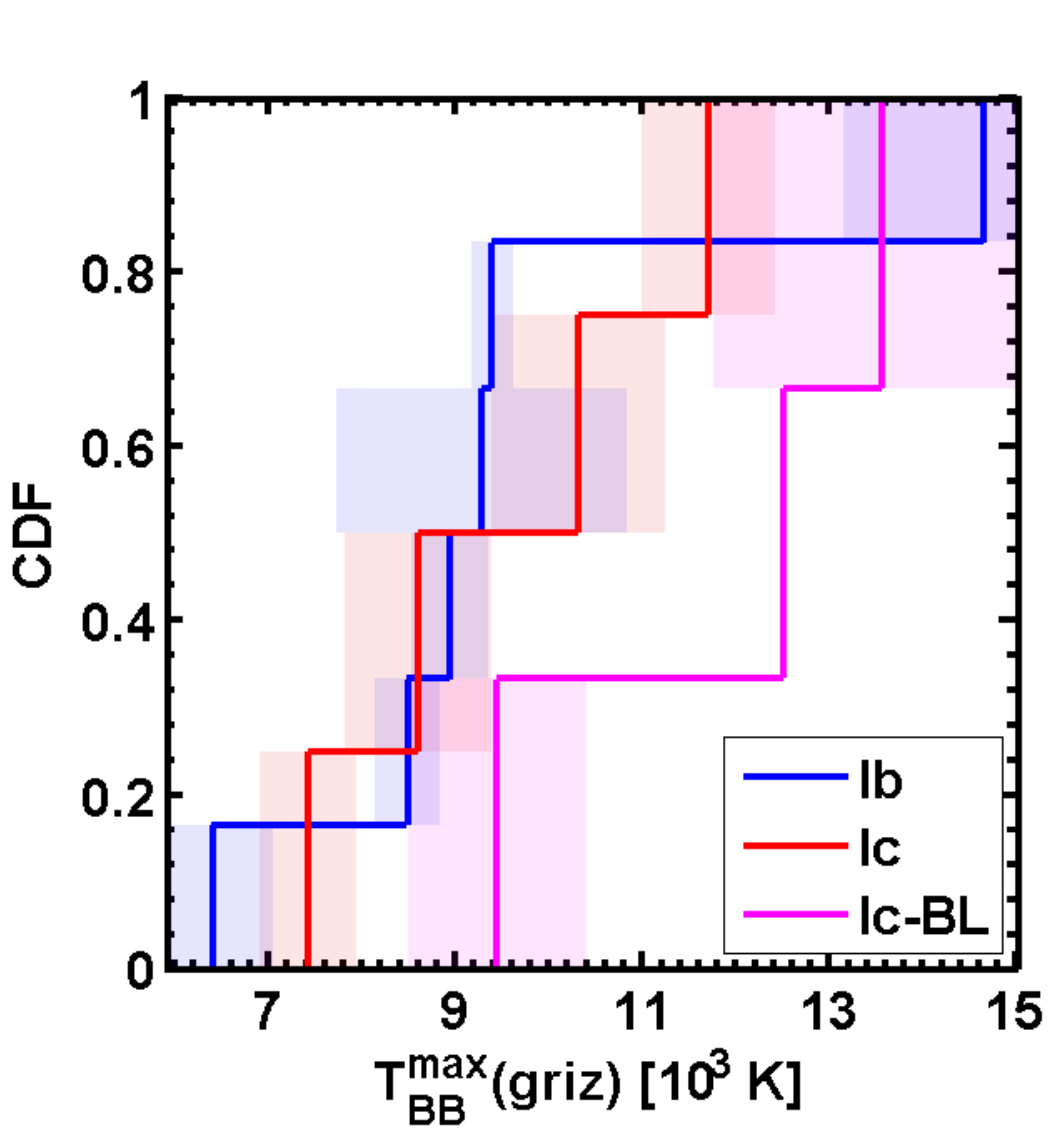}  
\end{array}$
 \caption{\label{T_noext_oneplot}({\it Lefthand panel:}) Temperature evolution for the 14 SDSS SNe~Ib/c with known L$_{\rm max}$ epoch. Here we include host-extinction corrections. The temperature evolution is similar for all the events. ({\it Righthand panel:}) Cumulative distributions for the temperatures at maximum light of SNe~Ib, Ic, and Ic-BL. Here we include host-extinction corrections. There is no significant difference among the three SN types. Shaded areas indicate the uncertainty for each SN.}
\end{figure*}

To compute the temperature (T$_{\rm BB}$) evolution for each object ,
we fitted the rest-frame
SEDs (including host-extinction correction) with a BB function.
We excluded the $u$ band from the fit, given the strong line-blanketing effect at 
these wavelengths (also $g$ band is partially affected), as well as the large statistical errors 
(due to poor sensitivity) for this pass-band for many SNe, as well as large extinction uncertainties.

 We show the T$_{\rm BB}$ evolution in 
Fig.~\ref{T_noext} for each object.
It is clear that T$_{\rm BB}$ generally decreases in the first $\sim$30 days, 
until it reaches an almost constant value. 
For a few objects (e.g., SNe~2006lc and~2007qx)  we also observe a smooth rise in 
the first days after detection, since we likely detected them soon after outburst.
In Fig.~\ref{T_noext_oneplot} (lefthand panel) we plot all the T$_{\rm BB}$ profiles to 
highlight their similar evolution. Around maximum light, 
the measured temperatures after host-extinction corrections are similar among the three SN classes (T$^{\rm max}_{\rm BB}$~$=$~7000$-$15000~K, only SNe~Ic-BL appear marginally hotter). This is a consequence of the assumption that all the SNe have the same intrinsic value of $g-r$ at just ten days past maximum. 
Figure~\ref{T_noext_oneplot} (righthand panel) displays the 
cumulative distribution functions (CDFs)
of T$^{\rm max}_{\rm BB}$ for SNe~Ib and Ic/Ic-BL.
The temperatures at tens of days after maximum light are less affected by the extinction corrections given that most of the flux at those epochs is emitted in the redder bands. Most of the objects show past-peak temperatures of $\sim$5000~K.

When we do not consider host-extinction correction, T$^{\rm max}_{\rm BB}$ ranges from 5500 to 11000~K, with SNe~Ic and Ic-BL that appear hotter than SNe~Ib. The range of temperatures can be either due to intrinsic differences and/or 
to different unaccounted reddening in the host galaxy. We assumed that the colou difference after maximum light is due to the host extinction to compute $E(B-V)_{\rm host}$ (see Sect.~\ref{sec:hostext}).

\subsubsection{Photospheric radius}
\label{sec:radius}

\begin{figure*}
\sidecaption
\centering
\includegraphics[width=12cm]{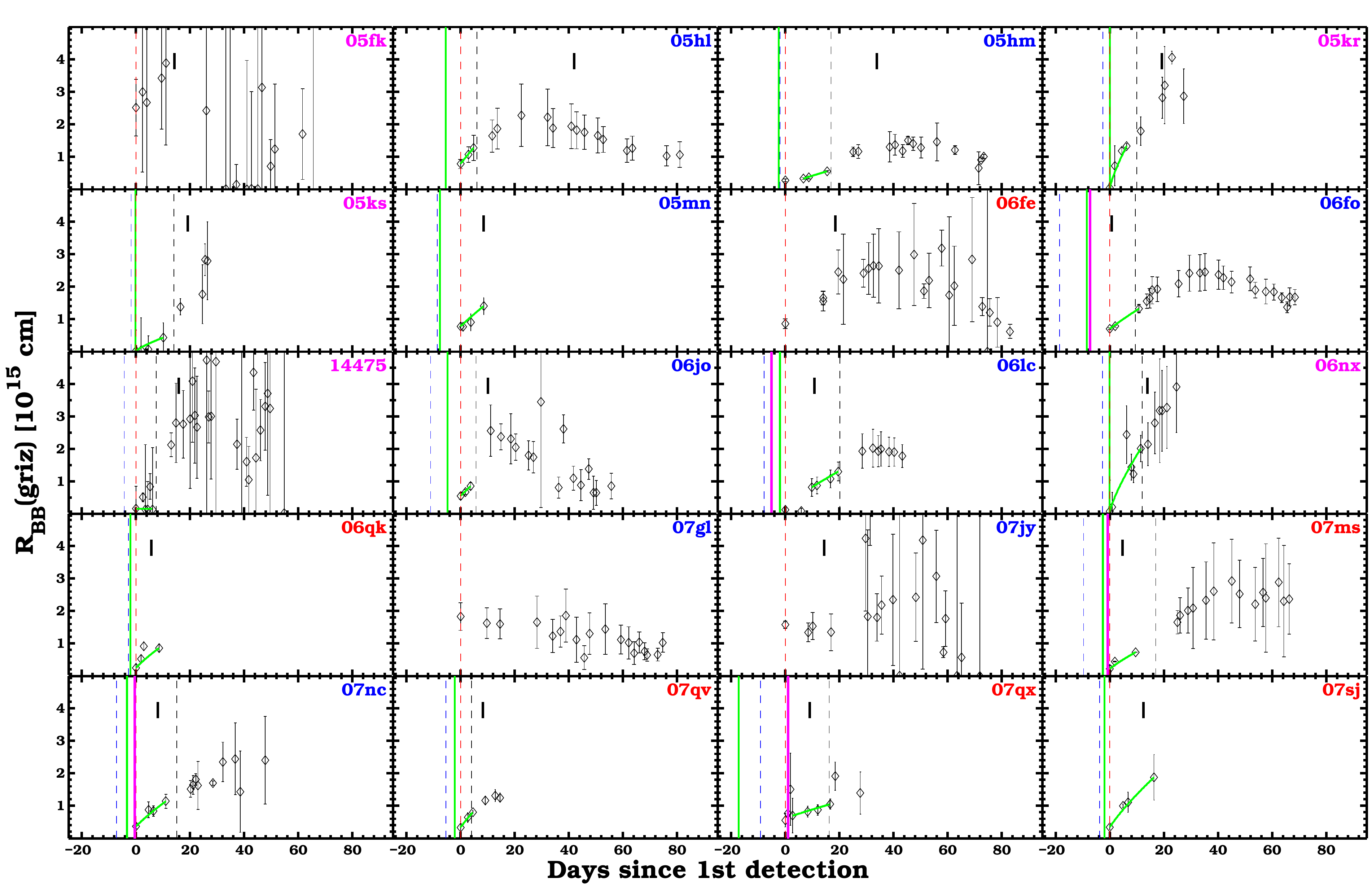}  
 \caption{\label{R_nohostext} Photospheric radius evolution for the 20 SDSS SNe~Ib/c from the BB fits on the $griz$ SEDs. Discovery epochs are marked by red dashed lines, the last non-detections by dashed blue lines. The epoch of the first spectrum is marked by a black segment. Here we include host-extinction corrections. The vertical green line indicates the explosion date as derived from the best PL fit (t$_{\rm expl}^{PL}$, see Sect.~\ref{sec:explo_error}) on the prepeak BB radii, which is shown with a green solid line. The PL-derived explosion dates show that most of the SNe were discovered at early epochs, when the photospheric radius was remarkably small, and thus that the $^{56}$Ni is likely to be present in the outer layers for most of them (see fig. 1 in PN13 and Sect.~\ref{sec:nimix}).
The vertical magenta line indicates the minimum explosion day t$_{\rm min}$
computed from the spectral velocities (see Sect.~\ref{sec:explo_error}). Both t$_{\rm min}$ and maximum luminosity epochs are marked by black vertical dashed lines. Blue, red, and magenta labels correspond to SNe~Ib, Ic, and Ic-BL, respectively.}
\end{figure*}

The BB fits on the $griz$ SEDs (corrected for the host extinction) combined with the known distances also allow an estimate of the photospheric radius 
evolution (Fig.~\ref{R_nohostext}). The radius linearly expands at early times, reaches a maximum, then slowly decreases.
This behavior is remarkably similar for all the objects. The maximum radius is reached around 20 days after peak luminosity for 
most of the events. Photospheric radii at the epoch of maximum light (R$^{\rm max}_{\rm BB}$) 
have typical values of 10$^{15}$~cm. The results on the radius are not significantly affected by the host-extinction corrections.

\begin{figure}
\centering
$\begin{array}{cc}
\includegraphics[width=4.5cm]{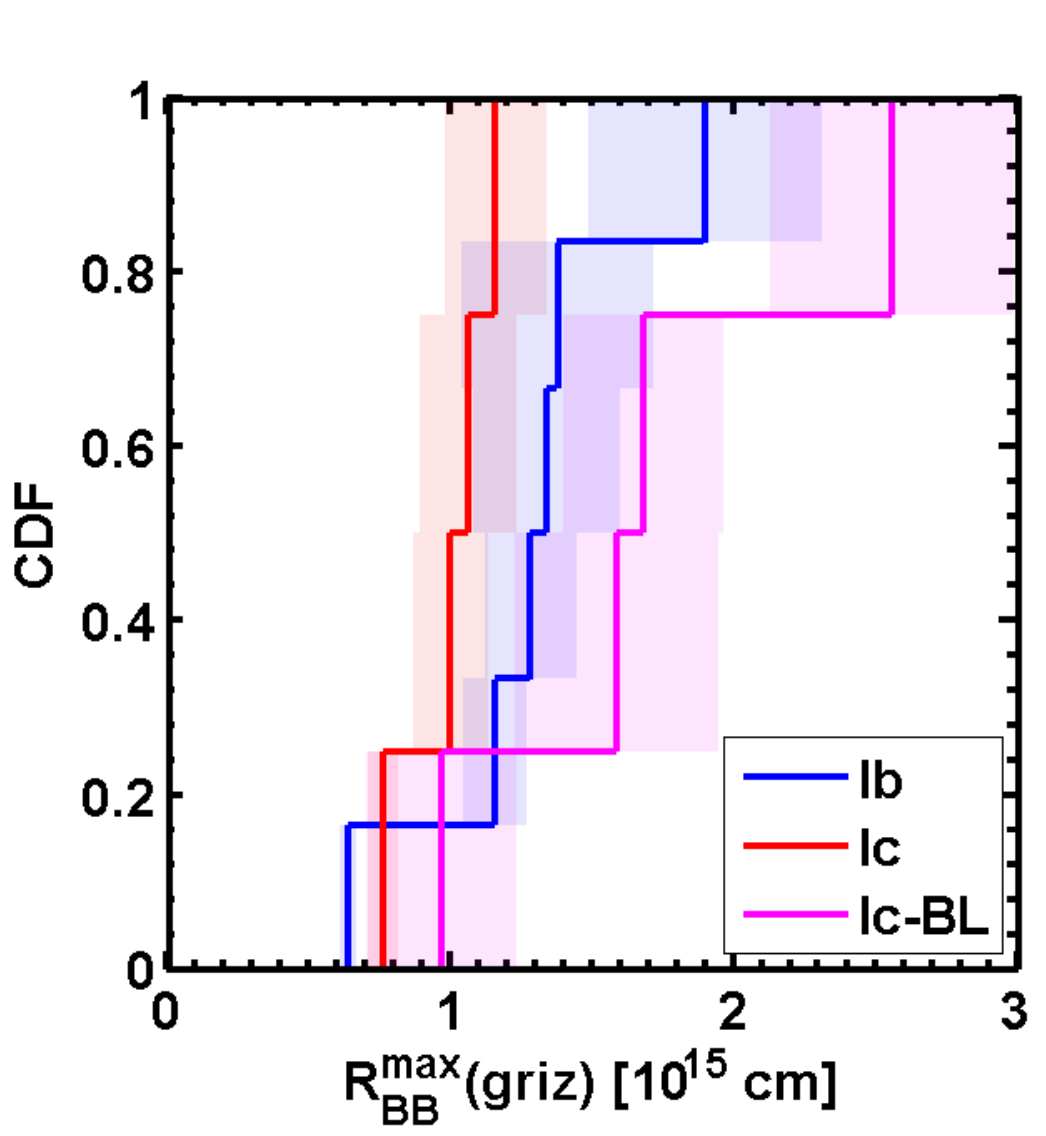} & 
\includegraphics[width=4.5cm]{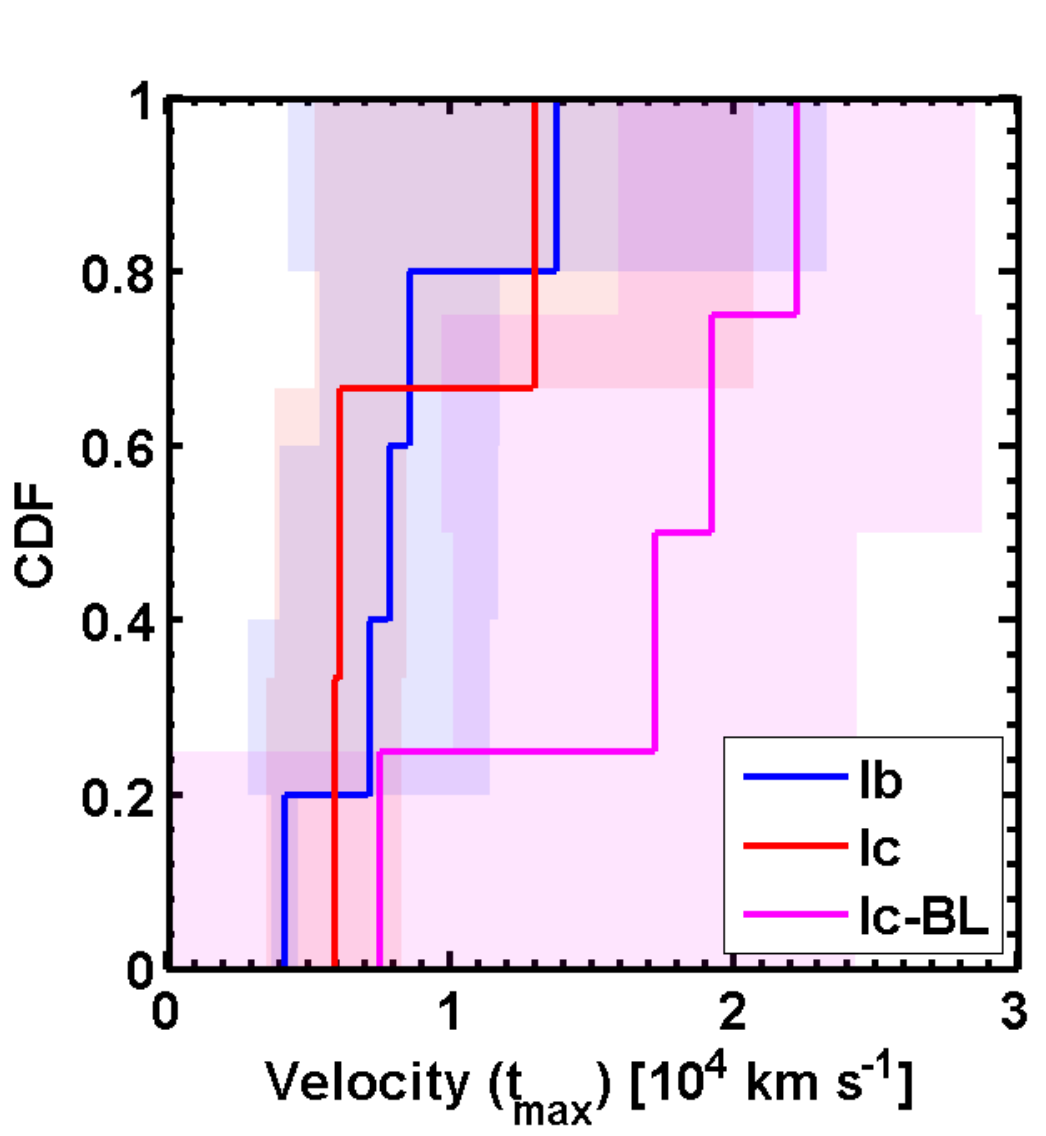} 
\end{array}$
 \caption{\label{R_noext_oneplot} 
 Cumulative distributions for the photospheric radii and the photospheric velocities at maximum light of SNe~Ib, Ic, and Ic-BL. 
Photospheric velocities are estimated from the photospheric radii divided by the bolometric light curve rise times. Here we include host-extinction corrections. Shaded areas indicate the uncertainty for each SN.}
\end{figure}

When comparing R$^{\rm max}_{\rm BB}$ for the different SN types, SNe~Ic tend 
to have a typical R$^{\rm max}_{\rm BB}$ that is slightly 
smaller than those of SNe~Ib and Ic-BL 
(lefthand panel of Fig.~\ref{R_noext_oneplot}).
Given that the rise times of SNe~Ic and Ic-BL are typically 
shorter than those of SNe~Ib,
this result implies that the photospheric velocities of SNe~Ib and Ic are similar at maximum light (although the uncertainties are large, mainly owing to the error on the explosion date), whereas the same line of reasoning demonstrates that the photospheric velocities of SNe~Ic-BL must be considerably higher (see the righthand panel of Fig.~\ref{R_noext_oneplot}). The latter is of course expected since SNe~Ic-BL have higher spectroscopical expansion velocities by definition.

We compared this result with the velocities that we measured from the spectra (from \ion{Si}{ii}~$\lambda$6355 and \ion{He}{i}~$\lambda$5876 absorption minima), which are reported in Table~\ref{tab:speclog}. A better approximation for the photospheric velocity would be what is obtained from the absorption minimum of \ion{Fe}{ii}~$\lambda$5169 \citep{branch02}. However, given the quality of the spectra, it is difficult to perform this measurement. For a few poor-quality spectra, we did not manage to properly measure the expansion velocity at all. Our spectral measurements confirm that SNe~Ic-BL show higher velocities than SNe~Ib and SNe~Ic. We do not find any clear difference between SN~Ib and Ic expansion velocities, but this spectral sample is not well suited to such a comparison, given the small number of high-quality spectra and the different phases at which the spectra were obtained.

\subsection{Progenitor and explosion properties}

The SDSS-II SN dataset allows us to derive (or at least place limits on) the properties of the SN progenitors and of the explosion parameters, based on the comparison between observed light curves and simple theoretical models at the photospheric phase.

\subsubsection{Explosion dates}
\label{sec:explo_error}

The assumption on the explosion date strongly affects the estimates of the SN progenitor properties, 
as shown by PN13 in the case of PTF10vgv. A precise knowledge of the time of explosion is useful for obtaining information about the $^{56}$Ni mixing in the SN, as well as putting strong limits on the pre-SN radius. Also other physical properties such as $^{56}$Ni mass, explosion energy, and ejecta mass can be obtained with higher precision if the explosion date is well established, since their estimates depend on the rise time.
We discuss several methods here for estimating the explosion epoch.

The most straight-forward approach to estimating the explosion date of a SN is to take the average
between the first detection and the last non-detection epochs and the uncertainty as half of the time interval 
determined by these two measurements. This is the explosion date 
used throughout this paper (coined t$_{\rm expl}$).
The precision of this method is largely affected by the cadence for the light curve sampling. 
With the cadence of the SDSS-II SN survey, this method implies a typical uncertainty on the explosion epoch of $\pm$2$-$3 days
for our SN sample. This uncertainty is small enough to exclude the existence of an initial long plateau \citep{dessart12_OriginIbc} in the bolometric light curves (see Sect.~\ref{sec:nimix}).
Also, the prediscovery images must be deep enough to ensure that the SN was 
not missed simply because it was too faint.  
We discuss this aspect in Sect.~\ref{sec:nimix}, where it is shown that, for most of our targets, the limits on the luminosity of the last non-detection are deep enough to exclude a dark plateau phase in the predicted luminosity range. 

\begin{figure*}
\centering
\includegraphics[width=18cm]{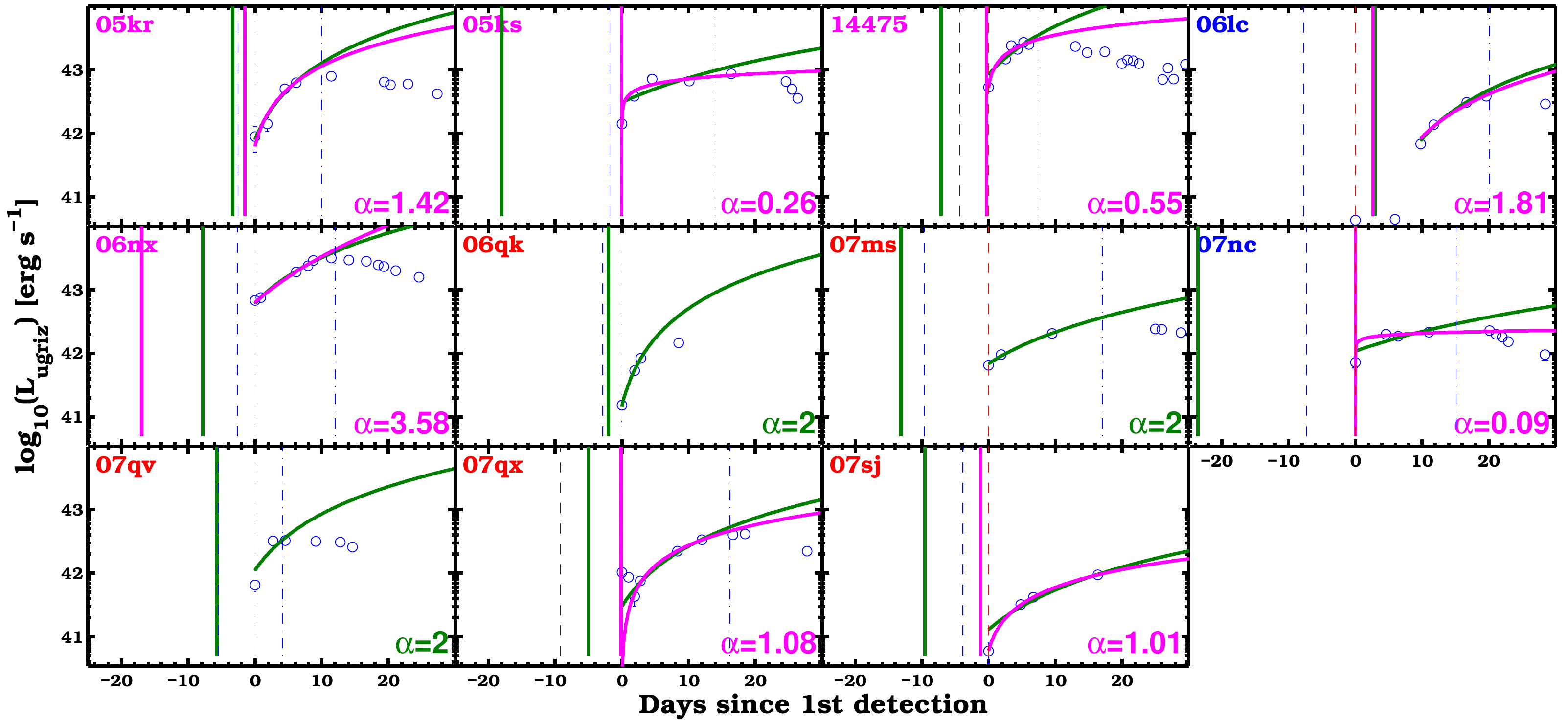}      
 \caption{\label{riseshape} Power-law (PL) fits on the early part of the pseudo-bolometric light curves of 11 SDSS SNe~Ib/c. Here we correct for host extinction.
 Green solid lines correspond to $\alpha=2$, magenta solid lines to $\alpha$ as a free parameter, whose value is reported in each sub-panel. Vertical green and magenta lines indicate the inferred explosion date from the PL fit. Blue and red dashed lines indicate the epochs of last non-detection and first detection, respectively. 
 Black dashed-dotted lines indicate the epoch of maximum light. Blue, red, and magenta labels belong to SNe~Ib, Ic, and Ic-BL, respectively.}
\end{figure*}

 Another possible way to define the explosion date is to extrapolate the light curve back in time when assuming some 
functional form, 
usually a power law (PL), $\rm L(t)=A(t-t_{\rm expl}^{PL})^{\alpha}$. \citet{cao13} applied a PL fit to the early 
light curve of iPTF13bvn, and the same was done by \citet{pignata11} for SN~2009bb. 
Both PL and broken PL fits have also been used to determine the explosion dates 
of SNe~Ia \citep[e.g., SN~2011fe and SN~2014J,][]{nugent11,zheng14,goobar14}.
Different PL shapes have been used to fit early-time light curves of SE~SNe. 
This approach assumes that the explosion coincides with the beginning of the light curve rise and therefore
disregards any dark plateau phase after collapse and before the light curve rise. We designate 
the power-law extrapolated explosion day $\rm t_{\rm expl}^{PL}$.
Figure~\ref{riseshape} shows our best PL fits to the early time bolometric light curves of those SNe that 
have enough data points before maximum light. We adopt $\alpha=$~2 (consistent with homologous expansion, e.g., \citealt{arnett82}) when we only have three points on the rise, whereas
we allow $\alpha$ to be a free parameter when we have at least four points on the rise (corresponding $\alpha$ values are reported in Table~\ref{fitparam_bolo}). 
The resulting t$_{\rm expl}^{\rm PL}$ from the best fits with $\alpha$ as a free parameter 
are consistent with the explosion dates t$_{\rm expl}$
estimated in this paper.
The lines that indicate the derived t$_{\rm expl}^{\rm PL}$($\alpha$) always fall between the 
last non-detection and the first detection, except in the case of SN~2006nx and, as expected, in that of SN~2006lc, where an early plateau is detected (see Sect.~\ref{sec:nimix}).
Conversely, when $\alpha\equiv$~2, 
we obtain too early explosion dates. 
The lines that indicate t$_{\rm expl}^{\rm PL}$($\alpha=2$) 
mark explosion dates prior to the last non-detection epoch in most cases.

The SDSS-II SN survey's multi-band coverage allowed us to derive temperatures and radii from the BB fit in 
Sects.~\ref{sec:temp} and \ref{sec:radius}. This information can also be used
to place additional constraints on the explosion date. PN13 (their equation 11) estimate the 
photospheric radius (r$_{\rm ph}$) evolution as 
$\propto$~$(\rm t-t_{\rm expl}^{\rm r_{ph}=0})^{0.78}$. Assuming that the photospheric 
radius is similar to the BB radius 
(basically assuming a constant dilution factor), 
we fit a PL with $\alpha=$~0.78 to the prepeak BB radii presented in Fig.~\ref{R_nohostext}, 
in order to derive the explosion epoch for each SN. 
We designate the explosion epoch derived from the PL fit with $\alpha=$~0.78 of the 
BB radius evolution as t$_{\rm expl}^{\rm r_{ph}=0}$.
For all SNe, t$_{\rm expl}^{\rm r_{\rm ph}=0}$ occurs between the last non-detection and the first detection (see Fig.~\ref{R_nohostext}, where the green vertical line is always included between the vertical blue and red lines), matching our simplest explosion date (t$_{\rm expl}$) estimates. The same result is obtained when we fit the BB radii from the host-extinction uncorrected SEDs.

We also use the spectral information at the epoch of the rise for some of our SNe in order to determine the minimum time of explosion, t$_{\rm min}$ as defined by PN13 in their equation~17. From the spectra we derive the velocity (see Table~\ref{tab:speclog}), whereas the temperature (T) and bolometric luminosity (L) was obtained by interpolating T$_{\rm BB}$ and R$_{\rm BB}$ at the spectral epoch and then assuming BB emission. For five events, t$_{\rm min}$ always occurs after the last non-detection and before discovery (Fig.~\ref{R_nohostext}). This is true regardless of the host-extinction corrections. The estimates of t$_{\rm min}$ linearly depend on the assumed velocity, and the spectral velocities are only an approximation of the actual photospheric velocity. However, we find that t$_{\rm min}$ occurs between last non-detection and discovery for the vast majority of our sample also when a variation in velocity of $\pm$20\% is allowed.

We conclude 
that the t$_{\rm expl}$ estimates presented in Sect.~\ref{sec:fit} and based on the first method presented in 
this section are in good agreement with the explosion dates inferred from the model for the 
evolution of the photospheric radius (t$_{\rm expl}^{\rm r_{ph}=0}$) and with t$_{\rm min}$ (PN13). 
Therefore the rise times shown in 
Fig.~\ref{rise_vs_lambda} should also correspond to the actual rise times, 
defined as the time interval between explosion and peak. This analysis also suggests that a long early plateau phase
is not common in SNe~Ib/c, since the light curves tend to rise soon after explosion  (see Sect.~\ref{sec:nimix}).

\subsubsection{$^{56}$Ni mass, explosion energy and ejecta mass}
\label{sec:NiEM}

\begin{figure*}
\centering
$\begin{array}{ccc}
\includegraphics[width=6cm]{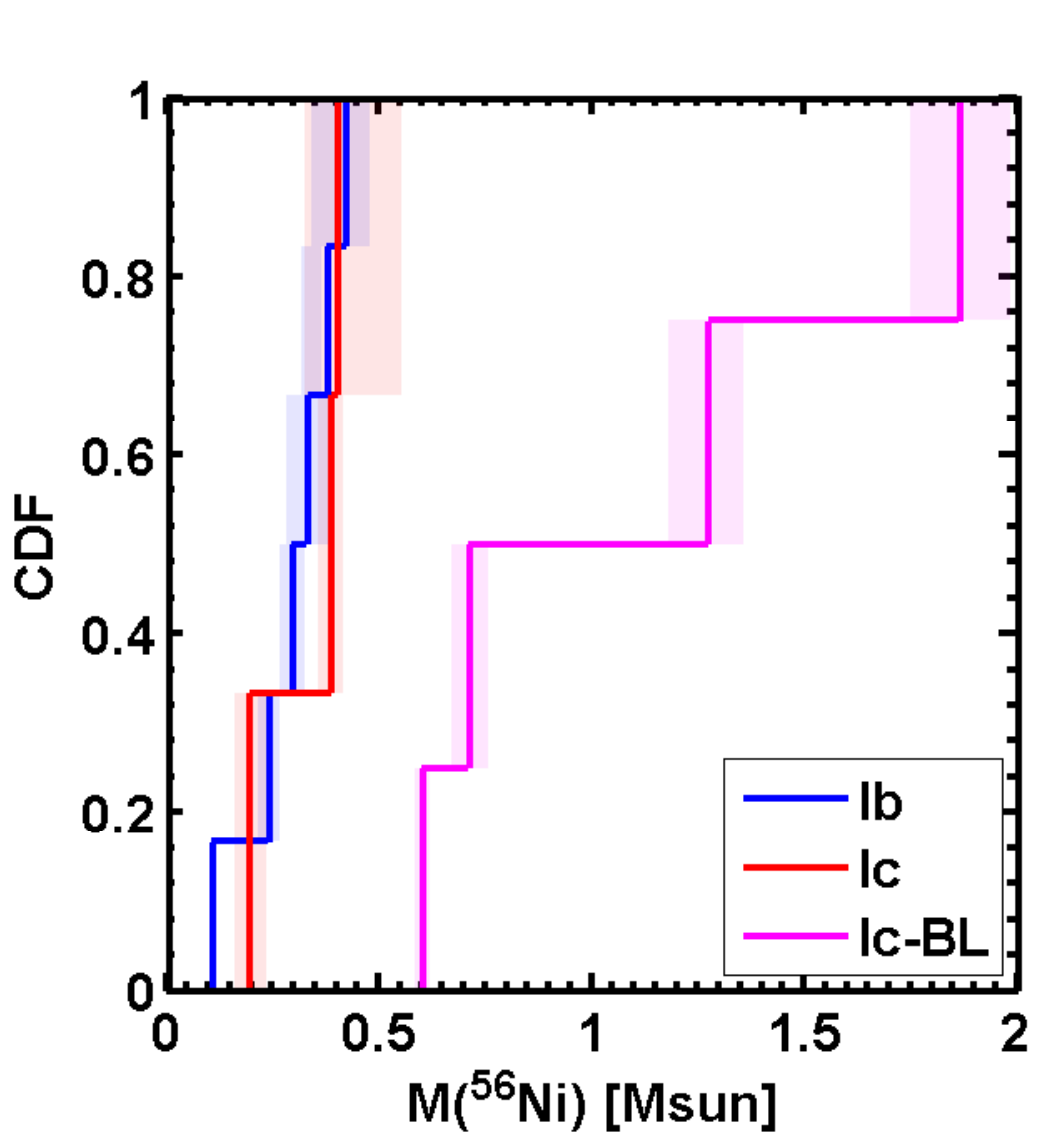}&
\includegraphics[width=6cm]{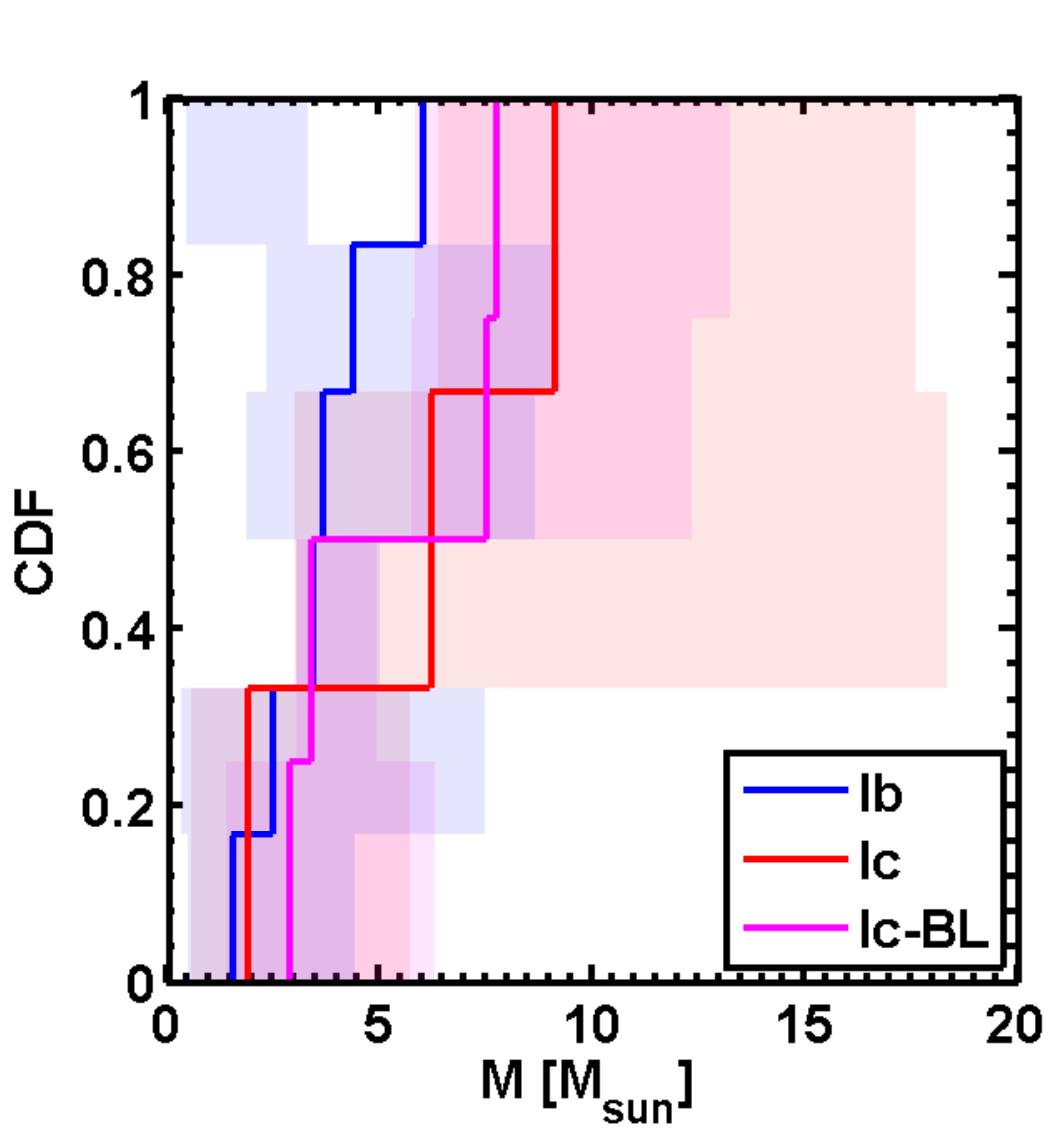}&
\includegraphics[width=6cm]{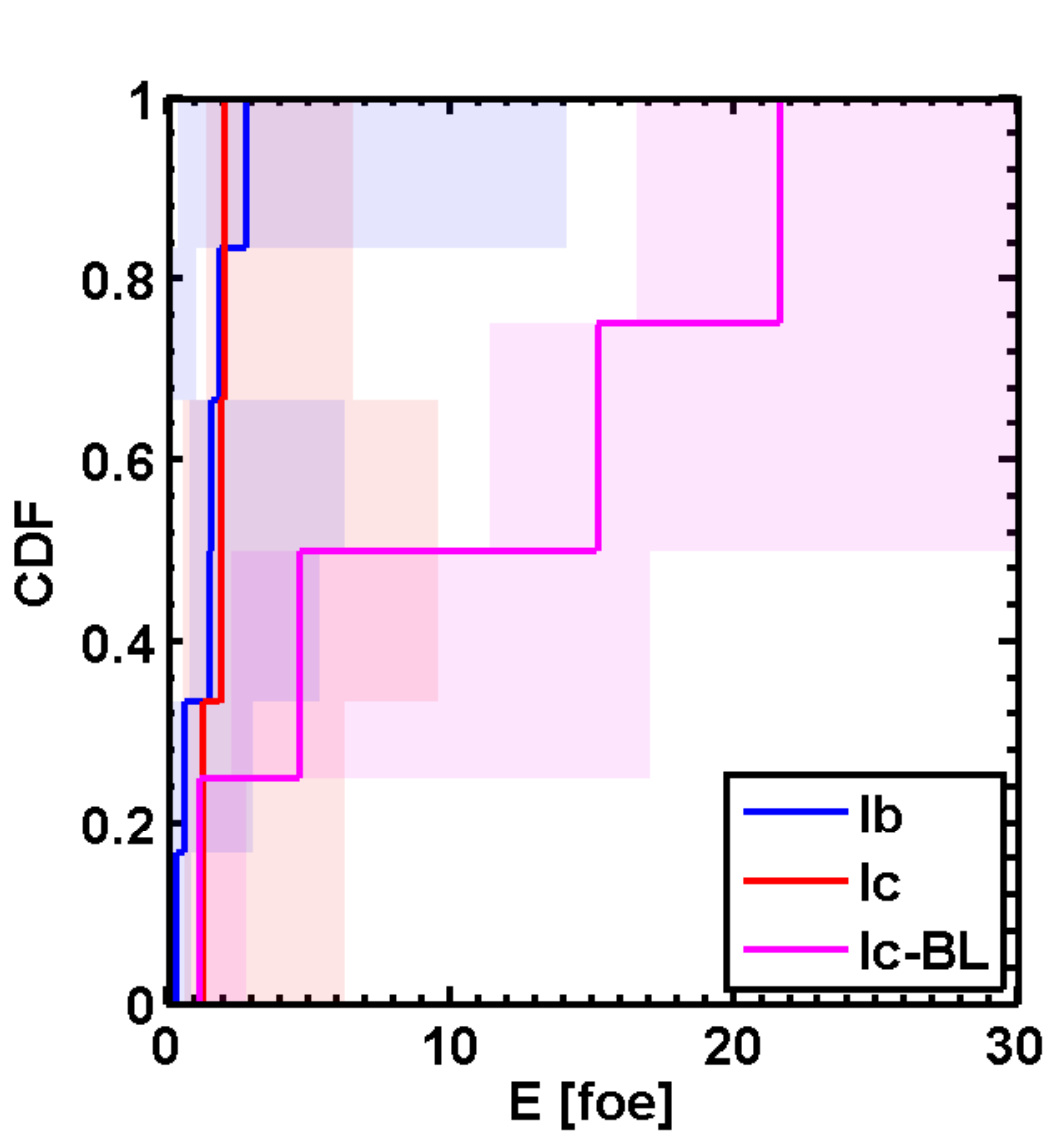}  
\end{array}$
 \caption{\label{Mni} ({\it Lefthand panel:}) $^{56}$Ni mass cumulative distributions for SNe~Ib, Ic, and Ic-BL observed at peak. The comparison reveals higher $^{56}$Ni masses for SNe~Ic-BL (host extinction included). Shaded areas indicate the uncertainties for each SN, caused by the error on the explosion day. ({\it Central panel:}) Ejecta mass (M) cumulative distributions for SNe~Ib, Ic, and Ic-BL. SNe~Ic and Ic-BL appear more massive than SNe~Ib, but the uncertainties are large.
 ({\it Righthand panel:}) Explosion energy (E$_{K}$) cumulative distributions for SNe~Ib, Ic, and Ic-BL. SNe~Ic-BL are clearly more energetic. We compare our results to those of \citet{cano13} and \citet{drout11} in Sect.~\ref{sec:progenitors}}.
\end{figure*}

The amount of $^{56}$Ni synthesized during the explosion, the ejecta mass (M$_{ej}$), and the explosion 
energy (E$_{K}$) of our SNe can be estimated by fitting the Arnett model \citep{arnett82} for SNe~I (see, e.g., equation~1 in \citealp{cano13}) to their bolometric light curves. This model can be applied in the photospheric phase, which is the case for our sample. Therefore we built optical$+$near-infrared (NIR) bolometric light curves by adding a near-infrared contribution to our host-extinction-corrected quasi-bolometric light curves (see Fig.~\ref{boloHEno_fit} and Sect.~\ref{sec:bolonoext}). This contribution is computed from the fraction of NIR ($\lambda>$~9000~\AA) BB flux to the total BB flux (redward of 3500~\AA) given the BB temperatures computed in Sect.~\ref{sec:temp} and shown in Fig.~\ref{T_noext}. At early times this correction is small for most SNe (the optical flux is $\sim$70$-$80\%), whereas after maximum light, the NIR contribution rises in importance as the temperature decreases. 
We also fixed the E$_{K}$/M$_{ej}$ ratio for each event by using the photospheric velocities at maximum light (v$_{ph}^2~=~\frac{10}{3}$E$_{K}$/M$_{ej}$, assuming the SN ejecta to be an expanding sphere with constant density) computed in Sect.~\ref{sec:radius} and shown in the righthand panel of Fig.~\ref{R_noext_oneplot}.
In the Arnett formula we adopted $^{56}$Ni and $^{56}$Co decay times ($\tau_{\rm Ni}$ and $\tau_{\rm Co}$), energy releases per second per gram ($\epsilon_{\rm Ni}$ and $\epsilon_{\rm Co}$), opacity ($\kappa~=$~0.07~cm$^{2}$~g$^{-1}$,
which is appropriate when the electron scattering dominates \citep[e.g.,][]{chevalier92}) and constant of integration ($\beta$) as in \citet{cano13}.
The assumed explosion dates ($t_{\rm expl}$) for each SN were those estimated and discussed in Sect.~\ref{sec:explo_error}.

The best fits to the bolometric light curves (shown in Fig.~\ref{boloHEno_fit}) give the values for $^{56}$Ni mass, M$_{ej}$, and E$_{K}$ that are listed in Table~\ref{tab:Mni}. The quoted uncertainty is due to the precision on the explosion date estimate. 
When we compare SNe~Ib and Ic, we do not find any statistically significant difference in the amount of ejected $^{56}$Ni (Fig.~\ref{Mni}, lefthand panel). On the other hand, SNe~Ic-BL appear richer in $^{56}$Ni than both SNe~Ib and Ic. When we compare (see Fig.~\ref{Mni}) M$_{ej}$ among SNe~Ib, Ic, and Ic-BL, we do not find any statistically significant difference. (We only have a few SNe~Ic, and the uncertainties are quite large.) However, SNe~Ic-BL and SNe~Ic appear slightly more massive than SNe~Ib.
The comparison of E$_{K}$ shows that this parameter is clearly larger for SNe~Ic-BL than for SNe~Ic and Ib. 
We compare our best estimates to those found by \citet{drout11} and \citet{cano13} in Sect.~\ref{sec:discussion}.

\subsubsection{Early plateau phase}
\label{sec:earlyplateau}

The $^{56}$Ni mass estimated in Sect.~\ref{sec:NiEM} can either be mixed out in the ejecta to the outermost layers, or be distributed more centrally. 
The mechanism that brings $^{56}$Ni near the ejecta surface is not clearly established, but jets might be a possible explanation (e.g., \citealp{maund07a,maund07b,maund09,couch09,couch11}). We note that the use of the word ``mixing" to define the presence of $^{56}$Ni in the outermost ejecta might erroneously suggest that this is the effect of small scale, quasi-homogeneous mixing, rather than large-scale plumes. Therefore, to study the $^{56}$Ni mixing is important for understanding the explosion mechanism of SN progenitors.

\begin{figure*}
\centering
$\begin{array}{ccc}
\includegraphics[width=5.7cm,angle=0]{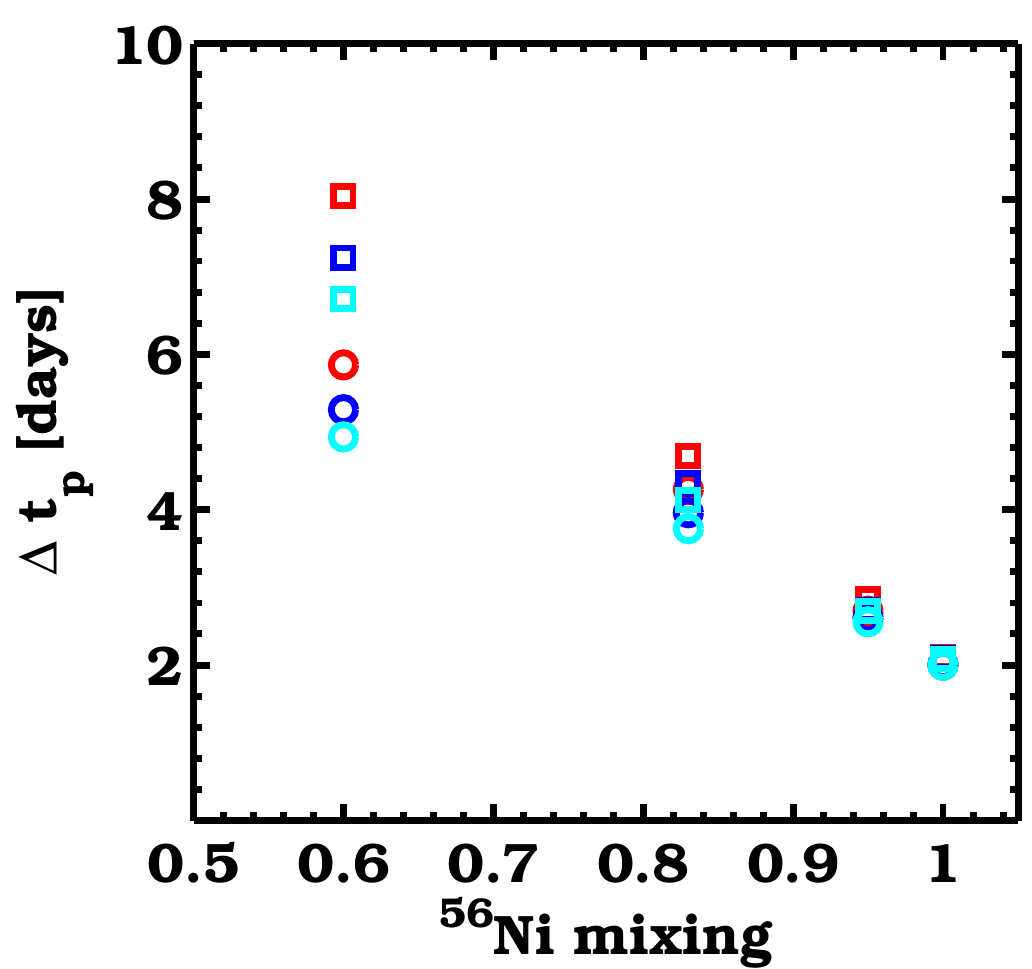}   &
\includegraphics[width=5.7cm,angle=0]{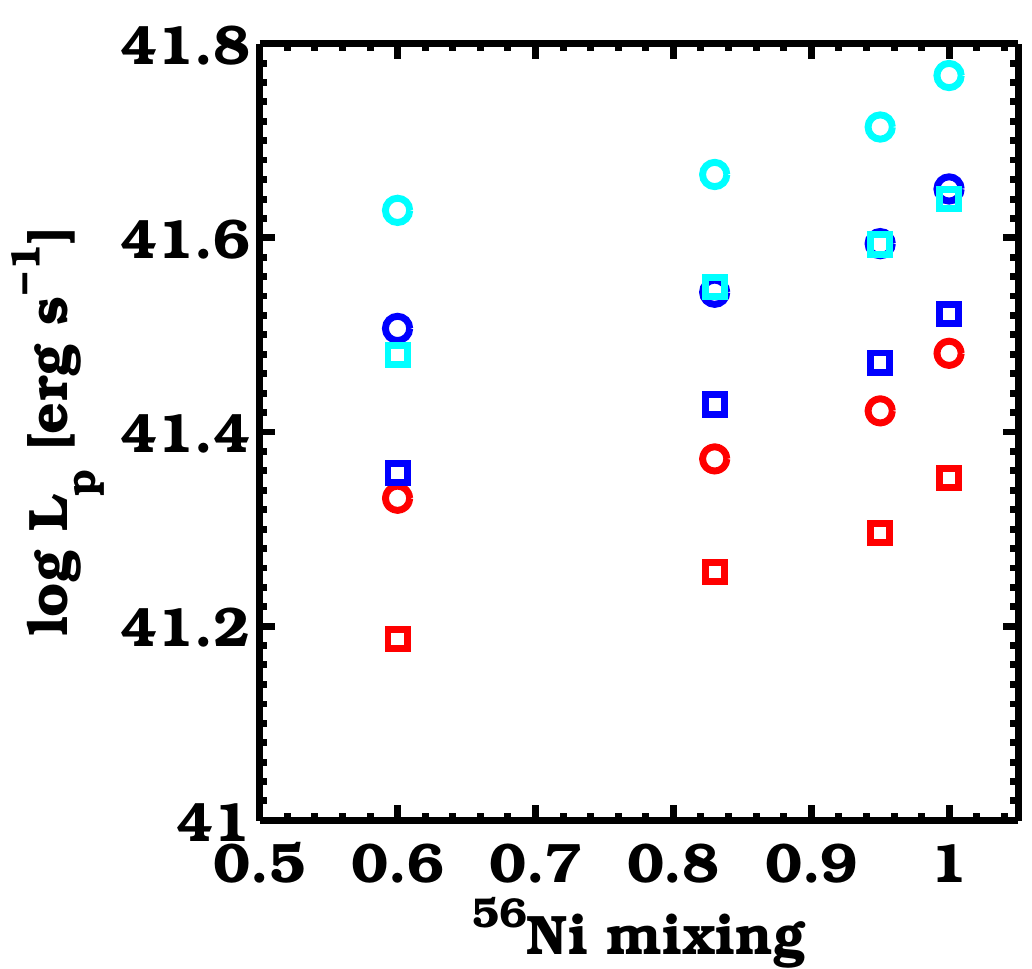}    &
\includegraphics[width=5.7cm,angle=0]{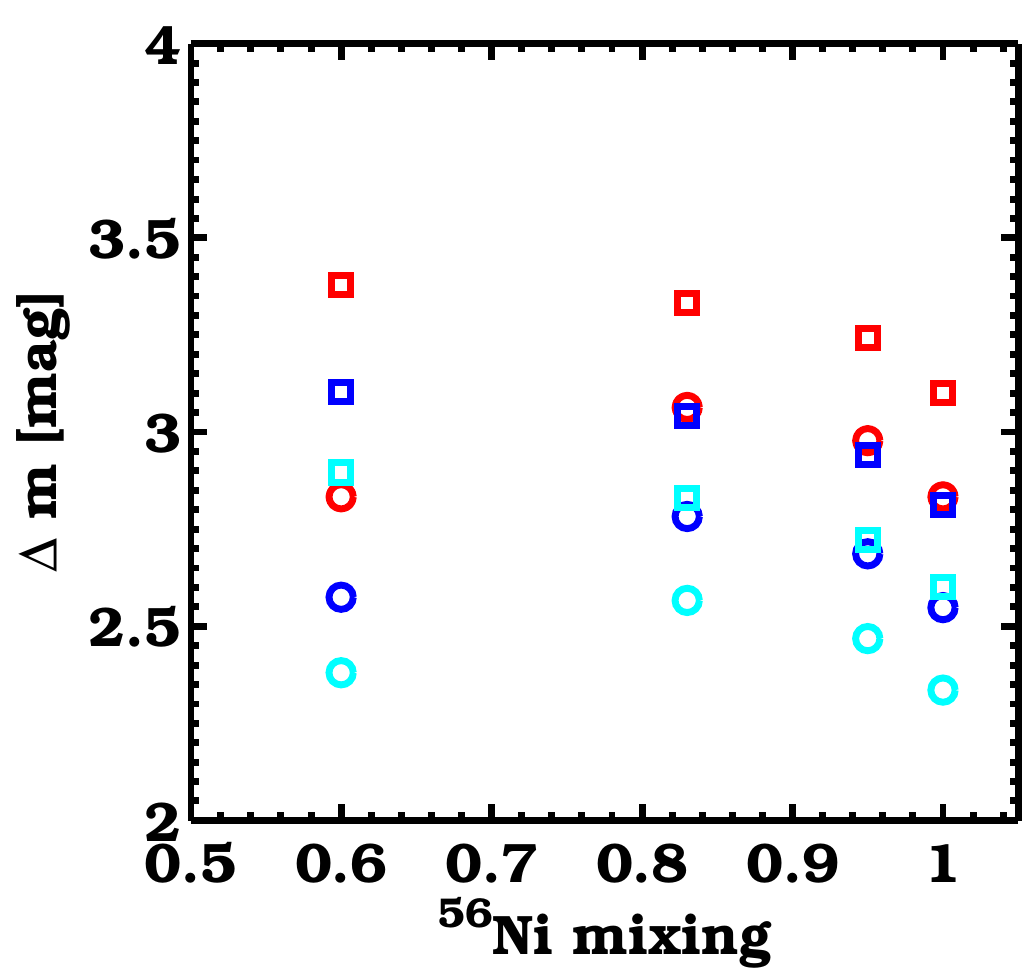}  
\end{array}$
 \caption{\label{melina_tp}Plateau length ($\rm \Delta t_p$), luminosity ($\rm L_p$), and difference in magnitude ($\Delta M$) between $L_p$ and the luminosity peak as functions of $^{56}$Ni mixing ($^{56}$Ni mass is mixed out to 0.60, 0.83, 0.95, and 1.00 of the total progenitor mass). Red, blue, and cyan symbols correspond to models with E~$=$~1,1.5,2~foe, respectively. Circles and squares correspond to models with ejecta mass and progenitor radius M$_{ej}$~$=$~1.7/R~$=$~2.5~R$_{\odot}$ and 2.2~M$_{\odot}$/R~$=$~2.3~R$_{\odot}$, respectively.}
\end{figure*}

The different degrees of mixing should affect the shape of the light curve, 
in particular in the early phases. 
For instance, the presence of an initial, low-luminous plateau between the shock breakout cooling tail and 
the rise of the light curve was noted as an outcome from hydrodynamical models of compact stars
by \citet[][their figure~1]{dessart12_OriginIbc}. 
They found that the duration of the initial plateau
was around ten days with a luminosity L$_{\rm p}$~$=$~10$^{40.6-41.3}$~erg~s$^{-1}$. 
Such a long plateau phase was produced because the $^{56}$Ni was assumed to be located close to the centre.
PN13 present a discussion (and a semi-analytical formula that describes the luminosity
and duration of this early plateau), which illustrates
that a deep $^{56}$Ni distribution leads to a long (a few days) early plateau, whereas a shallow 
(strongly mixed) $^{56}$Ni distribution 
produces a light curve that rises soon after shock breakout.
\citet{bersten13} confirm the presence of an initial plateau in compact stars and
studied the effect of $^{56}$Ni mixing applied to the particular
case of SN~2008D. That analysis concludes that $^{56}$Ni mixing is the main factor that determines the plateau duration.  
Figure~\ref{melina_tp} shows some results for the early plateau properties, which were obtained with the hydrodynamical code of \citet{bersten11}. We calculated a set of light curves for two 
initial models with ejecta masses and progenitor radii of 1.7~$M_{\odot}$, 2.5~R$_{\odot}$, and 
2.2~$M_{\odot}$, 2.3~R$_{\odot}$, respectively. We adopted a $^{56}$Ni mass of 0.2~$M_{\odot}$ in all the calculations. 
For each model we assumed four different degrees of $^{56}$Ni mixing, 0.60, 0.83, 0.95, and 1.00 of the progenitor mass ($^{56}$Ni is linearly distributed in mass coordinates, see \citealp{bersten12}) and three different explosion energies, E$_{K}$~$=$~1, 1.5, and 2~foe (1~foe~$=$~10$^{51}$~erg).
These assumed values for the $^{56}$Ni mass, E$_{K}$, and M$_{ej}$ are consistent with our findings in Sect.~\ref{sec:NiEM} and with those available in the literature for SNe~Ib/c \citep{drout11,cano13}. All models predict an initial plateau whose duration depends mainly on the mixing of $^{56}$Ni (lefthand panel of Fig.~\ref{melina_tp}), with short plateaus for strongly mixed models and long-duration plateaus for models with weak $^{56}$Ni mixing.
 The early plateau duration range is around two to ten days, 
with durations shorter than three to four days, implying an almost full $^{56}$Ni mixing ($\gtrsim$83--95\%).
 The luminosity of the plateau depends mainly on the explosion energy: the higher the energy, the brighter the plateau. There is also dependence on the ejecta mass: the higher the mass, the fainter the plateau. The difference in magnitude between peak and plateau luminosity ($\Delta$M) is large for low energy and/or for high mass.
In our models, the predicted range of plateau luminosity is $\sim$10$^{41.2-41.8}$ erg~s$^{-1}$ (central panel of Fig.~\ref{melina_tp}), and the $\Delta$M range is 2.3--3.4~mag (righthand panel of Fig.~\ref{melina_tp}).
 
\begin{figure*}
\centering
$\begin{array}{cc}
\includegraphics[width=8cm]{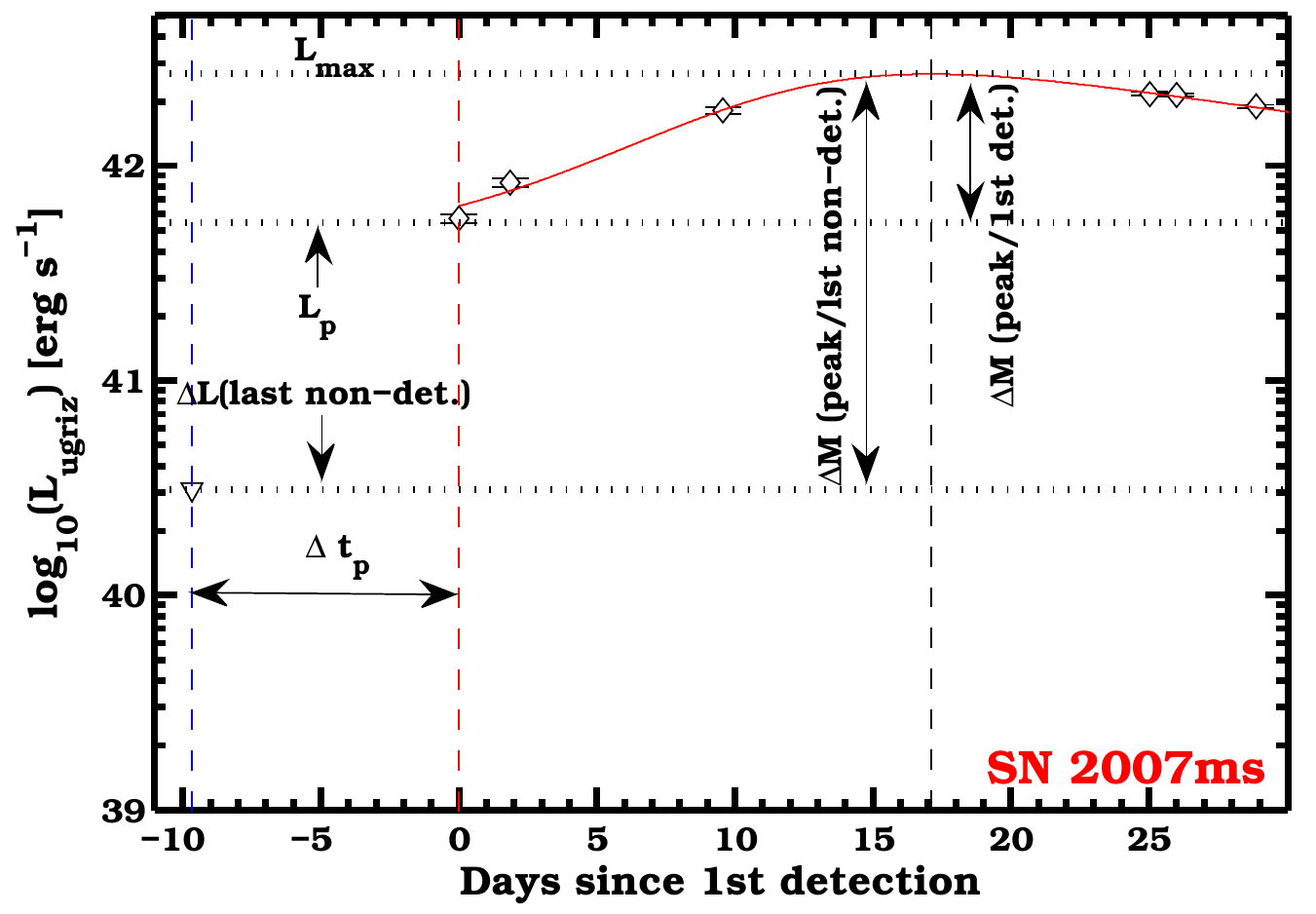}  
\includegraphics[width=8cm]{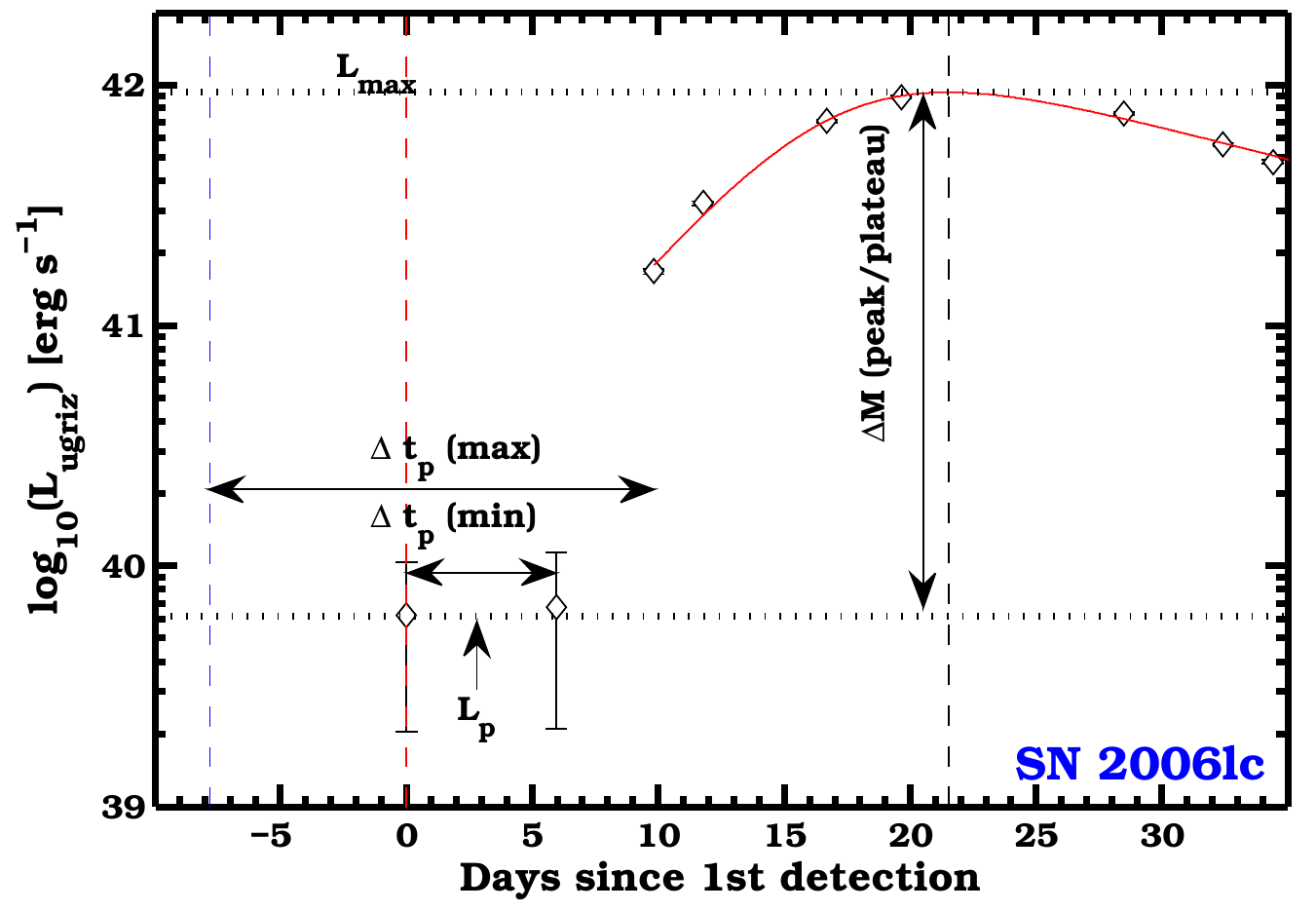} 
\end{array}$
 \caption{\label{param_plateau}Early-time pseudo-bolometric light curves (host extinction neglected) of SN~2007ms and SN~2006lc, where we label the parameters discussed in Sect.~\ref{sec:earlyplateau}.}
\end{figure*} 
 
\subsubsection{$^{56}$Ni mixing} 
\label{sec:nimix}

Since the early plateau phase is such an important diagnostic for the amount of $^{56}$Ni mixing, we studied the quasi-bolometric light curves of the SDSS~SNe~Ib/c to search for evidence of this phase.
Since we did not find any clear early plateau emission except in a single case (see below),
we examined further whether our observations had deep enough limits on luminosity and high enough cadence to actually detect this plateau.

Given the expected temperature
at the plateau phase ($\sim$7000$-$8000~K, \citealp{dessart12_OriginIbc}), the 
quasi-bolometric (optical) luminosities 
should range between $\sim$10$^{41.0-41.6}$ erg~s$^{-1}$ (assuming BB emission and the bolometric luminosity range from our models).
Table~\ref{tab:plateau} reports the last non-detection luminosity 1$\sigma$ limits ($\rm \Delta L_{last~non-det.}$). 
It is clear that we would have been able to detect a plateau at the expected luminosity range in most 
of our objects, since most of the detection limits are lower than the expected values provided by the models 
(7 out of 13 are $\le$~10$^{41}$~erg~s$^{-1}$). 

We can also put limits on the plateau duration ($\rm \Delta t_p$, see Table~\ref{tab:plateau}), assuming this is equal to
the time interval between the last non-detection (which is shown to be robust, see Sect.~\ref{sec:explo_error}) and the discovery of the SN, as exemplified in the lefthand panel of Fig.~\ref{param_plateau} for the case of SN~2007ms. 
In the specific cases of SN~2005hm and SN~2007qx, the limits on the duration of the plateau are set by the time interval between the first(second) and the second(fourth) detections, since these two SNe were very likely observed at the epoch of the shock breakout cooling tail (see Sect.~\ref{sec:radius}). Figure~\ref{param_plateau} also reports the other plateau parameters discussed in this section and presented for each SN in Table~\ref{tab:plateau}.
In three cases (SNe~2005ks, 2006qk, 2007qx) $\rm \Delta t_p$~$<$~2~days. Two of these objects are SNe~Ic, and the other is a SN~Ic-BL. Two additional SNe~Ic-BL (2005kr, 2006nx) show $\rm \Delta t_p$~$<$~3~days, SN~Ic 2007sj exhibits $\rm \Delta t_p$~$<$~4~days, and the SN~Ic-BL 14475 has a plateau duration that can be constrained to $\rm \Delta t_p$~$<$~4.4~days. We thus conclude that within the context of the aforementioned hydrodynamical models, five SNe must have $^{56}$Ni mixed into the outermost layers ($\rm \Delta t_p$~$<$~3~days), and that two more objects show strong $^{56}$Ni mixing (above $\sim$83\% of the progenitor mass). All of these events appear to be helium-poor.

We do not have equally tight constraints on the plateau duration for SNe~Ib, 
whose shortest limit is $\rm \Delta t_p$~$<$~6.8~days (SN 2005hm). 
For SN~Ib~2006lc we can actually observe what we interpret to be an early plateau phase, at faint luminosity 
(see the righthand panel of Fig.~\ref{param_plateau}) and T$_{\rm BB}$ temperature of 8000~K (with host-extinction correction).
 We have two detections 
at $\sim$81.3$\times$10$^{39}$~erg~s$^{-1}$, i.e., at 4.2~mag below peak luminosity. The duration of this plateau ranges from a minimum of 
$\rm \Delta t_p ({\rm min})\approx$5.9 days (the interval between the two detections) 
to a maximum of $\rm \Delta t_p ({\rm max})\approx$17.6 days (last non-detection epoch to first epoch on the rise). 
These results would imply a low degree of $^{56}$Ni mixing for SN~2006lc, lower than $\sim$60\% of the progenitor mass.

We also take this opportunity to review the few examples in the literature for which we know the 
explosion date with considerable accuracy and to investigate whether there is evidence for any early and faint plateau phase 
in these objects. 
Our knowledge of the time of explosion depends on the detection of the shock breakout or of its cooling phase.
SN~2008D \citep{soderberg08,malesani09,cowen10} presents $BVRI$ light curves at early times with a possible plateau lasting for around four days.
SN~1999ex \citep{stritzinger02}, a SN~Ib, shows an early dip in the optical light curves.
There is no clear plateau phase between the early light curve declining phase and the rise: any plateau can therefore not be longer than $\lesssim$2-3~days.
Two other SNe with good pre-explosion limits where the plateau is undetected 
are SN~Ic PTF10vgv \citep{corsi12} and SN~Ic-BL 2012gzk \citep{benami12}.
In the case of SNe associated to GRBs, such as SN~1998bw \citep{galama98} and SN~2006aj 
(\citealp{sollerman06}, \citealp{ferrero07}, \citealp{cowen10}), 
the early afterglow light could mask the possible plateau phase. However, these events are probably highly asymmetric explosions \citep{maeda08}, and we do expect significant mixing of radioactive nickel in such cases, hence a short or non-existent plateau phase as in the strongly-mixed models by \citet{dessart12_OriginIbc}. 
In summary, SNe~Ib/c with known explosion dates in the literature show short ($\lesssim$2$-$4) limits on the plateau duration, again suggesting that their $^{56}$Ni masses are strongly mixed in the ejecta.


The rising part of the light curves can also provide important information on the $^{56}$Ni mixing.
Within the context of the simple models provided by PN13, if we fit a 
PL [$\rm L(t)=A(t-t_{\rm expl}^{PL})^{\alpha}$] to the rising part of the light curves
and find an index $\alpha$~$<$~2, this would mean that the direct radioactive heating dominates, 
with the $^{56}$Ni significantly mixed out in the ejecta.
On the other hand, when the light curve is steeper than $\alpha$~$=$~2, this is interpreted by PN13 as
a diffusive $^{56}$Ni tail affecting its shape, owing to the central location of the $^{56}$Ni.
The contribution of the radioactive diffusive tail is as important as the direct radioactive heating until $\sim$9$-$12 days past explosion, after which it progressively vanishes as the light curve reaches the peak (see fig. 8 in PN13). Since it is crucial to have early observations to constrain the index $\alpha$, we have fit PLs with $\alpha$ as a free parameter only to those bolometric light curves where the first observed luminosity is at least 2~mag below maximum light. 
The fit was performed on all the prepeak epochs of the bolometric light curves. 
From Fig.~\ref{riseshape} it is clear that $\alpha=2$ is too steep to reproduce the early shape, and it also implies too early an explosion date. (The suggested explosion would occur even ten days before the last non-detection for some SNe.) 
The objects that were fit with $\alpha$ as a free parameter indicate that $\alpha$ is lower than 2, spanning 0.09~$<\alpha<$~1.42. The exceptions are SN~2006lc and SN~2006nx, whose early light curves are reproduced by $\alpha~\gtrsim~2$. In the context of the model by PN13, this means that the $^{56}$Ni is mixed throughout the ejecta in most of these objects. These conclusions also hold when we include the host-extinction corrections on the bolometric light curves.
A PL fit to the early light curve (first 4 days) of iPTF13bvn by \citet{cao13} gave $\alpha~=~1.54$, suggesting that the direct $^{56}$Ni radioactive heating dominates.

The temperature profiles in 
Figs.~\ref{T_noext} and \ref{T_noext_oneplot} 
confirm the $^{56}$Ni mixing in the outer layers of most of our objects. PN13 show that the temperature is expected to increase during the light curve rise if the $^{56}$Ni is centrally located, whereas it should stay roughly constant, or even drop, if the $^{56}$Ni is distributed more uniformly. In our SNe, we always observe constant or decreasing $\rm T_{\rm BB}$ at the prepeak epochs, with the exception of SN~2006lc, which is confirmed to have a steep $^{56}$Ni distribution.

In the literature, \citet{takaki13} and \citet{cano14} have recently estimated a high degree of $^{56}$Ni mixing for SNe~Ib 2012au and 1999dn, respectively.

\subsubsection{Progenitor radius}

We can also use the analytic model by PN13 for the early bolometric light curve plateau (see their equation~5) to put limits on the progenitor radius. The luminosity at the discovery epoch is used as an upper limit to the plateau luminosity, making a conservative assumption. 
We furthermore use values for E$_{K}$ and M$_{ej}$ derived in Sect.~\ref{sec:NiEM} and  listed in Table~\ref{tab:Mni} (for the SNe lacking E$_{K}$ and M$_{ej}$ estimates, the average values of their spectral class were used). 

We take into account that our derived bolometric luminosities only constitute a fraction of the total bolometric luminosities ($\sim$60\% if we assume a plateau temperature of 0.6~eV (6959~K) and BB emission as in PN13), since we integrated between 3500 and 9000~\AA.
The corresponding limits on the progenitor radii are listed in Table~\ref{tab:Mni}. 
Including host-extinction corrections, for SNe~Ic, Ic-BL, and Ib, the tightest upper limits are of 1.6, 6.0, and 1.7~R$_{\odot}$.
In the case of SN 2006lc, where we argue that something similar to the plateau phase is indeed detected, it is necessary to include the large and uncertain host-extinction correction in order to derive a reasonable radius.
 The main conclusion that we obtain from the SNe with the best limits on the plateau luminosity is that these transients arise from compact progenitors, with a size of no more than a few solar radii.

We stress that the limits on the progenitor radius are obtained in the framework of the analytic model by PN13, where the plateau luminosity is fixed to the specific value of the bolometric luminosity powered by the shock cooling when the temperature reaches 0.6~eV. The strong dependence of  L$_{\rm p}$  on the progenitor radius is not reproduced by hydrodynamical models (\citealp{bersten14}). In the future surveys (e.g., ZTF) we will be able to systematically detect the early emission from the shock cooling for SNe~Ib/c and will have better constraints on the explosion dates. These achievements will imply more accurate radius estimates.

Two objects, SNe~2007qx and 2005hm, 
show hints of the cooling that follows the shock breakout in their early light curves. 
An early dip in the light curves was detected for SN~2007qx, which suggests
that the explosion of this SN occurred immediately before the first detection.
The dip is clearly visible in the bluer bands and less pronounced in 
the redder ones, consistent with an initial cooling phase, as observed in, for example,
SN~1999ex \citep{stritzinger02}. Recently \citet{nakar14} have shown that the appearance
of an early maximum in the bluer optical bands, and the absence of it in the redder ones 
can be explained if the progenitor star has most of its mass next to the hydrostatic stellar radius, 
in an envelope whose mass M$_{\rm ext}$ is higher than that of the core M$_{\rm core}$. 
It is possible to estimate the progenitor radius of this object
by fitting the first two points of the quasi-bolometric light curve with the model described in equations 1 and 2 of PN13, assuming E$_{K}$ and M$_{ej}$ from the Arnett model fit presented in Sect.~\ref{sec:NiEM}. 
The quasi-bolometric luminosity model is derived from the total bolometric luminosity and the temperature profile by assuming BB emission and considering only the fraction of flux emitted in the wavelength range 
3500$-$9000~\AA.
 The analysis indicates that the explosion of 
SN~2007qx
occurred $\sim$5.5 days before first detection and that the progenitor radius is 16$-$32~R$_{\odot}$, including host-extinction corrections.

In the case of SN~2005hm, we observe a rather high luminosity at the first epoch (only 2 days after the last non-detection) if compared to the second epoch, which is already clearly on the rising part of the light curve (see Fig.~\ref{fitexample}). Unlike SN~2007qx, the first epoch is bright in all the optical bands.
This behavior is reminiscent of the early light curve of SN~2008D \citep{soderberg08,malesani09,modjaz09}, where a bright cooling phase dominated the light curve for about five days. To explain the fist peak, \citet{bersten13} proposed a scenario where a jet produced during the explosion has deposited $^{56}$Ni-rich material in the outer layers of the ejecta.
\citet{nakar14} suggest that the presence of an early maximum in the $r$ and $i$ bands
might arise from a non-standard stellar progenitor that has an extended low-mass envelope surrounding a more
massive core (M$_{\rm ext}$~$\ll$~M$_{\rm core}$). If we assume that the explosion date occurred exactly between last non-detection and discovery and that our first photometric observation corresponds to the early peak, equation 12 in \citet{nakar14} gives a radius of $\approx$8.3~R$_{\odot}$. Here we adopted energy and core mass from Table~\ref{tab:Mni}, and we corrected the quasi-bolometric luminosity of the first peak to take all of the BB emission into account.

With these two objects we confirm that SN~Ib/c progenitors are relatively compact.

\section{Discussion}
\label{sec:discussion}
In this paper we have presented the first study of a large, homogeneous sample of SNe~Ib/c observed in five optical passbands. 
When we compare absolute luminosities and light curve widths to those of the well studied sample of \citet{drout11}, we find similar trends (see Sect.~\ref{sec:dm15} and \ref{sec:lcpeak}). This suggests that, even though the SDSS-II SN survey was mainly aiming for SNe~Ia, the sample of SDSS SNe~Ib/c is not significantly biased, and we can use it to draw conclusions on the nature of SN~Ib/c progenitors.

\subsection{Implications for the SN progenitors}
\label{sec:progenitors}
\citet{dessart12_OriginIbc} show in their numerical models that the spectral difference between SNe~Ib and Ic might be largely due to different mixing. In SNe~Ib, the $^{56}$Ni could be mixed out to the outermost layers, where its decay can excite the helium lines that define the Type Ib class. In contrast, in SNe~Ic, $^{56}$Ni might be more centrally distributed; this would prevent the excitation of any helium layers and thus suppress the appearance of helium lines in their spectra.
The different degrees of $^{56}$Ni mixing should also have consequences for the light curves, in particular at early times, and thus allow investigation of whether this is the main parameter distinguishing between SNe Ib and Ic.

We have shown that for our SDSS SNe~Ic and for other SNe~Ic in the literature, the limits on the plateau length are consistent with a shallow $^{56}$Ni distribution. Also the early light curve shape and the temperature evolution suggest that $^{56}$Ni is thoroughly mixed in most of our SNe, including SNe~Ic.
Moreover, a strongly mixed SN has a shorter light curve rise time than a poorly mixed one, as reported by \citet{dessart12_natureof}.
Since the light-curve rise times are shorter for helium-poor transients than for SNe~Ib/IIb, one could suspect that
mixing is actually stronger in SNe~Ic. The lack of mixing of radioactive elements in the outer layers of helium does not appear to cause the difference between SNe~Ib and SNe Ic.  
The difference in rise time could also be produced by a larger E$_{K}$/M$_{ej}$ ratio for SNe~Ic than for SNe~Ib, but there is no evidence to support this conclusion (see also \citealt{cano13}).
We conclude that $^{56}$Ni mixing is important for the early light curves of SNe Ib/c, but that the difference in degree of mixing does not play a pivotal role in the spectroscopic difference between SNe~Ib and Ic. 
The driving factor of this difference is therefore more likely a real lack of helium in SNe~Ic \citep[e.g.,][]{wheeler87}. Also \citet{cano14} have recently found a strong mixing for SN~1999dn and two other SNe~Ib, whereas in the single case of  SN~Ic 2007gr, the mixing seems to be poor. Another possible explanation for the absence of helium lines in SNe~Ic can be found in a strongly asymmetric $^{56}$Ni mixing \citep{dessart12_natureof} in these SNe, which would produce the absence of helium lines if the SN is observed under specific viewing angles.

In our work we found that SNe~Ic-BL have more $^{56}$Ni mass and higher energy than SNe~Ib and Ic. These results were also found by \citep{drout11} and \citet{cano13}. Our estimates for the $^{56}$Ni masses are more like those of 
\citet{drout11} than those of \citet{cano13}. For SNe~Ib, Ic, and IcBL, we and \citet{drout11} found 
M($^{56}$Ni)~$=$~0.3, 0.3, and 0.6$-$1.1~M$_{\odot}$, whereas \citet{cano13} estimated M($^{56}$Ni)~$=$~0.2, 0.2, and 0.4~M$_{\odot}$. In terms of ejecta masses, we found M$_{ej}$~$=$~3.6, 5.7, 5.4~M$_{\odot}$ for SNe~Ib, Ic and IcBL. The two other works
estimated M$_{ej}$~$=$~2.0, 1.7, 4.7~M$_{\odot}$ \citep{drout11} and M$_{ej}$~$=$~4.7, 4.5, 5.4~M$_{\odot}$ \citep{cano13}. However, the errors on the ejecta masses are large in all the works, and the differences are not significant. Concerning the explosion energy, 
we obtained E$_{K}$~$=$~1.5, 1.7, 10.7~foe, whereas \citet{drout11} reports E$_{K}$~$=$~1.2, 1.0, 11.0~foe and \citet{cano13} E$_{K}$~$=$~3.3, 3.3, 12.6~foe; i.e. all the works report similar values (within the errors, see Table~\ref{tab:Mni})\footnote{During the refereeing process, a paper by \citet{lyman14} was published with similar results to those of our paper. In particular from their analysis, it emerges that SNe Ic-BL have higher $^{56}$Ni masses and the kinetic energies than those of the other subtypes. Furthermore, SNe~Ib and Ic show very similar explosion parameters, i.e. a likely similarity in progenitors.}.

With the early time light curves, we have been able to constrain the
radius of the SN progenitors. Our limits are based on the analytic model of PN13. For the SNe with the earliest observations and the tightest constraints on the plateau luminosity, we found that these objects are compact (R~$\lesssim$~1.6(Ic), 6.0(Ic-BL), 1.7(Ib)~R$_{\odot}$).
 For the Type Ib SN~2006lc, whose early time plateau was observable, we estimated a radius of $\approx$~1.7~R$_{\odot}$. For SN~Ic~2007qx, the estimate from the possible early shock breakout cooling tail gives a radius of $\approx$~16$-$32~R$_{\odot}$. For SN~Ib 2005hm, the first peak along with the known energy and mass reveal a radius of $\approx$~8.3~R$_{\odot}$.
These radii agree with the values expected for hot and massive (M$_{ZAMS}~>~$~25~M$_{\odot}$) Wolf-Rayet stars, which are stripped of their hydrogen-rich envelopes by strong line-driven winds, but also with the radii of stars that were initially less massive and were stripped by a companion \citep{yoon10}.
In both cases the expected final mass for the progenitors is around 4$-$8~M$_{\odot}$ (the smaller the radius, the higher the final mass). These values are compatible with our findings on the ejecta masses from the Arnett model.
SNe~IIb, like SN~2011dh, have been found to arise from extended ($\sim200$~R$_{\odot}$) yellow supergiant progenitors \citep{maund11,bersten12,ergon14,ergon14b}. 
We note that, using the PN13 model, we have not explored the possibility that, at early times, there is some circumstellar interaction contributing to the light curve. 
This is not unreasonable because SE~SN progenitors should suffer large mass losses before collapse. We have also neglected that a substantial part of the gamma-ray deposition is non-local. This is probably a significant effect when $^{56}$Ni is transported to near the surface. Another assumption is the constant opacity, used in the Arnett model.

\subsection{Outlook for ongoing and future surveys (iPTF, ZTF)}
 The ongoing iPTF survey and, even more, the future ZTF survey will discover and follow hundreds of 
new SE~CC~SNe. These surveys aim to scan large portions of the sky with high cadence in order to find young transients. The same fields will be imaged multiple times per night, obtaining exceptionally dense light curve sampling. The study of the early light curves of SE~CC~SNe will obviously benefit from these observations. 

Given the compact nature of their progenitors, SNe~Ib/c are expected to emit a short burst of high-energy (soft $\gamma$-rays, X-ray, see \citealp{nakar10}) radiation at the moment of shock breakout, which should last no more than $\sim$15~minutes. Thereafter, the cooling will bring the emission into the UV-optical range, which can be detected by iPTF/ZTF. This phase should last at most a few hours, typically less than a day. A cadence of a few hours per field would thus allow the shock breakout cooling tail of such SNe to be systematically detected. These early observations will be crucial for deriving the progenitor radius with good precision (see, e.g., \citealt{bersten11,bersten13}). Our analysis of SDSS-II SNe has demonstrated the importance of observing in several bands in order to constrain the temperature and to construct bolometric light curves that can be compared to theoretical models. iPTF/ZTF have the potential to observe at early times in both the $r$ and $g$ bands and soon will be able to obtain prompt spectra via the SED machine \citep[e.g.,][]{galyam14}.

After one day, if the $^{56}$Ni is not significantly mixed, a plateau phase at relatively low luminosity and with 
optical temperatures should be observable, lasting about four to ten days, according to several theoretical predictions.
A cadence of about a day would be enough to observe at least $\gtrsim$~4$-$10 epochs on the early plateau. 
The detection of the dark plateau phase requires that the observations are deep enough to detect luminosities of $\sim$10$^{41}$~erg~s$^{-1}$ (or M~$\sim~-$13~mag). 
This means limiting apparent magnitudes of $\sim$~22~mag for objects at distances of $\sim$~100~Mpc. 

For iPTF and ZTF, the best strategy is to increase the cadence rather than the individual exposure times to reach
a combined limiting magnitude of $r\sim$23~mag, required
to discover (or rule out) more dark plateaus of SNe~Ib/c. The observations of the early plateau, with better constraints on its duration and luminosity, will be crucial to estimate the $^{56}$Ni mixing. SN~2006lc is an example of what iPTF/ZTF will systematically observe in the near future, making it possible to precisely disentangle a shallow $^{56}$Ni distribution (immediate rise after cooling) from a centrally concentrated $^{56}$Ni distribution (delayed rise after a dark plateau).

 Finally, after the plateau, the bolometric light curve will rise powered by the $^{56}$Ni, reaching its peak after $\sim$15$-$25 days. To also properly estimate other explosion/progenitor parameters, such as E$_{K}$, M$_{ej}$, it will be necessary for future surveys to follow the early discovered SNe~Ib/c until at least a few weeks after peak, to properly fit an analytical or hydrodynamical model and to obtain information on the photospheric velocity evolution from the spectra (see, e.g., \citealp{bersten13}). With our sample we have not explored the late and very late-time epochs (up to 300$-$400 days after maximum), which would be useful for comparing the late-time light curve slopes \citep{clocchiatti96}, thus obtaining information on SN properties, such as the gamma-ray deposition fraction \citep{chatz12}.

High-cadence spectral coverage with high S/N data will also allow for distinguishing between helium-rich and helium-poor SNe with high accuracy. As a matter of fact, premaximum spectra obtained soon after discovery may not allow distinguishing between the two groups, since traces of helium can also be present in the early spectra of SN~Ic \citep{branch03}. In addition, SNe~IIb and Ib are impossible to distinguish if there is insufficient spectral coverage \citep{milisav13}.
In this paper we investigated the differences between SE~CC~SNe sub-classes. Most of our targets were spectroscopically observed before and after maximum, making the classification possible. However, when only limited spectral information was available, the classification was problematic. Future surveys will have to take this into account if they aim to investigate the differences among SE~SN sub-groups better.


\section{Conclusions}
\label{sec:conclusions}

In this paper we have presented a systematic analysis of the SDSS-II SN~Ib/c light curves and an 
extended comparison among the different SE~CC sub-types. Although several of the noted trends are 
not statistically significant for this sample, our work clearly provides valuable hints for future studies 
of SN~Ib/c samples with high cadence. The main results of this work are:

\begin{itemize}
\item{We have enlarged the sample of well-studied SNe~Ib/c, analyz\begin{description}
\item[label] 
text
\end{description}ing the light curves of 20 events (9 SNe~Ib, 6 SNe~Ic, 5 SNe~Ic-BL ).}
\item{All SNe~Ib/c light curves peak first in the bluer and later in the redder optical bands.}
\item{Helium-rich SNe (Ib, IIb) tend to rise to peak in a longer time than helium-poor events (Ic, Ic-BL).}
\item{All the SE~SNe subtypes show similar distributions of $\Delta$m$_{15}$.}
\item{SNe~Ic-BL are brighter than Ib and Ic. The last two show different luminosity distributions only if we neglect host-extinction corrections based on the temperature after maximum (SNe~Ic then appear brighter and hotter).}
\item{$^{56}$Ni mass estimates of SNe~Ib/Ic/Ic-BL give typical values of 0.3, 0.3, 1.1~M$_{\odot}$, respectively. 
Energies are on the order of 1.5$-$1.7~foe for SNe~Ib and Ic and about one order of magnitude larger for SNe~Ic-BL. 
Typical ejecta masses for SNe~Ib/Ic/Ic-BL are 3.6, 5.7, and 5.4~M$_{\odot}$.}
\item{Limits on the early-plateau duration, light curve shape on the rise and early temperature evolution imply strong $^{56}$Ni mixing for most SNe~Ic and Ic-BL. Therefore, the cause of the spectral difference between SNe~Ic and Ib subtypes is 
not explained well by nickel mixing, but is understood
better in the context of a real lack of helium in SN~Ic progenitors. We note that the number of SNe~Ic is still rather small.}
\item{We detect an early plateau in the case of SN~Ib~2006lc and radiation from the shock cooling tail in the early light curves of SNe~2005hm and 2007qx.}
\item{The best limits from PN13 model on the progenitor radii suggest that SNe~Ib, Ic, and Ic-BL arise from compact objects.}
\item{Future surveys will observe numerous young SE~SNe with high cadence. We discuss the importance of obtaining multiband photometry and spectra (until a few weeks after peak) in order to constrain the temperature and the photospheric velocity and thus the most important progenitor properties (R, $^{56}$Ni mixing, E$_{K}$, M$_{ej}$). Deep photometry ($r\sim$~23~mag) is needed to detect early time plateaus. Good spectral coverage is required to allow distinguishing different SN sub-classes.}
\end{itemize}

\begin{acknowledgements}
We thank Melina C. Bersten for providing us with the data from her hydrodynamical models, which allowed us to quantify the
properties of the early-time plateau as a function of $^{56}$Ni mixing, explosion energy, and ejecta mass.
  The Oskar Klein Centre is funded by the Swedish Research Council.
F. T. and M. D. S. acknowledge funding provided by the Instrument Center for Danish Astrophysics (IDA).
M.~D. S.  gratefully acknowledges generous support provided by the Danish Agency for Science and Technology and Innovation  realized through a Sapere Aude Level 2 grant. 
Support for L. G. is provided by the Ministry of Economy, Development, and Tourism's Millennium Science Initiative through grant IC12009, awarded to The Millennium Institute of Astrophysics, MAS. L. G. acknowledges support by CONICYT through FONDECYT grant 3140566. The Dark Cosmology Centre is funded by the Danish National Research Foundation. 
Funding for the creation and distribution of the SDSS and SDSS-II has been provided by the Alfred P. Sloan Foundation, the Participating Institutions, the National Science Foundation, the U.S. Department of Energy, the National Aeronautics and Space Administration, the Japanese Monbukagakusho, the Max Planck Society, and the Higher Education Funding Council for England. The SDSS Web site is \verb9http://www.sdss.org/9.
  The SDSS is managed by the Astrophysical Research Consortium for the
  Participating Institutions. The Participating Institutions are the American
  Museum of Natural History, Astrophysical Institute Potsdam, University of
  Basel, Cambridge University, Case Western Reserve University, University of
  Chicago, Drexel University, Fermilab, the Institute for Advanced Study, the
  Japan Participation Group, Johns Hopkins University, the Joint Institute for
  Nuclear Astrophysics, the Kavli Institute for Particle Astrophysics and
  Cosmology, the Korean Scientist Group, the Chinese Academy of Sciences
  ({\small LAMOST}), Los Alamos National Laboratory, the Max-Planck-Institute
  for Astronomy ({\small MPIA}), the Max-Planck-Institute for Astrophysics
  ({\small MPA}), New Mexico State University, Ohio State University,
  University of Pittsburgh, University of Portsmouth, Princeton University,
  the United States Naval Observatory, and the University of Washington.
  The Hobby-Eberly Telescope HET is a joint project of the University of
  Texas at Austin, the Pennsylvania State University, Stanford University,
  Ludwig-Maximillians-Universit\"at M\"unchen, and Georg-August-Universit\"at
  G\"ottingen.  The HET is named in honor of its principal benefactors,
  William P. Hobby and Robert E. Eberly. 
  The Subaru Telescope is operated by the National
  Astronomical Observatory of Japan.  
  The William Herschel Telescope is
  operated by the Isaac Newton Group, and the Nordic Optical Telescope is
  operated jointly by Denmark, Finland, Iceland, Norway, and Sweden, both on
  the island of La Palma in the Spanish Observatorio del Roque de los
  Muchachos of the Instituto de Astrofisica de Canarias.  
  Observations at the
  {\small ESO} New Technology Telescope at La Silla Observatory were made
  under program {\small ID}s 77.A-0437, 78.A-0325, and 79.A-0715.  
  Kitt Peak
  National Observatory, National Optical Astronomy Observatory, is operated by
  the Association of Universities for Research in Astronomy, Inc. ({\small
    AURA}) under cooperative agreement with the National Science Foundation.
This research made use of the ``K-corrections calculator'' service available at http://kcor.sai.msu.ru/.
This research has made use of the NASA/IPAC Extragalactic Database (NED) which is operated by the Jet Propulsion Laboratory, California Institute of Technology, under contract with the National Aeronautics and Space Administration.  
\end{acknowledgements}

\bibliographystyle{aa}  

 \onecolumn

\onltab{1}{
\clearpage
\begin{deluxetable}{lcclcrcccc}
\tabletypesize{\scriptsize}
\tablewidth{0pt}
\tablecaption{SDSS sample of 20 SNe~Ib, Ic and Ic-BL.\label{sample}}
\tablehead{
\colhead{SN} &
\colhead{RA (J2000)} &
\colhead{DEC (J2000)} &
\colhead{Type} &
\colhead{Redshift} &
\colhead{Distance} &
\colhead{Host galaxy} &
\colhead{$E(B-V)_{\rm MW}$}&
\colhead{$E(B-V)_{\rm host}$\tablenotemark{*}}&
\colhead{$M_g^{gal}$\tablenotemark{**}} \\
\colhead{} &
\colhead{(hh:mm:ss)} &
\colhead{(dd:mm:ss)} &
\colhead{} &
\colhead{} &
\colhead{(Mpc)} &
\colhead{} &
\colhead{(mag)} &
\colhead{(mag)} &
\colhead{(mag)}}
\startdata
2005fk & $21:15:19.84$ & $-00:22:58.6$ &Ic-BL & 0.264  & 1341.1 & A211519-0022              &0.054 &\ldots &-19.90\\  
2005hl & $20:55:19.79$ & $+00:32:34.7$ &Ib    & 0.023  & 100.4  & A205519+0032              &0.073 &0.533 &-20.23 \\  
2005hm & $21:39:00.65$ & $-01:01:38.7$ &Ib    & 0.035  & 151.6  & A213900-0101              &0.048 &     0&-15.51 \\  
2005kr & $03:08:29.66$ & $+00:53:20.2$ &Ic-BL & 0.134  & 627.1  & A030829+0053              &0.087 &     0.075&-17.57 \\  
2005ks & $21:37:56.56$ & $-00:01:56.9$ &Ic-BL & 0.099  & 451.5  & A213756-0001              &0.050 &0.537 &-19.39 \\  
2005mn & $03:49:18.44$ & $-00:41:31.4$ &Ib    & 0.047  & 209.2  & A034918-0041              &0.166 &\ldots &-18.86 \\ 
2006fe & $20:52:09.10$ & $-00:30:39.3$ &Ic    & 0.070  & 316.3  & SDSS J205209.10-003039.2  &0.098 &0.132 &-20.56 \\  
2006fo & $02:32:38.89$ & $+00:37:03.0$ &Ib    & 0.021  & 89.4   & UGC 2019                  &0.026 &0.301 &-20.31 \\  
14475  & $22:24:30.90$ & $+00:12:12.3$ &Ic-BL & 0.149  & 705.3  & SDSS J222430.86+001212.3  &0.072 &0.650 &-18.05 \\  
2006jo & $01:23:14.72$ & $-00:19:46.7$ &Ib    & 0.077  & 345.8  & A012314-0019              &0.032 &0.460 &-20.81 \\  
2006lc & $22:44:24.48$ & $-00:09:53.5$ &Ib    & 0.016  & 69.7   & NGC7364                   &0.057 &0.510 &-21.20 \\  
2006nx & $03:33:30.63$ & $-00:40:38.2$ &Ic-BL & 0.137  & 641.9  & A033330-0040              &0.108 &0.461 &-19.19 \\  
2006qk & $22:25:32.38$ & $+00:09:15.1$ &Ic    & 0.058  & 259.5  & A222532+0009              &0.075 &\ldots &-17.96 \\ 
2007gl & $03:11:33.21$ & $-00:44:46.7$ &Ib    & 0.028 &  122.6  &  KUG 0309-009   & 0.059 &  \ldots &   -19.51      \\
2007jy & $20:51:21.43$ & $+00:23:57.8$ &Ib    & 0.180  & 869.0  & A205121+0023              &0.095 &\ldots &-19.86 \\ 
2007ms & $20:32:18.34$ & $-01:00:53.1$ &Ic    & 0.039  & 170.9  & A203218-0100              &0.184 &0.040 &-17.76 \\  
2007nc & $00:01:09.30$ & $+01:04:06.5$ &Ib    & 0.087  & 393.9  & A000109+0104              &0.025 &0.227 &-20.21 \\  
2007qv & $22:35:07.91$ & $-01:06:37.5$ &Ic    & 0.095  & 433.5  & A223507-0106              &0.048 &     0&-19.61 \\  
2007qx & $00:27:41.78$ & $+01:13:59.7$ &Ic    & 0.080  & 363.3  & A002741+0113              &0.023 &0.371 &-20.20 \\  
2007sj & $00:10:39.63$ & $-00:03:10.2$ &Ic    & 0.039  & 170.1  & A001039-0003              &0.032 &\ldots &-21.24 \\ 
\enddata
\tablenotetext{*}{See Sect.~\ref{sec:hostext} for details.}
\tablenotetext{**}{The absolute magnitude in $g$ band for each host galaxy is corrected for the Milky Way extinction at the position of the host center and K-corrected.}
\tablecomments{Typical errors on the distance are on the order of 5$-$10\%. Redshifts are usually known
with a precision down to $\delta z~\sim$~0.001 \citep{zheng08}.}
\end{deluxetable}}

\onltab{2}{
\begin{deluxetable}{lllll|lllll}
\tabletypesize{\scriptsize}
\tablewidth{0pt}
\tablecaption{Spectral log for the SDSS SNe~Ib/c.\label{tab:speclog}}
\tablehead{
\colhead{SN}&
\colhead{Type} &
\colhead{Spectral epoch}&
\colhead{Telescope}&
\colhead{Velocity}&
\colhead{SN} &
\colhead{Type} &
\colhead{Spectral epoch}&
\colhead{Telescope}&
\colhead{Velocity}\\
\colhead{} &
\colhead{} &
\colhead{(MJD-53000)}&
\colhead{}&
\colhead{(km~s$^{-1}$)}&
\colhead{}&
\colhead{} &
\colhead{(MJD-53000)} &
\colhead{}&
\colhead{(km~s$^{-1}$)}}
\startdata
2005fk  &Ic-BL  &640.10*  & HET              &  11700 (Si)& 2006lc  &Ib  &  1030.17  & H &  6080 (Si)   \\ 
2005hl  &Ib  &  665.12*  &MGH &  5450 (He) &       &  &  1035.19  & H &  11261 (He)   \\ 
2005hm  &Ib  &  663.19*  & MGH &  9470 (He) &       &  &  1043.14* & H &  9050 (He)    \\ 
2005kr  &Ic-BL  &  696.19*  & HET              &  12200 (Si)&       &  &  1059.13  & H &  8246 (He)    \\ 
2005ks  &Ic-BL  &  696.06*  & HET              &  15500 (Si)&       &  &  1082.08  & ESO-NTT      &  7141 (He)    \\
2005mn  &Ib  &  705.29*    &HET             &11900 (He)           &       &  &  1069.06  & H &              \\ 
      &  &  708.35 &  ARC           &  &       &  &  1072.61  & H &              \\ 
      &  &  711.27   & HET              &  11000 (He)& 2006nx  &Ic-BL  &  1056.19  & NOT          &  15400 (Si)   \\ 
      &  &  767.13   & HET              &           &       &  &  1057.33  & KP         &              \\ 
      &  &  767.13   & HET              &           &       &  &  1058.36  & H &              \\ 
      &  &  767.13   & HET              &           &       &  &  1063.31* & ESO-NTT      &  14190 (Si)   \\ 
2006fe  &Ic  &  994.02*  & ESO-NTT          & 4760 (Si)  &       &  &  1064.32  & ESO-NTT      &  13190 (Si)   \\ 
      &  &  995.26   & SUBARU           & 5340 (Si)  & 2006qk  &Ic  &  1065.11* & ESO-NTT      &  13740 (He)        \\
2006fo  &Ib  &  995.16   & ESO-NTT          & 10480 (He) &       &  &  1072.06  & H &              \\ 
      &  & 1004.40  & KP             & 10080 (He)  &       &  &  1073.07  & H &              \\ 
      &  &  1010.44* & ARC        & 9530 (He) &       &  &  1085.04  & ESO-NTT      &              \\ 
      &  &  1015.38  & H     & 9280 (He)  & 2007jy  &Ib  &  1363.09* & ESO-NTT      &              \\ 
      &  &  1018.41  & H     & 9380 (He)  & 2007ms  &Ic  &  1361.03  & ESO-NTT      & 11410 (He)             \\ 
      &  &  1024.36  & ESO-NTT          & 9030 (He)  &       &  &  1387.05* & ESO-NTT      & 7130 (Si)     \\ 
      &  &  1027.36  & H     & 8680 (He)  &       &  &  1389.04  & ESO-NTT      &              \\ 
      &  &  1031.31  & H     & 8430 (He)  & 2007nc  &Ib  &  1390.16* & ESO-NTT      & 12680 (Si)    \\ 
      &  &  1037.32  & H     & 8030 (He)  & 2007qv  &Ic  &  1413.10* & H                  &              \\  
      &  &  1059.33  & H     &                  &       &  &  1415.09  & H &              \\
      &  &  1064.27  & ESO-NTT & 7380 (He)  & 2007qx  &Ic  &  1417.18* & ESO-NTT      & 11790 (He)             \\
14475 &Ic-BL  &  1023.31* & SUBARU           & 18700 (Si) & 2007sj  &Ic  &  1429.13  & H &      \\ 
2006jo  &Ib  &  1019.30  & H     &           &       &  &  1441.14* & ESO-NTT      &      \\ 
      &   & 1023.26* &  ESO-NTT         & 14360 (He) &       &  &             &              &              \\
\enddata

\tablenotetext{{\rm (He)}}{Velocity measured from \ion{He}{i}~$\lambda$5876.}
\tablenotetext{{\rm (Si)}}{Velocity measured from \ion{Si}{ii}~$\lambda$6355.}
\tablenotetext{*}{Spectra plotted in Fig.~\ref{spec_class}.}
\tablecomments{All the spectra were released by \citet{sako14}. ARC~$=$~Astrophysical Research Consortium 3.5m telescope at the Apache Point Observatory (New Mexico), ESO-NTT~$=$~New Technology 3.6m Telescope in La Silla (Chile), HET~$=$~Hobby-Eberly 9.2m Telescope at McDonald Observatory (Texas), H~$=$~Hiltner 2.4m telescope at the Michigan-Dartmouth-MIT observatory (Arizona), KP$=$Kitt Peak National Observatory Mayall 4m telescope (Arizona), MGH~$=$~McGraw-Hill 1.3m telescope at the Michigan-Dartmouth-MIT observatory (Arizona), NOT~$=$~Nordic Optical 2.5m Telescope in La Palma (Spain), SUBARU~$=$~8.2m Subaru Telescope at the National Astronomical Observatory of Japan (Hawaii).}
\end{deluxetable}}

\onltab{3}{
\begin{deluxetable}{l | ccccc | cccccccc}
\rotate
\tabletypesize{\scriptsize}
\tablewidth{0pt}
\tablecaption{Parameters from the fit of the SDSS Ib/c SN light curves.\label{fitparam}}
\tablehead{
\colhead{SN} &
\colhead{A} &
\colhead{$\tau_{\rm rise}$} &
\colhead{$\tau_{\rm fall}$} &
\colhead{t$_{0}$} &
\colhead{C} &
\colhead{t$_{\rm max}$} &
\colhead{F$_{\rm max}$} &
\colhead{$\Delta$m$_{15}$} &
\colhead{$\Delta$m$_{-10}$} &
\colhead{t$_{\rm rise}$}&
\colhead{t$_{\rm expl}$}&
\colhead{m$_{\rm max}$} &
\colhead{M$_{\rm max}$} \\
\colhead{} &
\colhead{($\mu$Jy)} &
\colhead{(days)} &
\colhead{(days)} &
\colhead{(MJD)} &
\colhead{($\mu$Jy)} &
\colhead{(MJD)} &
\colhead{($\mu$Jy)}&
\colhead{(mag)} &
\colhead{(mag)} &
\colhead{(days)}&
\colhead{(MJD)} &
\colhead{(mag)} &
\colhead{(mag)}}
\startdata
\multicolumn{14}{c}{\bf $\mathbf{u}$ band}\\
2005hm & 113.38(11.15) & 2.54(0.48) & 5.61(1.67) & 53641.53(3.47) & 2.46(1.10) & 53642.04(2.24) & 59.41(5.21) & 1.98(0.44) & 1.38(0.46) & 14.35(5.98) &    &  19.47(0.10)  &                     $-$16.67      \\ 
2006fo & 180.59(27.50) & 2.95(0.17) & 3.88(0.57) & 54004.40(1.54) & 4.68(0.84) & 54000.90(0.34) & 108.93(6.81) & 2.26(0.16) & 0.56(0.20) & 15.66(9.43)&    &  18.81(0.07)  &                     $-$17.57      \\ 
2006lc & 97.47(6.67) & 1.74(0.16) & 3.75(0.73) & 54038.61(0.80) & 5.40(1.29) & 54038.85(0.46) & 54.25(3.16) & 2.21(0.15) & 1.81(0.16) & 23.23(4.14)   &    &  19.56(0.06)  &                     $-$17.46      \\  
2007ms & 86.54(16.93) & 3.90(0.84) & 6.90(4.67) & 54370.38(6.65) & 2.48(1.36) & 54369.33(3.79) & 46.11(8.56) & 1.33(0.53) & 0.61(0.96) & 17.48(19.19) &    &  19.74(0.20)  &                     $-$17.52      \\ 
2007qv & 35.33(7.22) & 0.49(0.33) & 7.41(2.17) & 54410.25(0.78) & 6.66(3.81) & 54411.67(0.68) & 34.36(3.43) & 1.28(0.24) & \ldots     & 5.89(3.16)    &    &  20.06(0.11)  &                     $-$18.36      \\ 
\hline                                                                                                                                                                                                       
\multicolumn{14}{c}{\bf $\mathbf{g}$ band}\\
2005hl & 145.58(13.52) & 5.29(0.31) & 6.73(0.92) & 53634.11(1.86) & 17.99(0.58) & 53627.06(0.62) & 104.64(1.21) & 0.72(0.07) & 0.18(0.05) & \ldots      &  &  18.85(0.01)  &                     $-$18.46       \\ 
2005hm & 123.94(2.81) & 3.21(0.08) & 8.80(0.63) & 53643.65(0.67) & 7.71(0.69) & 53645.49(0.32) & 72.01(0.96) & 1.11(0.05) & 0.94(0.05) & 17.69(1.06)    &  &  19.26(0.01)  &                     $-$16.82       \\ 
2005kr & 35.52(4.74) & 2.22(0.28) & 13.16(2.01) & 53681.39(0.74) & 0.00(0.90) & 53685.41(0.57) & 22.56(1.75) & 1.04(0.15) & 2.19(0.25) & 11.03(1.60)    &  &  20.52(0.08)  &                     $-$19.12       \\ 
2005ks & 20.30(5.60) & 1.83(0.48) & 14.55(5.94) & 53680.48(2.11) & 0.00(0.37) & 53684.38(1.59) & 13.91(2.37) & 0.97(0.36) & 2.97(1.62) & 9.33(3.43)     &  &  21.04(0.18)  &                     $-$19.66       \\ 
2006fo & 437.58(13.43) & 4.44(0.14) & 8.01(0.29) & 54004.47(0.87) & 43.17(0.66) & 54003.49(0.22) & 263.24(4.67) & 0.89(0.02) & 0.43(0.04) & 18.20(9.37) &  &  17.85(0.02)  &                     $-$18.15       \\ 
14475 & 11.85(1.89) & 1.38(0.47) & 13.32(2.61) & 54008.68(0.79) & 0.15(0.31) & 54012.10(1.07) & 8.64(0.78) & 1.07(0.17) & 3.74(3.04) & 8.15(3.34)       &  &  21.56(0.10)  &                     $-$20.80       \\ 
2006jo & 49.47(3.46) & 1.85(0.16) & 6.20(0.67) & 54011.58(0.68) & 2.13(0.41) & 54013.29(0.35) & 29.03(1.27) & 1.80(0.13) & 2.12(0.26) & 10.16(5.68)     &  &  20.24(0.05)  &                     $-$19.44       \\ 
2006lc & 338.98(4.58) & 2.62(0.04) & 11.62(0.33) & 54036.59(0.24) & 0.00(1.17) & 54039.87(0.15) & 198.77(1.80) & 1.13(0.02) & 1.67(0.05) & 24.23(3.96)  &  &  18.15(0.01)  &                     $-$18.20       \\ 
2006nx & 26.11(5.65) & 1.40(0.85) & 8.60(1.20) & 54049.87(1.04) & 4.78(1.85) & 54052.47(0.82) & 21.52(1.16) & 1.04(0.18) & 1.58(0.18) & 11.94(2.00)     &  &  20.57(0.06)  &                     $-$20.87       \\ 
2007ms & 144.64(5.63) & 3.76(0.17) & 11.21(0.77) & 54371.13(0.59) & 7.47(0.89) & 54373.80(0.34) & 83.87(3.45) & 0.88(0.05) & 0.76(0.06) & 21.78(4.95)   &  &  19.09(0.04)  &                     $-$17.91       \\ 
2007nc & 18.91(2.52) & 3.75(0.72) & 12.40(4.12) & 54389.81(4.16) & 0.00(0.66) & 54393.21(1.91) & 10.24(0.78) & 0.93(0.33) & 0.88(0.37) & 14.59(7.32)    &  &  21.37(0.08)  &                     $-$17.63       \\ 
2007qv & 32.16(2.34) & 0.61(0.17) & 32.82(4.35) & 54409.19(0.42) & 0.00(1.59) & 54411.83(0.64) & 29.33(0.59) & 0.48(0.07) & \ldots        & 6.03(3.11)  &  &  20.23(0.02)  &                     $-$18.13       \\ 
2007qx & 19.69(3.04) & 2.55(0.44) & 14.55(4.29) & 54415.30(1.13) & 1.25(0.79) & 54419.57(0.81) & 13.62(0.64) & 0.79(0.22) & 1.39(0.23) & 19.68(5.25)    &  &  21.06(0.05)  &                     $-$18.34         \\ 
\hline                                                                                                                                                                                           
\multicolumn{14}{c}{\bf $\mathbf{r}$ band}\\
2005fk & 16.54(3.75) & 2.80(0.85) & 7.93(3.92) & 53627.85(6.55) & 0.63(0.42) & 53629.99(4.75) & 9.27(1.49) & 1.36(0.71) & 1.25(1.35) & \ldots            &  \ldots         &  21.48(0.17)  &     $-$19.30        \\ 
2005hl & 334.02(3.34) & 4.65(0.16) & 15.26(0.62) & 53627.63(0.46) & 47.51(1.37) & 53631.56(0.17) & 228.14(1.72) & 0.51(0.01) & 0.44(0.03) & \ldots       &  \ldots         & 18.00(0.01)  &      $-$18.67         \\ 
2005hm & 115.86(3.32) & 3.32(0.12) & 14.04(0.84) & 53644.66(0.42) & 16.40(1.02) & 53648.69(0.01) & 83.47(1.68) & 0.64(0.03) & 0.77(0.04) & 20.78(0.96)   &  53627.18(0.99) & 19.10(0.02)  &      $-$16.94      \\ 
2005kr & 44.83(5.11) & 2.34(0.29) & 25.45(6.34) & 53681.50(0.87) & 0.00(0.11) & 53687.58(0.01) & 32.97(1.89) & 0.54(0.11) & 1.76(0.32) & 12.95(1.28)     &  53672.90(1.45) & 20.10(0.06)  &      $-$19.35        \\ 
2005ks & 33.67(4.81) & 3.58(0.52) & 14.84(7.83) & 53687.32(3.12) & 0.00(0.52) & 53691.83(0.01) & 19.38(1.39) & 0.80(0.28) & 0.95(0.39) & 16.11(0.91)     &  53674.12(1.00) & 20.68(0.08)  &      $-$19.33        \\ 
2005mn & \ldots      & \ldots     & \ldots      & \ldots         & \ldots     & \ldots         & \ldots      & \ldots    & \ldots  & \ldots              &  53691.87(4.53) & \ldots       &      \ldots       \\
2006fe & 50.78(2.72) & 3.89(1.25) & 18.36(3.42) & 53979.18(2.89) & 5.17(1.08) & 53984.65(2.74) & 35.47(5.81) & 0.52(0.06) & 0.63(0.70) & \ldots          &  \ldots         & 20.03(0.18)  &      $-$18.12       \\ 
2006fo & 592.21(4.20) & 4.77(0.09) & 12.63(0.44) & 54005.57(0.52) & 107.07(1.75) & 54008.00(0.01) & 412.24(1.99) & 0.53(0.01) & 0.38(0.01) & 22.62(9.32) &  53984.92(9.51) & 17.36(0.01)  &      $-$18.30      \\ 
14475 & 24.49(1.42) & 2.70(0.73) & 19.71(1.82) & 54011.60(1.35) & 0.00(0.01)    & 54017.31(0.01) & 16.42(0.84) & 0.67(0.04) & 1.47(0.50) & 12.69(2.18)   &  54002.73(2.51) & 20.86(0.06)  &      $-$20.62       \\ 
2006jo & 79.24(3.56) & 2.24(0.12) & 8.10(0.69) & 54013.57(0.48) & 4.44(0.52) & 54015.89(0.01) & 48.39(1.59) & 1.35(0.08) & 1.66(0.13) & 12.58(5.55)      &  54002.35(5.98) & 19.69(0.04)  &      $-$19.44        \\ 
2006lc & 733.85(11.82) & 3.36(0.04) & 14.37(0.35) & 54037.99(0.22) & 0.00(0.48) & 54042.04(0.01) & 426.02(4.12) & 0.85(0.02) & 1.07(0.02) & 26.37(3.93)  &  54015.24(1.97) & 17.33(0.01)  &      $-$18.44        \\ 
2006nx & 61.13(4.60) & 3.57(0.19) & 16.16(3.31) & 54051.16(1.25) & \ldots     & 54056.27(0.01) & 36.05(0.72) & 0.74(0.13) & 0.94(0.07) & 15.29(1.33)     &  54038.89(1.51) & 20.01(0.02)  &      $-$20.75          \\ 
2006qk & \ldots      & \ldots     & \ldots      & \ldots         & \ldots     & \ldots         & \ldots      & \ldots    & \ldots  & \ldots              &  54058.15(1.01) & \ldots       &      \ldots          \\ 
2007ms & 165.18(2.91) & 4.02(0.36) & 26.90(1.37) & 54369.48(0.41) & 9.53(2.89) & 54376.75(0.01) & 117.87(3.44) & 0.39(0.02) & 0.58(0.09) & 24.62(4.83)   &  54351.17(5.02) & 18.72(0.03)  &      $-$18.05        \\ 
2007nc & 25.49(2.49) & 3.39(0.48) & 25.13(4.83) & 54387.68(1.43) & 0.00(1.26) & 54394.52(0.01) & 17.17(0.77) & 0.49(0.07) & 0.91(0.21) & 15.79(3.66)     &  54377.36(3.97) & 20.81(0.05)  &      $-$17.90         \\ 
2007qv & 30.32(2.96) & 1.04(0.27) & \ldots & 54409.66(0.44) & 0.00(2.31) & 54414.83(0.01) & 28.61(0.72) & 0.15(0.07) & \ldots & 8.78(2.70)               &  54405.22(2.96) & 20.26(0.03)  &      $-$18.06           \\ 
2007qx & 42.16(9.09) & 4.61(0.74) & \ldots & 54420.04(3.44) & 0.55(1.77) & 54424.39(0.01) & 23.55(1.53) & 0.66(0.36) & 0.58(0.17) & 24.15(4.59)          &  54398.30(4.96) & 20.47(0.07)  &      $-$18.48          \\ 
2007sj & \ldots      & \ldots     & \ldots      & \ldots         & \ldots     & \ldots         & \ldots      & \ldots    & \ldots  & \ldots              &  54414.24(2.00) & \ldots       &      \ldots          \\ 
\hline                                                                                                                                                                                           
\multicolumn{14}{c}{\bf $\mathbf{i}$ band}\\
2005fk & 17.12(4.16) & 2.12(0.85) & 12.41(5.07) & 53625.43(6.43) & 0.67(0.65) & 53629.66(3.87) & 11.51(1.94) & 1.00(0.61) & 1.96(1.12) & \ldots           & & 21.25(0.18)  &                     $-$19.50       \\ 
2005hl & 376.90(5.43) & 5.02(0.29) & 18.95(0.89) & 53628.36(0.58) & 48.04(2.39) & 53633.60(0.24) & 259.41(2.47) & 0.42(0.01) & 0.39(0.04) & \ldots        & & 17.87(0.01)  &                     $-$18.41      \\ 
2005hm & 114.03(3.60) & 3.41(0.23) & 18.09(1.30) & 53645.37(0.43) & 18.57(1.79) & 53650.52(0.30) & 88.86(2.23) & 0.50(0.03) & 0.70(0.06) & 22.55(1.05)    & & 19.03(0.03)  &                     $-$16.97     \\ 
2005kr & 64.93(8.67) & 3.96(0.44) & 15.56(9.31) & 53687.40(2.62) & 0.00(0.00)     & 53692.23(1.08) & 36.83(1.78) & 0.74(0.23) & 0.79(0.15) & 17.04(2.45)  & & 19.98(0.05)  &                     $-$19.36       \\ 
2006fo & 628.76(6.50) & 4.90(0.10) & 19.42(0.82) & 54004.23(0.49) & 113.66(3.86) & 54009.66(0.24) & 471.03(2.22) & 0.38(0.01) & 0.37(0.01) & 24.24(9.38)  & & 17.22(0.01)  &                     $-$18.22        \\ 
14475 & 25.72(1.77) & 3.21(0.52) & 25.69(6.84) & 54013.01(3.33) & 0.00(1.77) & 54020.19(1.75) & 17.65(0.80) & 0.49(0.13) & 0.99(0.39) & 15.20(5.26)       & & 20.78(0.05)  &                     $-$20.23    \\ 
2006jo & 85.60(4.46) & 2.45(0.19) & 8.57(0.96) & 54014.98(0.73) & 5.72(0.80) & 54017.40(0.36) & 52.78(2.32) & 1.22(0.09) & 1.42(0.16) & 13.98(5.69)       & & 19.59(0.05)  &                     $-$19.22       \\ 
2006lc & 819.56(17.18) & 2.83(0.14) & 15.50(0.58) & 54038.27(0.17) & 38.80(8.57) & 54042.59(0.15) & 548.14(10.00) & 0.75(0.02) & 1.23(0.06) & 26.91(3.96) & & 17.05(0.02)  &                     $-$18.34           \\ 
2006nx & 46.91(4.75) & 2.75(0.31) & 32.35(13.88) & 54048.30(0.96) & 0.00(0.20) & 54055.72(0.61) & 35.08(1.10) & 0.41(0.12) & 1.21(0.19) & 14.81(1.69)     & & 20.04(0.03)  &                     $-$20.35         \\ 
2007nc & 30.87(3.52) & 4.33(0.80) & 20.79(5.33) & 54391.71(2.95) & 0.00(1.59) & 54397.99(1.41) & 18.51(1.01) & 0.54(0.12) & 0.63(0.20) & 18.98(5.64)      & & 20.73(0.06)  &                     $-$17.82        \\ 
2007qv & 31.58(1.14) & 0.99(0.17) &\ldots & 54409.72(0.24) & 0.00(0.00)   & 54414.67(3.37) & 29.80(0.85) & 0.16(0.05) & \ldots & 8.63(14.05)              & & 20.21(0.03)  &                     $-$18.07          \\ 
2007qx & 45.70(9.94) & 4.30(1.33) &\ldots & 54420.19(3.49) & 0.00(2.55) & 54423.76(2.27) & 24.48(2.55) & 0.80(1.55) & 0.69(0.35) & 21.61(9.74)            & & 20.43(0.11)  &                     $-$18.27     \\ 
\hline                                                                                                                                                                                                       
\multicolumn{14}{c}{\bf $\mathbf{z}$ band}\\
2005hl & 299.31(14.31) & 4.75(0.66) & 39.43(6.86) & 53623.11(1.27) & 32.91(17.20) & 53632.77(0.75) & 240.14(4.83) & 0.24(0.02) & 0.34(0.08) & \ldots       & & 17.95(0.02)  &                    $-$17.97       \\ 
2005hm & 93.60(9.55) & 4.23(1.18) & 43.47(11.28) & 53640.36(2.24) & 0.00(8.28) & 53650.11(2.00) & 68.03(7.34) & 0.27(0.09) & 0.47(0.41) & 22.15(4.96)      & & 19.32(0.12)  &                    $-$16.65        \\ 
2006fe & 60.21(9.27) & 5.68(2.52) & 38.16(10.74) & 53980.90(11.04) & 0.00(3.03) & 53991.51(8.37) & 39.54(5.04) & 0.27(0.49) & 0.30(4.89) & \ldots          & & 19.91(0.14)  &                    $-$17.95      \\ 
2006fo & 412.58(15.29) & 3.93(0.28) & 34.26(4.91) & 54000.41(1.15) & 141.81(18.95) & 54008.61(0.76) & 430.81(3.52) & 0.22(0.01) & 0.37(0.04) & 23.21(9.90) & & 17.31(0.01)  &                    $-$17.93        \\ 
2006jo & 67.27(13.17) & 2.39(0.78) & 11.41(6.74) & 54012.88(3.87) & 8.87(3.99) & 54016.30(1.68) & 49.15(6.85) & 0.85(0.28) & 1.25(0.61) & 12.96(8.39)      & & 19.67(0.15)  &                    $-$18.83       \\ 
2006lc & 749.56(25.35) & 3.78(0.18) & 23.76(1.52) & 54037.90(0.41) & 0.00(13.44) & 54044.30(0.22) & 483.79(7.35) & 0.50(0.03) & 0.77(0.05) & 28.59(3.98)   & & 17.19(0.02)  &                    $-$17.87          \\ 
2007ms & 174.71(10.99) & 5.32(1.24) & 53.65(5.06) & 54369.26(1.75) & 0.00(6.41) & 54381.45(2.81) & 126.48(7.54) & 0.20(0.02) & 0.27(0.33) & 29.14(12.70)   & & 18.65(0.06)  &                    $-$17.85         \\ 
\hline
\enddata
\tablecomments{A, $\tau_{\rm rise}$, $\tau_{\rm fall}$, t$_{0}$ and C are the parameters included in eq.~\ref{fitformula}. t$_{\rm max}$, F$_{\rm max}$, $\Delta$m$_{15}$, $\Delta$m$_{-10}$, t$_{\rm rise}$ and t$_{\rm expl}$ are defined in Sect.~\ref{sec:fit} and \ref{sec:risetime}.
 m$_{\rm max}$ is the peak apparent magnitude, M$_{\rm max}$ the peak absolute magnitude (host-extinction corrected). The error on M$_{\rm max}$ is about 0.15~mag, which is mainly due to the uncertainty on the distance ($\sim$7\%).}
\end{deluxetable}}

\onltab{4}{
\begin{deluxetable}{llccl}
\tabletypesize{\scriptsize}
\tablewidth{0pt}
\tablecaption{Well-constrained $R$/$r$-band light-curve rise-times and $\Delta$m$_{-10}$ for 4, 13, 10 and 9 SNe~IIb, Ib, Ic and Ic-BL in the literature.\label{tab:risetimeR}}
\tablehead{
\colhead{SN} &
\colhead{Type} &
\colhead{t$_{\rm rise}$} &
\colhead{$\Delta$m$_{-10}$} &
\colhead{Ref.}\\
\colhead{} &
\colhead{} &
\colhead{(days)} &
\colhead{(mag)}&
\colhead{}}
\startdata
1993J    &IIb  & 22.53(3.24)   & 0.77(0.15)   &  \citet{kumar13,richmond94}\\
2008ax   &IIb  & 21.50(0.40)   & 0.63(0.15)   &  \citet{taubenberger11}\\
2011dh   &IIb  & 21.86(0.23)   & 0.61(0.10)   &  \citet{ergon14}\\
2011fu   &IIb  & 26.40(2.90)   & 0.42(0.15)   &  \citet{kumar13}\\
\hline  
1998dt   &Ib   & 17.60(3.00)   & 1.00(0.10)   & \citet{matheson01}\\                               
1999ex   &Ib   & 20.49(0.52)   & 0.38(0.10)   & \citet{stritzinger02};\\
2004dk   &Ib   & 23.75(1.55)   & 0.50(0.10)   & \citet{drout11}       \\
2005hm   &Ib   & 21.06(1.17)   & 0.76(0.03)   & This paper  \\
2006fo   &Ib   & \ldots        & 0.38(0.01)   &  This paper \\
2006lc   &Ib   & 26.41(5.18)   & 1.08(0.02)   &  This paper \\
2007Y    &Ib   & 21.00(0.50)   & 0.67(0.10)   & \citet{stritzinger09} \\
2007nc   &Ib   & \ldots        & 0.80(0.15)   & This paper  \\
2007uy   &Ib   & 21.72(2.50)   & \ldots       & \citet{roy13}, explosion date from modelling of radio data\\
2008D    &Ib   & 20.75(1.00)   & 0.31(0.10)   & \citet{soderberg08,malesani09}   \\
2009jf   &Ib   & 24.50(1.00)   & 0.34(0.10)   & \citet{valenti11}, good pre-explosion limit\\
2011ei   &Ib   & 19.50(2.50)   & 0.69(0.20)   & \citet{milisav13}     \\
iPTF 13bvn & Ib &   18.55(0.70)  &    0.57(0.10)  &\citep{fremling14}\\
\hline                                 
1994I    &Ic   & 10.01(1.02)   & 2.50(0.50)   & \citet{richmond96}\\
2004aw   &Ic   & \ldots        & 0.27(0.05)   & \citet{taubenberger06}\\
2004dn   & Ic  & \ldots        & 0.60(0.15)   & \citet{drout11}   \\
2004fe   & Ic  & \ldots        & 1.20(0.10)   & \citet{drout11}   \\
2007gr   &Ic   & 15.50(3.02)   & 0.73(0.04)   & \citet{hunter09}\\
2007ms   &Ic   & \ldots        & 0.56(0.08)   & This paper\\
2007qv   &Ic   & 8.72(3.08)    & \ldots       & This paper\\
2007qx   &Ic   & \ldots        & 0.62(0.17)   & This paper\\
PTF 10vgv   &Ic   & 11.91(0.85)   & 2.63(0.50)   & \citet{corsi12}\\
2013dk   &Ic   & \ldots        & 1.33(0.20)   & \citet{eliasrosa13}, $\Delta$m$_{-10}$ scaled to $r$ from $V$\\
\hline                                 
1998bw   &Ic-BL& 17.50(0.50)   &0.68(0.15)    & \citet{clocchiatti11}\\
2003jd   &Ic-BL& 16.20(1.00)   & \ldots       & \citet{valenti08}, only error on max epoch\\
2005kr   &Ic-BL& 15.40(1.45)   &1.18(0.16)    & This paper\\
14475       &Ic-BL& 14.36(2.58)   &1.15(0.31)    & This paper\\  
2006aj   &Ic-BL& 12.30(0.50)   &2.40(0.50)    & \citet{ferrero07}, $\Delta$m$_{-10}$ is extrapolated \\
2006nx   &Ic-BL& 14.99(1.41)   &1.10(0.12)    & This paper\\            
2009bb   &Ic-BL& 14.65(1.25)   &1.37(0.20)    & \citet{pignata11}\\
2010bh   &Ic-BL& 8.00(1.00)    &2.00(0.50)    & \citet{bufano12}, $\Delta$m$_{-8}$ \\
PTF 12gzk   &Ic-BL& 20.74(0.62)   &0.45(0.10)    & \citet{benami12} \\
\enddata
\end{deluxetable}}

\onltab{5}{
\begin{deluxetable}{l | ccccc | ccccc | c}
\rotate
\tabletypesize{\scriptsize}
\tablewidth{0pt}
\tablecaption{Parameters from the fit of the 14 SDSS Ib/c SN pseudo-bolometric light curves (host-extinction corrections included).\label{fitparam_bolo}}
\tablehead{
\colhead{SN} &
\colhead{A} &
\colhead{$\tau_{\rm rise}$} &
\colhead{$\tau_{\rm fall}$} &
\colhead{t$_{0}$} &
\colhead{C} &
\colhead{t$_{\rm max}$} &
\colhead{L$_{\rm max}$} &
\colhead{$\Delta$m$_{15}$} &
\colhead{$\Delta$m$_{-10}$} &
\colhead{t$_{\rm rise}$}&
\colhead{$\alpha$} \\
\colhead{} &
\colhead{(10$^{41}$erg~s$^{-1}$)} &
\colhead{(days)} &
\colhead{(days)} &
\colhead{(MJD)} &
\colhead{(10$^{41}$erg~s$^{-1}$)} &
\colhead{(MJD)} &
\colhead{(10$^{41}$erg~s$^{-1}$)}&
\colhead{(mag)} &
\colhead{(mag)} &
\colhead{(days)}&
\colhead{}}
\startdata
\hline
2005hl & 75.24(0.22)  & 5.79(0.09) & 11.75(0.30) & 53628.26(0.39) & 9.37(0.10) & 53628.44(0.10) & 46.99(0.12) & 0.54(0.01) & 0.28(0.01) & \ldots      & \ldots  \\ 
2005hm & 16.27(0.76)  & 3.22(0.19) & 13.04(1.42) & 53641.96(0.94) & 1.77(0.21) & 53645.68(0.55) & 11.08(0.25) & 0.73(0.05) & 0.87(0.09) & 17.87(1.26) & \ldots  \\ 
2005kr & 106.37(11.61)& 1.93(0.47) & 25.79(12.48)& 53680.21(1.02) & 0.00(2.86) & 53685.70(1.23) & 81.56(4.66) & 0.55(0.13) & 2.47(0.97) & 11.29(2.78) & 1.42  \\ 
2005ks & 164.92(25.52)& 4.46(1.30) & 12.12(13.90)& 53687.82(8.73) & 0.00(0.01) & 53690.47(5.29) & 85.43(9.37) & 0.87(2.99) & 0.64(1.41) & 14.88(28.91)& 0.26  \\ 
2006fe & 47.85(3.88)  & 0.37(0.23) & 16.73(2.44) & 53974.58(0.28) & 4.32(0.63) & 53976.10(0.88) & 47.30(3.17) & 0.82(0.09) & 2.60(0.19) & \ldots      & \ldots  \\ 
2006fo & 54.61(0.22)  & 4.51(0.07) & 12.36(0.16) & 54001.54(0.24) & 7.93(0.07) & 54004.09(0.09) & 36.26(0.20) & 0.60(0.01) & 0.45(0.01) & 18.78(9.33) & \ldots  \\ 
14475  & 309.09(6.87) & 8.32(0.33) & 10.18(2.04) & 54028.05(0.49) & 0.00(0.65) & 54013.71(0.63) & 192.16(2.76)& 0.35(0.06) & 0.09(0.76) & 9.56(2.59)  & 0.55  \\ 
2006jo & 170.23(4.16) & 2.38(0.09) & 7.21(0.39)  & 54012.67(0.39) & 10.10(0.54)& 54014.48(0.17) & 100.36(1.64)& 1.43(0.05) & 1.46(0.09) & 11.27(5.58) & \ldots  \\ 
2006lc & 62.70(0.33)  & 2.74(0.01) & 13.66(0.13) & 54036.58(0.05) & 0.93(0.23) & 54040.43(0.02) & 38.91(0.16) & 0.91(0.00) & 1.45(0.01) & 12.99(1.93) & 1.81  \\ 
2006nx & 464.77(19.19)& 2.48(0.49) & 12.13(2.32) & 54050.21(0.71) & 37.51(7.55)& 54054.04(0.46) & 317.56(3.58)& 0.89(0.11) & 1.38(0.18) & 13.32(1.54) & 3.58  \\ 
2007ms & 42.32(1.15)  & 3.59(0.22) & 17.65(1.28) & 54368.79(0.42) & 4.67(0.59) & 54373.88(0.30) & 30.22(1.06) & 0.54(0.04) & 0.71(0.07) & 21.86(4.92) & \ldots  \\ 
2007nc & 37.43(4.90)  & 3.12(0.92) & 4.80(5.67)  & 54400.04(4.49) & 6.14(3.27) & 54397.93(1.81) & 25.74(3.28) & 1.19(0.31) & 0.53(0.12) & 18.93(6.92) & 0.09 \\ 
2007qv & 36.21(3.54)  & 0.70(0.25) & 52.79(15.17)& 54409.36(0.55) & 0.00(2.35) & 54412.67(0.93) & 33.74(1.23) & 0.29(0.10) & 10.57(3.49)& 6.80(3.56)  & \ldots  \\ 
2007qx & 65.95(7.89)  & 2.94(0.40) & 11.35(2.50) & 54417.48(1.37) & 4.53(1.89) & 54420.81(0.64) & 41.75(1.19) & 0.92(0.20) & 1.10(0.15) & 20.84(5.00) & 1.08  \\ 
2007sj & \ldots       & \ldots     & \ldots      & \ldots         & \ldots     & \ldots         & \ldots      & \ldots     & \ldots     & \ldots      & 1.01  \\
\hline
\enddata
\tablecomments{A, $\tau_{\rm rise}$, $\tau_{\rm fall}$, t$_{0}$ and C are the parameters included in eq.~\ref{fitformula}. t$_{\rm max}$, F$_{\rm max}$, $\Delta$m$_{15}$, $\Delta$m$_{-10}$ and t$_{\rm rise}$ are defined in Sect.~\ref{sec:fit} and \ref{sec:risetime}. $\alpha$ is the exponent of the best PL fit to the early quasi-bolometric light curves, see Sect.~\ref{sec:explo_error} and Fig.~\ref{riseshape}.}
\end{deluxetable}}

\onltab{6}{
\begin{deluxetable}{llcccl}
\tabletypesize{\scriptsize}
\tablewidth{0pt}
\tablecaption{$^{56}$Ni mass, ejecta mass, explosion energy and progenitor radius for the SDSS sample of SNe~Ib/c.\label{tab:Mni}}
\tablehead{
\colhead{SN} &
\colhead{Type}&
\colhead{M$_{^{56}\rm Ni}$} &
\colhead{M$_{ej}$} &
\colhead{E$_{K}$}&
\colhead{R}  \\
\colhead{} &
\colhead{} &
\colhead{(M$_{\odot}$)}&
\colhead{(M$_{\odot}$)}&
\colhead{(foe)}& 
\colhead{(R$_{\odot}$)}}
\startdata

  2005fk & Ic-BL          &  \ldots                 &  \ldots                 & \ldots                      & $<$38.1   \\
 2005hl & Ib             &  0.33$^{+0.05}_{-0.05}$ &  1.56$^{+2.88}_{-1.05}$ & 0.64$^{+2.40}_{-0.43}$       & $<$357.4  \\
 2005hm & Ib             &  0.11$^{+0.01}_{-0.01}$ &  3.45$^{+1.54}_{-0.40}$ & 0.36$^{+0.50}_{-0.04}$       & $<$57.2  \\
 2005kr & Ic-BL          &  0.71$^{+0.04}_{-0.04}$ &  7.75$^{+5.48}_{-1.94}$ & 15.18$^{+28.00}_{-3.80}$     & $<$6.0  \\
 2005ks & Ic-BL          &  0.60$^{+0.02}_{-0.02}$ &  3.39$^{+1.53}_{-0.32}$ & 1.17$^{+1.65}_{-0.11}$       & $<$87.9  \\
 2005mn & Ib             &  \ldots                 &  \ldots             &     \ldots                        & $<$25.4  \\
 2006fe & Ic             &  \ldots                 &  \ldots               &   \ldots                        & $<$116.0  \\
 2006fo & Ib             &  0.38$^{+0.01}_{-0.06}$&  6.04$^{+5.57}_{-2.73}$  & 1.90$^{+1.65}_{-0.86}$       & $<$172.2  \\
 14475  & Ic-BL          &  1.27$^{+0.08}_{-0.09}$ &  2.90$^{+3.38}_{-1.48}$ & 4.71$^{+12.29}_{-2.41}$      & $<$93.1  \\
 2006jo & Ib             &  0.42$^{+0.05}_{-0.08}$ &  2.51$^{+4.98}_{-2.14}$ & 2.83$^{+11.24}_{-2.42}$      & $<$152.1  \\
 2006lc & Ib             &  0.30$^{+0.03}_{-0.03}$ &  3.67$^{+4.96}_{-1.79}$ & 1.60$^{+4.68}_{-0.78}$       & 1.7  \\
 2006nx & Ic-BL          &  1.86$^{+0.12}_{-0.12}$ &  7.52$^{+4.83}_{-1.74}$ & 21.60$^{+37.53}_{-5.01}$     & $<$53.9  \\
 2006qk & Ic             &  \ldots                 &  \ldots                 & \ldots                       & $<$5.1  \\
 2007ms & Ic             &  0.39$^{+0.03}_{-0.03}$ &  9.12$^{+8.45}_{-2.76}$ & 2.05$^{+4.54}_{-0.62}$       & $<$40.7  \\
 2007nc & Ib             &  0.24$^{+0.02}_{-0.03}$ &  4.38$^{+4.87}_{-2.02}$ & 1.54$^{+3.88}_{-0.71}$       & $<$34.4  \\
 2007qv & Ic             &  0.20$^{+0.04}_{-0.03}$ &  1.91$^{+3.81}_{-1.34}$ & 1.92$^{+7.64}_{-1.34}$       & 11.5     \\
 2007qx & Ic             &  0.40$^{+0.15}_{-0.08}$ &  6.21$^{+12.10}_{-3.20}$& 1.28$^{+5.02}_{-0.66}$       & $<$28.7  \\
 2007sj & Ic             &    \ldots               &  \ldots                 & \ldots                       & $<$1.6  \\
  $<$Ib$>$    &          & 0.30$\pm$0.05           &     3.60$\pm$0.63       &  1.48$\pm$0.36               &   \\
 $<$Ic$>$     &          & 0.33$\pm$0.07           &     5.75$\pm$2.09       &  1.75$\pm$0.24               &   \\
 $<$Ic-BL$>$  &          & 1.11$\pm$0.29           &     5.39$\pm$1.30       &  10.66$\pm$4.70              &   \\

\enddata
\tablecomments{The reported errors are due to the uncertainty on the rise time. Typical uncertainties on the $^{56}$Ni mass due to the error on the distance are $\sim$7\%.}
\end{deluxetable}}

\onltab{7}{
\begin{deluxetable}{lccccccc}
\tabletypesize{\scriptsize}
\tablewidth{0pt}
\tablecaption{Limits on the early plateau parameters and peak luminosity values for the SDSS sample of SNe~Ib/c (host extinction included). \label{tab:plateau}}
\tablehead{
\colhead{SN} &
\colhead{Type} &
\colhead{$\rm \Delta t_{p}$} &
\colhead{log$_{10}$($\rm \Delta L_{last~non-det.}$)} & 
\colhead{log$_{10}$($\rm L_p$)} & 
\colhead{log$_{10}$($\rm L_{\rm max}$)}& 
\colhead{$\Delta$M (peak/last non-det.)} &
\colhead{$\Delta$M (peak/1st det.)} \\
\colhead{} &
\colhead{} &
\colhead{(days)} &
\colhead{(log$_{10}$[~erg~s$^{-1}$])} &
\colhead{(log$_{10}$[~erg~s$^{-1}$])} &
\colhead{(log$_{10}$[~erg~s$^{-1}$])} &
\colhead{(mag)}&
\colhead{(mag)}}
\startdata
   2005hl &Ib    &  \ldots                     & \ldots     &   $<$42.63                  & 42.67   &  6.08   &  0.10    \\
   2005hm &Ib    &  $<$6.8\tablenotemark{*}    & $<$40.35   &   $<$41.56                  & 42.04   &  4.23   &  1.21   \\
   2005kr &Ic-BL &  $<$2.6                     & $<$41.57   &   $<$41.95                  & 42.91   &  3.36   &  2.41   \\
   2005ks &Ic-BL &  $<$1.8                     & $<$41.21   &   $<$42.15                  & 42.93   &  4.29   &  1.97   \\
   2005mn &Ib    &  $<$8.6                     & $<$40.99   &   $<$41.80                  & \ldots  &  \ldots &  \ldots   \\
   2006fe &Ic    &  \ldots                     & \ldots     &   $<$42.24                  & 42.67   &  4.26   &  1.10   \\
   2006fo &Ib    &  $<$18.6                    & $<$39.90   &   $<$42.39                  & 42.56   &  6.65   &  0.42   \\
   14475  &Ic-BL &  $<$4.4                     & $<$41.63   &   $<$42.73                  & 43.28   &  4.14   &  1.40   \\
   2006jo &Ib    &  $<$11.1                    & $<$40.99   &   $<$42.75                  & 43.00   &  5.03   &  0.62   \\
   2006lc &Ib    &  5.9$-$17.6                 & \ldots     &   40.91                     & 42.59   &  \ldots &  4.20     \\
   2006nx &Ic-BL &  $<$2.6                     & $<$41.41   &   $<$42.83                  & 43.50   &  5.24   &  1.68   \\
   2006qk &Ic    &  $<$1.9                     & $<$40.98   &   $<$41.18                  & \ldots  &  \ldots &  \ldots   \\
   2007ms &Ic    &  $<$9.7                     & $<$40.51   &   $<$41.81                  & 42.48   &  4.93   &  1.67   \\
   2007nc &Ib    &  $<$7.3                     & $<$41.51   &   $<$41.86                  & 42.42   &  2.27   &  1.42   \\
   2007qv &Ic    &  $<$5.4                     & $<$41.51   &   $<$41.81                  & 42.53   &  2.55   &  1.79   \\
   2007qx &Ic    &  $<$1.8\tablenotemark{**}   & \ldots     &   41.63\tablenotemark{***}  & 42.62   &  \ldots &  2.47  \\
   2007sj &Ic    &  $<$3.8                     & $<$40.53   &   $<$40.78                  & \ldots  &  \ldots &  \ldots   \\
   
\enddata
\tablenotetext{*}{Time interval between the first two epochs.} 
\tablenotetext{**}{Time interval between the second and the fourth epoch.} 
\tablenotetext{***}{Luminosity of the third epoch.} 
\tablecomments{$\rm \Delta t_p$ corresponds to the time interval between last non-detection and first detection in the rest frame, $L_p$ to the luminosity of the 1st detection. The limit on the luminosity of the last non-detection ($\Delta L_{last~non-detec.}$) corresponds to its 1$\sigma$ error. The last two columns ($\Delta$M) report the difference in magnitude between the peak and the last non-detection limit, and the difference in magnitude between peak and the 1st-detection luminosity.}
\end{deluxetable}}



\end{document}